\documentclass[11pt,epsf]{article}
 \usepackage{amsmath}
 \usepackage{graphicx}
 \usepackage[merge,numbers,compress]{natbib}
 \usepackage[T1]{fontenc}
 \usepackage{booktabs}
 \usepackage{xcolor} 
 \usepackage{xspace}
 \usepackage{dcolumn}
 \usepackage{placeins}
 \usepackage[colorlinks=true,citecolor=blue!50!black,linkcolor=black]{hyperref}
 \usepackage{caption}
 \usepackage
 [subrefformat=parens,position=top,skip=-15pt,margin=15pt,justification=justified,singlelinecheck=false]
 {subcaption}

\newcommand{\change}[1]{#1}

\newcommand{\Pl}{\ell}
\newcommand{\fb}{{\ensuremath\unskip\,\text{fb}}\xspace}
\def\refeq#1{\mbox{(\ref{#1})}}

\def\refta#1{\mbox{Table~\ref{#1}}}
\def\reftas#1{\mbox{Tables~\ref{#1}}}
\def\refse#1{\mbox{Section~\ref{#1}}}

\def\citere#1{\mbox{Ref.~\cite{#1}}}
\def\citeres#1{\mbox{Refs.~\cite{#1}}}

\newcommand{\rd}{\mathrm d}

\newcommand{\ie}{\emph{i.e.}\ }

\def\be{\begin{equation}}
\def\ee{\end{equation}}

\newcommand{\PH}{\ensuremath{\text{H}}\xspace}
\newcommand{\Pj}{\ensuremath{\text{j}}\xspace}
\newcommand{\Pp}{\ensuremath{\text{p}}\xspace}
\newcommand{\Pe}{\ensuremath{\text{e}}\xspace}
\newcommand{\Pb}{\ensuremath{\text{b}}\xspace}
\newcommand{\Pq}{\ensuremath{{q}}\xspace}
\newcommand{\Pt}{\ensuremath{\text{t}}\xspace}
\newcommand{\Pu}{\ensuremath{\text{u}}\xspace}
\newcommand{\Pd}{\ensuremath{\text{d}}\xspace}
\newcommand{\Ps}{\ensuremath{\text{s}}\xspace}
\newcommand{\Pc}{\ensuremath{\text{c}}\xspace}
\newcommand{\Pg}{\ensuremath{\text{g}}\xspace}

\newcommand{\PW}{\ensuremath{\text{W}}\xspace}
\newcommand{\PZ}{\ensuremath{\text{Z}}\xspace}

\newcommand{\Mt}{\ensuremath{m_\Pt}\xspace}

\newcommand{\MWOS}{\ensuremath{M_\PW^\text{OS}}\xspace}
\newcommand{\MW}{\ensuremath{M_\PW}\xspace}
\newcommand{\MZOS}{\ensuremath{M_\PZ^\text{OS}}\xspace}
\newcommand{\MZ}{\ensuremath{M_\PZ}\xspace}
\newcommand{\Mb}{\ensuremath{m_\Pb}\xspace}

\newcommand{\GH}{\ensuremath{\Gamma_\PH}\xspace}

\newcommand{\GZOS}{\ensuremath{\Gamma_\PZ^\text{OS}}\xspace}

\newcommand{\GWOS}{\ensuremath{\Gamma_\PW^\text{OS}}\xspace}

\newcommand{\GeV}{\ensuremath{\,\text{GeV}}\xspace}
\newcommand{\TeV}{\ensuremath{\,\text{TeV}}\xspace}

\newcommand{\alphas}{\ensuremath{\alpha_\text{s}}\xspace}
\newcommand{\order}[1]{\ensuremath{\mathcal{O}{\left(#1\right)}}\xspace}

\newcommand{\GF}{\ensuremath{G_\mu}}

\newcommand{\ptsub}[1]{\ensuremath{p_{\text{T},#1}}\xspace}

\newcommand{\MVOS}{\ensuremath{M_{\text{V}}^\text{OS}}\xspace}%
\newcommand{\GVOS}{\ensuremath{\Gamma_{\text{V}}^\text{OS}}\xspace}%

\newcommand{\vh}{\ensuremath{\vphantom{\int_A^A}}}

\newcommand{\newc}{\newcommand}
\newc{\bi}{\begin{itemize}}
\newc{\ei}{\end{itemize}}
\newc{\benu}{\begin{enumerate}}
\newc{\eenu}{\end{enumerate}}
\newc{\bc}{\begin{center}}
\newc{\ec}{\end{center}}
\newc{\bfig}{\begin{figure}}
\newc{\efig}{\end{figure}}
\newc{\qbar}{\bar{q}}
\newc{\go}{\tilde{g}}
\newc{\PB}{\textsc{Powheg-Box}}

\newcommand{\Recola}{{\sc Recola}\xspace}
\newcommand{\Sherpa}{{\sc Sherpa}\xspace}
\newcommand{\SherpaRecola}{{\sc Sherpa\!{+}Recola}\xspace}
\newcommand{\MoCaNLO}{{\sc MoCaNLO}\xspace}
\newcommand{\MoCaNLORecola}{{\sc  MoCaNLO{+}Recola}\xspace}

\newcommand{\Comix}{C\protect\scalebox{0.8}{OMIX}\xspace}

\newcommand{\collier}{{\sc Collier}\xspace}

\newcommand{\rT}{{\mathrm{T}}}
\newcolumntype{.}{D{.}{.}{-1}}
\newcolumntype{d}[1]{D{.}{.}{#1}}

\newcommand{\QED}{\ensuremath{\text{QED}}}
\newcommand{\QCD}{\ensuremath{\text{QCD}}}
\newcommand{\EW}{\ensuremath{\text{EW}}}
\newcommand{\EWapprox}{\ensuremath{\text{EW}_\text{approx}}}
\newcommand{\QCDmEW}{{\ensuremath{\text{QCD--\EW}}}}
\newcommand{\QCDpEW}{{\ensuremath{\QCD+\EW}}}
\newcommand{\QCDtEW}{{\ensuremath{\QCD\times\EW}}}
\newcommand{\QCDpEWapprox}{{\ensuremath{\QCD+\EWapprox}}}
\newcommand{\QCDtEWapprox}{{\ensuremath{\QCD\times\EWapprox}}}

\newcommand{\MEPS}{\text{\textsc{MePs}}\xspace}
\newcommand{\MEPSatLO}{\text{\textsc{MePs@Lo}}\xspace}
\newcommand{\MEPSatNLO}{\text{\textsc{MePs@Nlo}}\xspace}
\newcommand{\MEPSatNLOQCD}{\MEPSatNLO\ \QCD\xspace}
\newcommand{\MEPSatNLOQCDpEWapprox}{\MEPSatNLO\ \ensuremath{\QCD+\EWapprox}\xspace}
\newcommand{\MEPSatNLOQCDtEWapprox}{\MEPSatNLO\ \ensuremath{\QCD\times\EWapprox}\xspace}
\newcommand{\MENLOPS}{\text{\textsc{MeNloPs}}\xspace}

\newcommand{\MCatNLO}{\text{\textsc{Mc@Nlo}}\xspace}

\newcommand{\mr}[1]{\ensuremath{\mathrm{#1}}}
\newcommand{\mc}[1]{\ensuremath{\mathcal{#1}}}
\newcommand{\muR}{\ensuremath{\mu_{\mr{R}}}}
\newcommand{\muF}{\ensuremath{\mu_{\mr{F}}}}
\newcommand{\muQ}{\ensuremath{\mu_{\mr{Q}}}}
\newcommand{\muCKKW}{\ensuremath{\mu_{\text{CKKW}}}}
\newcommand{\done}{\ensuremath{\mr{d}}}
\newcommand{\Qcut}{\ensuremath{Q_\text{cut}}}
\newcommand{\nmax}{\ensuremath{n_\text{max}}}
\newcommand{\nmaxnlo}{\ensuremath{n_\text{max}^\text{NLO}}}
\newcommand{\mucore}{\ensuremath{\mu_\text{core}}}
\newcommand{\njet}{\ensuremath{n_\text{jet}}}
\newcommand{\Bbar}{\ensuremath{\overline{\mr{B}}}}
\newcommand{\deltaEWapprox}{\ensuremath{\delta_\EW^\text{approx}}}
\newcommand{\ETWmean}{\ensuremath{\overline{E}_{\mr{T},\PW}}}

\colorlet{tableoverheadcolor}{gray!37.5}
\colorlet{tableheadcolor}{gray!25}
\colorlet{tablerowcolor}{gray!12.5}

\newcommand{\lsim}
{\;\raisebox{-.3em}{$\stackrel{\displaystyle <}{\sim}$}\;}

\newlength{\width}
\newlength{\height}

\def\draftdate{\relax}
\def\mda{\relax}
\def\mua{\relax}
\def\mla{\relax}
\def\draft{
\def\thtystars{******************************}
\def\sixtystars{\thtystars\thtystars}
\typeout{}
\typeout{\sixtystars**}
\typeout{* Draft mode!
         For final version remove \protect\draft\space in source file *}
\typeout{\sixtystars**}
\typeout{}
\def\draftdate{\today}
\def\mua{\marginpar[\boldmath\hfil$\uparrow$]%
                   {\boldmath$\uparrow$\hfil}\color{black}%
                    \typeout{marginpar: $\uparrow$}\ignorespaces}
\def\mda{\color{red}\marginpar[\boldmath\hfil$\downarrow$]%
                   {\boldmath$\downarrow$\hfil}%
                    \typeout{marginpar: $\downarrow$}\ignorespaces}
\def\mla{\marginpar[\boldmath\hfil$\rightarrow$]%
                   {\boldmath$\leftarrow $\hfil}%
                    \typeout{marginpar: $\leftrightarrow$}\ignorespaces}
\def\Mua{\marginpar[\boldmath\hfil$\Uparrow$]%
                   {\boldmath$\Uparrow$\hfil}\color{black}%
                    \typeout{marginpar: $\uparrow$}\ignorespaces}
\def\Mda{\color{red}\marginpar[\boldmath\hfil$\Downarrow$]%
                   {\boldmath$\Downarrow$\hfil}%
                    \typeout{marginpar: $\downarrow$}\ignorespaces}
\def\Mla{\marginpar[\boldmath\hfil\textcolor{red}{$\Rightarrow$}]%
                   {\boldmath\textcolor{red}{$\Leftarrow $}\hfil}%
                    \typeout{marginpar: $\leftrightarrow$}\ignorespaces}
\overfullrule 5pt
\oddsidemargin 15mm
\marginparwidth 29mm
}

\begin{document}

\title{\hfill ~\\[-30mm]
\phantom{h} \hfill\mbox{\small Cavendish-HEP 20/05, IPPP/20/19, MCNET-20-14}
\\[1cm]
\vspace{13mm}   \textbf{Fixed-order and merged parton-shower predictions for $\PW\PW$ and $\PW\PW\Pj$
  production at the LHC including NLO \QCD\ and \EW\ corrections}}

\date{}
\author{
Stephan Br\"auer$^{1\,}$\footnote{E-mail: \texttt{stephan.braeuer@phys.uni-goettingen.de}},
Ansgar Denner$^{2\,}$\footnote{E-mail:
  \texttt{ansgar.denner@physik.uni-wuerzburg.de}},
Mathieu Pellen$^{3\,}$\footnote{E-mail:
  \texttt{mpellen@hep.phy.cam.ac.uk}},
Marek Sch\"onherr$^{4\,}$\footnote{E-mail:
  \texttt{marek.schoenherr@durham.ac.uk}},
Steffen Schumann$^{1\,}$\footnote{E-mail: \texttt{steffen.schumann@phys.uni-goettingen.de}}
\\[9mm]
{\small\it
$^1$ Georg-August Universit\"at G\"ottingen, %
Institut f\"ur Theoretische Physik,} \\ %
{\small\it Friedrich-Hund-Platz 1, \linebreak %
        37077 G\"ottingen, %
        Germany}\\[3mm]
{\small\it
$^2$ Universit\"at W\"urzburg, %
        Institut f\"ur Theoretische Physik und Astrophysik,} \\ %
{\small\it Emil-Hilb-Weg 22, \linebreak %
        97074 W\"urzburg, %
        Germany}\\[3mm]
{\small\it
$^3$ University of Cambridge, Cavendish Laboratory,} \\ %
{\small\it 19 JJ Thomson Avenue, Cambridge CB3 0HE, United Kingdom}\\[3mm]
{\small\it
$^4$ Institute for Particle Physics Phenomenology, Durham University,} \\ %
{\small\it Durham DH1 3LE, United Kingdom}\\[3mm]
        }
\maketitle

\begin{abstract}
\noindent

First, we present a combined analysis of $\Pp\Pp \to \mu^+ \nu_\mu \Pe^- \bar \nu_\Pe$ and $\Pp\Pp \to \mu^+ \nu_\mu \Pe^- \bar \nu_\Pe \Pj$ at next-to-leading order, 
including both \QCD\ and electroweak corrections.
Second, we provide all-order predictions for $\Pp\Pp \to \mu^+ \nu_\mu
\Pe^- \bar \nu_\Pe+\mathrm{jets}$ using merged parton-shower
simulations that also include approximate \EW\ effects.
A fully inclusive sample for $\PW\PW$ production is compared to the fixed-order computations for \change{exclusive} zero- and one-jet selections.
The various higher-order effects are studied in detail at the level of cross sections and differential distributions for realistic experimental set-ups.
Our study \change{confirms} that merged predictions are significantly more stable than the fixed-order ones in particular regarding ratios between the two processes.

\end{abstract}
\thispagestyle{empty}
\vfill
\newpage
\setcounter{page}{1}

\tableofcontents
\newpage

\section{Introduction}

The Large Hadron Collider (LHC) is entering a precision era with the analysis of the full run~2 data set.
Many processes will be measured with an unprecedented accuracy and, in that respect, the consideration and
evaluation of all possible theoretical effects is mandatory.

Measurements of $\PW^+\PW^-$ production have been long on-going,
leading to very precise results~\cite{Aaboud:2019nkz}.  They are largely
motivated by the search for anomalous triple gauge-boson couplings \cite{Sirunyan:2017bey}
and in turn provide stringent tests of the Standard Model.  However, so far,
only a single measurement of di-boson production in
association with a jet has been published~\cite{Aaboud:2016mrt}.  Such a measurement is complementary to the
di-boson ones as it probes similar physics effects in a
different kinematics.

On the theoretical side, many higher-order computations have been
performed for $\PW^+\PW^-$ production in order to match the
experimental precision.  It started many years ago with the
calculation of next-to-leading order (NLO) \QCD\ corrections for the production of two W bosons
\cite{Ohnemus:1991kk,Baur:1995uv,Campbell:1999ah}.  These have been
subsequently matched to parton-shower simulations
\cite{Frixione:2002ik,Hamilton:2010mb}.  Electroweak (\EW) corrections
have been computed over several years
\cite{Kuhn:2011mh,Bierweiler:2012kw,Baglio:2013toa,Gieseke:2014gka,Biedermann:2016guo,Kallweit:2017khh}.
The NNLO \QCD\ corrections have been obtained a few years ago
\cite{Gehrmann:2014fva,Grazzini:2016ctr}.  These have been recently
combined with \EW\ corrections \cite{Kallweit:2019zez} and with
parton-shower corrections \cite{Re:2018vac}.  Resummed computations
\cite{Grazzini:2015wpa} as well as the gluon--gluon loop-induced
contribution \cite{Caola:2015rqy,Grazzini:2020stb} are also available.
Very recently, a combination of fixed-order predictions with resummed
ones has been presented in \citere{Kallweit:2020gva} for vetoed cross 
sections and transverse observables.
Concerning $\PW\PW\Pj$ production, far fewer results are available.
Owing to the higher multiplicity the NNLO \QCD\ corrections
have not been evaluated yet.  However, the NLO \QCD\ corrections are known
\cite{Dittmaier:2009un,Cascioli:2013gfa}, and merged predictions based on the {\sc{MiNLO}}
prescription have been presented \cite{Hamilton:2016bfu}.  The NLO \EW\ 
corrections for on-shell W bosons have been computed recently~\cite{Li:2015ura}.

The present work focuses on the computation and the combination of NLO corrections of \QCD\ and \EW\ type for the
processes $\Pp\Pp \to \mu^+ \nu_\mu \Pe^- \bar \nu_\Pe$ and $\Pp\Pp \to \mu^+
\nu_\mu \Pe^- \bar \nu_\Pe \Pj$ at the LHC.  For the first time,  NLO
\QCD\ and NLO \EW\ corrections for the off-shell production of both
$\PW\PW$ and $\PW\PW\Pj$ are presented together.  All off-shell,
non-resonant, and interference contributions are taken into account.
Subsequently, all-order predictions based on multi-jet merged
parton-shower simulations as implemented in the \Sherpa framework are provided.
These predictions also include \EW\ effects by combining them with 
the NLO \QCD\ merged predictions of different jet multiplicity using the 
virtual approximation of the \EW\ corrections \cite{Kallweit:2015dum}
applied to the dominant part of the QCD corrections. 
The fully inclusive merged sample for $\PW\PW$ production can be used in combination with the zero- and one-jet selections.
We also compare these sub-samples using two merging prescriptions
against fixed-order predictions, hence providing a deeper insight in
the merging procedure. 

{\sloppy
All results presented in this work have been obtained with the fully automated
framework \SherpaRecola \cite{Biedermann:2017yoi} in realistic
experimental set-ups.  In particular, vetoes on extra jets are applied
for both processes in order to avoid large $K$~factors.
Hadronisation and underlying-event effects are not included in the
present study but can easily be incorporated thanks to the
\Sherpa framework.
The results are presented in the form of cross sections and differential
distributions.  Given the similarity of the $\PW\PW$ and $\PW\PW\Pj$
production processes, we provide
ratios of cross sections and differential distributions between the
two processes. They deliver useful information concerning the
correlations between the two channels.
We also state for reference the cross sections of the loop-induced contributions \cite{Caola:2015rqy,Grazzini:2020stb,Cascioli:2013gfa},
which can be treated completely independently and simply be added to our results.}

This article is organised as follows: in \refse{se:features}, the features of
the calculations are explained.  In particular, the various
contributions included and the methods used are reviewed. Technical
details and the set-ups of the calculations are provided.
Section \ref{se:results}  is devoted to
the numerical results and their discussion.
It is divided into two parts: in \refse{se:results:fo} the fixed-order
predictions are displayed, and in \refse{sec:results:merged} results
based on multi-jet merging are presented.
In each section, various cross sections and a wide range of differential
distributions are discussed.  Finally,  \refse{se:conclusion}
contains a short summary and concluding remarks.

\section{Features of the calculations}
\label{se:features}

\change{In the present computation, we have opted for the 4-flavour scheme.
Thus, bottom quarks are treated as massive, and contributions with bottom
quarks in the initial state do not appear.  Moreover, partonic processes
with bottom quarks in the final state are omitted.}

\subsection{Born contributions}

In this work, we consider two hadronic processes corresponding to $\PW\PW$ and $\PW\PW\Pj$ production at the LHC.
The first one,
\begin{equation}
 \Pp\Pp \to \mu^+ \nu_\mu \Pe^- \bar \nu_\Pe ,
\end{equation}
describes the production of two off-shell W bosons that decay
leptonically.  The leading-order (LO) cross section is of order
$\mathcal{O} \left( \alpha^4 \right)$.  \change{In the 4-flavour
  scheme, the} contributing partonic
channels have initial states $\Pq \bar \Pq$ with $\Pq = \Pu, \Pd, \Pc,
\Ps$ and $\gamma\gamma$.
\change{However, the photon-induced contribution has not been included in our computations.}

The second process involves in addition an extra \QCD\ jet,
\begin{equation}
 \Pp\Pp \to \mu^+ \nu_\mu \Pe^- \bar \nu_\Pe \Pj .
\end{equation}
The dominant partonic channels contribute to the cross section at
order $\mathcal{O} \left( \alpha_{\rm s} \alpha^4 \right)$, where
besides the $\Pq \bar \Pq$ channels also contributions from
$\Pg\Pq$ and $\Pg\bar\Pq$ initial states appear.  Sample diagrams are
shown in Fig.~\ref{fig:diagLO}.  Subleading contributions of order
$\mathcal{O} \left( \alpha^5 \right)$ originate from initial states
$\gamma\Pq$ and $\gamma\bar\Pq$, where again $\Pq = \Pu, \Pd, \Pc, \Ps$.
While always considering the full off-shell production, in the following,
both processes are sometimes referred to as $\PW\PW$ and $\PW\PW\Pj$, respectively.
\bfig[t!]
  \center
  \includegraphics[width=0.30\textwidth]{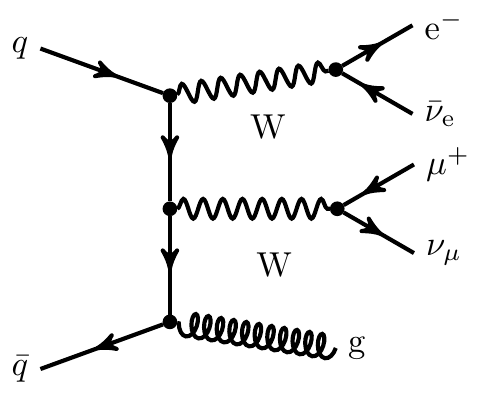}\hspace*{50pt}
  \qquad\includegraphics[width=0.30\textwidth]{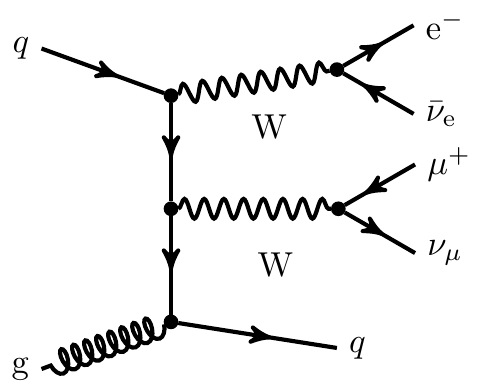}
  \caption{\label{fig:diagLO} Sample Feynman diagrams for the channels $\Pq\bar\Pq \to \mu^+ \nu_\mu \Pe^- \bar \nu_\Pe \Pg$ (left) and $\Pg \Pq \to \mu^+ \nu_\mu \Pe^- \bar \nu_\Pe \Pq$ (right), respectively.
  } 
\efig

In addition to tree-level contributions, there are also loop-induced contributions with two gluons in the initial state,
$\Pg\Pg \to \mu^+ \nu_\mu \Pe^- \bar \nu_\Pe$ and $\Pg\Pg \to \mu^+ \nu_\mu \Pe^- \bar \nu_\Pe \Pg$ for $\PW\PW$ and $\PW\PW\Pj$, respectively.
In \refse{se:results}, their LO fiducial cross sections are given for reference but no in-depth analysis is presented.
Such contributions are known at NLO for $\PW\PW$ \cite{Caola:2015rqy,Grazzini:2020stb} and 
have also been studied in detail for $\PW\PW\Pj$ \cite{Cascioli:2013gfa} \change{where $\Pg\Pq \to \mu^+ \nu_\mu \Pe^- \bar \nu_\Pe \Pq$ contributions have been included}.

\subsection{\QCD\ corrections}

The \QCD\ corrections to the cross section for $\PW\PW$ production are
of order $\mathcal{O} \left( \alpha_{\rm s} \alpha^4 \right)$, while
those for
$\PW\PW\Pj$ production  are of order $\mathcal{O} \left(
  \alpha^2_{\rm s} \alpha^4 \right)$.  They consist of real and
virtual contributions.  
\change{The use of the  4-flavour scheme} \change{and the exclusion of final states
  containing bottom quarks} \change{avoids contributions such as
  $\Pg\Pb \to \PW^+\PW^-\Pb$ or $\Pg\Pg \to 
\PW^+\PW^-\Pb \bar \Pb$ which are dominated by $\PW\Pt$ and $\Pt\bar\Pt$ production.}
\change{Finally, the $\Pg\Pg$ loop-induced contributions are not included in our definition of the NLO QCD predictions.}

\subsection{\EW\ corrections}

The \EW\ corrections to $\PW\PW$ and $\PW\PW\Pj$ production are of order
$\mathcal{O} \left( \alpha^5 \right)$ and $\mathcal{O} \left(
  \alpha_{\rm s} \alpha^5 \right)$, respectively.  For both processes,
\emph{usual} \EW\ corrections are included, consisting of virtual
corrections as well as real-photon radiation.  A sample virtual diagram
with the insertion of neutral \EW\ gauge bosons is shown in Fig.~\ref{fig:diagNLO}
(left).
\bfig[t!]
  \center
  \includegraphics[width=0.30\textwidth]{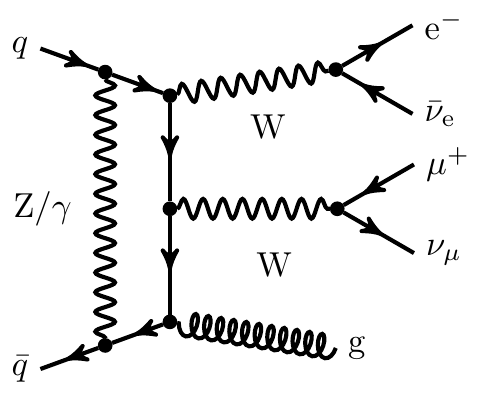}\hspace*{50pt}
  \qquad\includegraphics[width=0.30\textwidth]{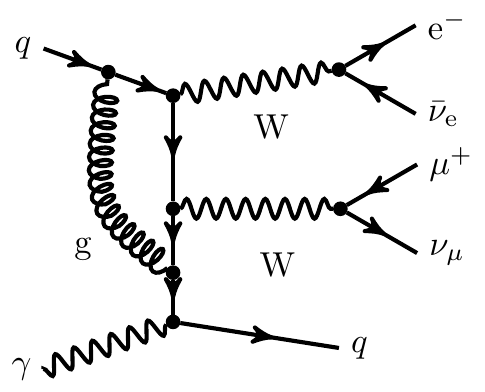}
  \caption{\label{fig:diagNLO} Sample Feynman diagrams representing
    \emph{usual} \EW\ corrections of order $\mathcal{O} \left(
      \alpha_{\rm s} \alpha^5 \right)$ in the channel $\Pq\bar\Pq \to
    \mu^+ \nu_\mu \Pe^- \bar \nu_\Pe \Pg$ (left) and
   \QCD\ corrections of order $\mathcal{O} \left( \alpha_{\rm s}
     \alpha^5 \right)$ in the channel $\gamma \Pq/\bar\Pq \to \mu^+
   \nu_\mu \Pe^- \bar \nu_\Pe \Pq/\bar\Pq$ (right). 
 }
\efig

Given that a recombination algorithm is used to cancel IR
divergences differentially, soft and collinear photons are recombined
with jets.  At NLO \EW, these jets are either made of a single gluon or
a single quark.  This opens the possibility to recombine a soft gluon
with a hard photon into a jet which suffers from IR singularities
related to soft gluons.  These singularities are, by definition, not
cancelled by the virtual \EW\ corrections but by virtual \QCD\ corrections
to $\Pq\bar\Pq \to \PW^+\PW^-\gamma$.\footnote{Note that in \Sherpa the
  corresponding \QCD\ dipoles are always included by default.  These
  have an underlying Born configuration of the form $a+b\to \mu^+ \nu_\mu \Pe^- \bar
  \nu_\Pe \gamma$ with $a$ and $b$ \QCD\ partons.  However, such
  configurations would never be accepted by the selector function that
  requires at least one \QCD\ jet in the final state.}
To deal with such configurations properly, prescriptions for
photon--jet separation are needed.  For the processes studied here,
the related effects are, however, rather suppressed.%
\footnote{In practice, the IR singularities are regulated by technical
  cuts, but the dependence on these cuts is very small.}  Therefore,
for the results presented here no prescriptions for photon--jet
separation have been used.  To justify this procedure, we provide in
the appendix a comparison of results obtained without any such
prescriptions and a fully consistent approach employing a photon--jet
separation based on jet-energy fractions and fragmentation functions
following
\citeres{Denner:2009gj,Denner:2010ia,Denner:2011vu,Denner:2014ina}.

For $\PW\PW\Pj$ production, another type of contributions appears in
the real corrections, namely interferences between diagrams of orders
$\mathcal{O} \left( g_{\rm s}^2 g^4 \right)$ and $\mathcal{O}
\left(g^6 \right)$ for $\Pq\bar\Pq \to \mu^+ \nu_\mu \Pe^- \bar
\nu_\Pe \Pq'\bar\Pq'$.  These contributions are IR-finite and have two
(anti-)quarks in both the initial and final state.  An example of
such an interference term is shown in Fig.~\ref{fig:diagint}.
\bfig[t!]
  \center
  \includegraphics[width=0.55\textwidth]{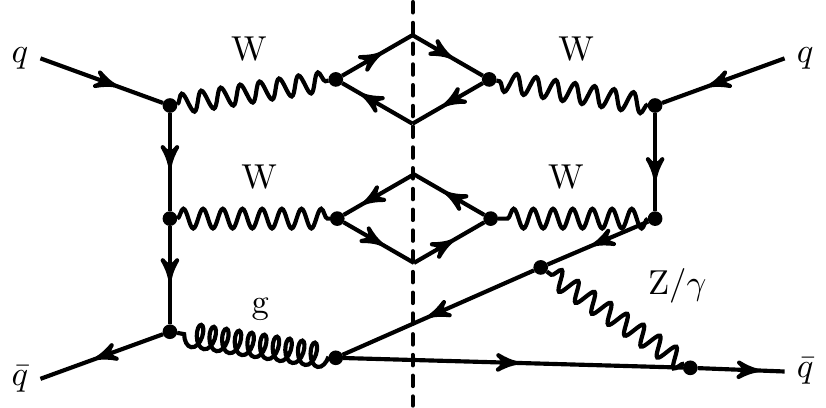}
  \caption{\label{fig:diagint} A squared sample diagram representing
    interference contributions in the real corrections at order $\mathcal{O} \left( \alpha_{\rm s} \alpha^5 \right)$
  in the channel $\Pp\Pp \to \mu^+ \nu_\mu \Pe^- \bar \nu_\Pe \Pj\Pj$.
 }
\efig

Also, the \QCD\ corrections to the photon-induced processes in $\Pp\Pp \to
\mu^+ \nu_\mu \Pe^- \bar \nu_\Pe \Pj$ are counted as part of the \EW\ 
corrections.  Indeed, even if these are \QCD\ corrections to the
underlying process of order $\mathcal{O} \left( \alpha^5 \right)$,
$\gamma\Pq \to \mu^+ \nu_\mu \Pe^- \bar \nu_\Pe \Pq$, they give rise to
contributions of order $\mathcal{O} \left( \alpha_{\rm s} \alpha^5
\right)$, \ie of the same order as the \EW\ corrections to the dominant
LO contributions.  One example of such contributions is shown in
Fig.~\ref{fig:diagNLO} (right).

Finally, there are further \EW\ corrections for $\PW\PW\Pj$ of order
$\mathcal{O} \left( \alpha^6 \right)$.  They result from pure \EW\ 
contributions to partonic processes $\Pq\bar\Pq \to \mu^+ \nu_\mu
\Pe^- \bar \nu_\Pe \Pq'\bar\Pq'$. They include IR-singular
contributions from the splitting $\gamma^* \to \Pq \bar \Pq$, which
have been first encountered in the computation of \QCD\ corrections to
WZ vector-boson scattering \cite{Denner:2019tmn}.  They can be
treated using the photon-to-jet conversion function introduced in
\citere{Denner:2019zfp}, where a numerical study for Z+j production
has been presented.  This study showed that the corrections of
relative order $\mathcal{O} \left( \alpha^2/\alpha_\mathrm{s} \right)$
are of the order of a per cent for the fiducial cross section and
reach up to $10\%$ for large transverse momenta.  Owing to their reduced
numerical size, the $\mathcal{O} \left( \alpha^6 \right)$ corrections have
been neglected in the present work.

\subsection{Merged predictions with virtual \EW\ approximation}
\label{se:merged_predictions}

Besides calculating $\PW$-pair production in association with zero and 
one jet at fixed-order, we also match both calculations to the 
parton shower and build a multi-jet merged event simulation that 
incorporates exact NLO \QCD\ corrections and approximate \EW\ corrections 
for both the $\PW\PW$ and $\PW\PW\Pj$ final states, 
based on the \MEPS method \cite{Hoeche:2009rj,Hoche:2010kg,Hoeche:2012yf,Gehrmann:2012yg} 
implemented in the \Sherpa Monte Carlo event generator. 
The aim of the method is to generate an inclusive event sample 
wherein the hardest $n=0,1,\ldots,\nmax$ associated \QCD\ jets are
described by $n$-jet matrix elements of the desired LO or NLO accuracy. 
A measure $Q_n$ and a resolution criterion \Qcut\ are introduced 
such that $Q_1>\ldots>Q_n>\Qcut>Q_{n+1}>\ldots$ defines the 
$n$-jet process and, thus, separates the $n$-jet region 
from the $n+1$-jet region. 
The measure $Q_n$ is only required to be the clustering scale of an infrared-safe 
jet algorithm but otherwise arbitrary. In practice we choose it to 
coincide with the parton-shower branching scale $t_n$ at the 
reconstructed splitting.

At LO \cite{Hoeche:2009rj}, \MEPSatLO, the exclusive
cross section with 
$n<n_\text{max}$ jets reads
\begin{equation}
  \begin{split}
    \done\sigma_n^\MEPSatLO
    =&\;\done\Phi_n\,\mr{B}_n(\Phi_n)\,\Theta(Q_n-\Qcut)\,\mc{F}_n(\muQ^2;<\Qcut)\,.
  \end{split}
\end{equation}
Herein, $\mr{B}_n$ is the Born matrix element of the $n$-jet process, 
including all PDF, flux and symmetry/averaging factors, while 
$\Phi_n$ is the $n$-jet phase-space configuration.
The $\Theta$-function ensures that all $n$ jets are resolved 
under the jet criterion \Qcut.
The parton-shower generating functional $\mc{F}_n(\muQ^2;<\Qcut)$ 
\cite{Hoeche:2014rya}
applies a truncated vetoed parton shower to the $n$-jet configuration 
starting at the hard scale $\muQ$ 
and ensures that all further emissions fall into the unresolved region, \emph{i.e.}\
$Q_{n+k}<\Qcut$ ($k>0$).
For the highest multiplicity, $n=\nmax$, this veto is increased to 
$Q_{\nmax}$ to render the highest multiplicity fully inclusive with
respect to additional emissions.
The application of this veto also supplies the respective Sudakov form 
factors to the $n$-jet configuration, resumming the hierarchy of 
reconstructed parton-shower branching scales $t_1,\ldots,t_n$. 
In concert with the CKKW scale choice $\muR=\muCKKW$, 
defined through \cite{Catani:2001cc},
\begin{equation}\label{eq:muRCKKW}
  \begin{split}
    \alpha_s^n(\muCKKW^2)
    =&\;\alpha_s(t_1)\cdots\alpha_s(t_n)\;,
  \end{split}
\end{equation}
and the factorisation and shower-starting scales fixed by the scale
of the core process, \emph{i.e.}\ 
$\muF=\muQ=\mucore$, a smooth transition across \Qcut\ 
is ensured. 
With these definitions, $\mucore$ is the only free scale of the 
CKKW algorithm and fixes the other relevant perturbative scales.
In analogy to the fixed-order calculations, the core scale 
for the reconstructed $\Pp\Pp\to\mu^+\nu_\mu\Pe^-\bar\nu_\Pe$ process
is chosen as 
\begin{equation}\label{eq:mQCKKW}
  \begin{split}
    \mucore
    ={}&\tfrac{1}{2}\left(E_{\mr{T},\PW^+}+E_{\mr{T},\PW^-}\right)
    \equiv\tfrac{1}{2}\,\overline{E}_{\mr{T},\PW}\,.
  \end{split}
\end{equation}
This construction can now be lifted to NLO 
accuracy in \QCD, the \MEPSatNLO method \cite{Hoeche:2012yf,Gehrmann:2012yg}. 
Its exclusive $n$-jet cross section, for $n<\nmaxnlo$, 
based on the \MCatNLO expression in \citeres{Hoeche:2011fd,Hoeche:2012ft,Hoche:2012wh}, 
is defined as 
\begin{equation}\label{eq:mepsnlo}
  \begin{split}
    \done\sigma_n^{\MEPSatNLOQCD}
    ={}&\Theta(Q_n-\Qcut)\left[
          \done\Phi_n\Bbar_n^\QCD(\Phi_n)\,\widetilde{\mc{F}}_n(\muQ^2;<\Qcut)
          \vphantom{\int}
        \right.\\
     &\left.{}
          +\done\Phi_{n+1}\,\mr{H}_n^\QCD(\Phi_{n+1})\,\Theta(\Qcut-Q_{n+1})\,
           \mc{F}_{n+1}(\muQ^2;<\Qcut)
           \vphantom{\int}
        \right]
 \,.
  \end{split}
\end{equation}
Here, $\Bbar_n^\QCD$ describes configurations 
with $n$ resolved emissions with $Q_n>\Qcut$,  
and takes the form 
\begin{equation}
  \begin{split}
    \Bbar_n^\QCD(\Phi_n)
    ={}&\;\mr{B}_n(\Phi_n)+\tilde{\mr{V}}_n^\QCD(\Phi_n)
        +\int\done\Phi_1\,\mr{D}_n^\QCD(\Phi_n,\Phi_1)\,\Theta(\muQ^2-t_{n+1})\,.
  \end{split}
\end{equation}
It contains \QCD\ NLO renormalised virtual corrections 
including initial-state mass-factorisation counterterms, 
$\tilde{\mr{V}}_n^\QCD$, and the integral over the real-emission \QCD\ 
corrections described by splitting functions in $\mr{D}_n^\QCD$. 
The functions $\mr{D}_n^\text{QCD}$ are, by construction, also the emission 
kernels of the fully colour- and spin-correlated parton shower 
$\widetilde{\mc{F}}_n$ \cite{Hoeche:2014rya,Hoeche:2011fd} generating the $(n+1)$-th emission. 
The $\mr{H}_n^\QCD$ term corrects its approximate emission pattern 
to the exact NLO \QCD\ expression. It takes the form
\begin{equation}
  \begin{split}
    \mr{H}_n^\QCD(\Phi_{n+1})
    =&\;\mr{R}_n^\QCD(\Phi_{n+1})-\mr{D}_n^\QCD(\Phi_n,\Phi_1)\,\Theta(\muQ^2-t_{n+1})\;,
  \end{split}
\end{equation}
with the NLO \QCD\ real-emission matrix element $\mr{R}_n^\QCD$.
\change{The quantity $\mr{H}_n$ is thus a real-subtracted contribution.}

When $\nmax>\nmaxnlo$, \emph{i.e.}\ when only the first \nmaxnlo\ 
emissions can be described at NLO \QCD\ accuracy, the additional 
$\nmax-\nmaxnlo$ emissions are added at LO accuracy. 
In this case the \MENLOPS method \cite{Hoche:2010kg,Gehrmann:2012yg,Hoeche:2014rya} is used for 
$n=\nmax^\text{NLO}+k$ ($k>0$), 
\begin{equation}
  \begin{split}
    \done\sigma_n^\MENLOPS
    =&\;\done\Phi_n\,k_{\nmaxnlo}(\Phi_{\nmaxnlo}(\Phi_n))\,
        \mr{B}_n(\Phi_n)\,\Theta(Q_n-\Qcut)\,
        \mc{F}_n(\muQ^2;<\Qcut)\,.
  \end{split}
\end{equation}
It thus furnishes a local $K$~factor, defined on the highest 
multiplicity phase space for which NLO 
corrections are available, $\Phi_{\nmaxnlo}$, 
\begin{equation}
  \begin{split}
    k_n(\Phi_n)
    =&\;\frac{\Bbar_n(\Phi_n)}{\mr{B}_n(\Phi_n)}
        \left(1-\frac{\mr{H}_n(\Phi_{n+1})}{\mr{B}_{n+1}(\Phi_{n+1})}\right)
        +\frac{\mr{H}_n(\Phi_{n+1})}{\mr{B}_{n+1}(\Phi_{n+1})}\;,
  \end{split}
\label{eq:local_k-factor}
\end{equation}
to the \MEPSatLO expression of that multiplicity. 
\change{The $K$ factor expands to $1+\mc{O}(\alphas)$ and, thus,} retains both the NLO accuracy of 
the $\nmaxnlo$-parton process and the LO accuracy of the $n$-parton 
process, while simultaneously guaranteeing a smooth transition 
across the merging parameter \Qcut\ \change{for all multiplicities.}\footnote{%
  \change{%
  Reference \cite{Danziger:2715727} explored the possibility to
  substitute the local $K$~factor $k_n$
  defined on the highest NLO multiplicity by the $K$~factor defined on the 
  lowest NLO multiplicity, \ie the core process.
  For the considered processes it found a negligible 
  dependence on this choice for most observables. 
  In the context of employing the virtual EW approximation, however, 
  the higher multiplicity LO processes would then not directly inherit approximate 
  EW corrections for kinematic quantities like the leading jet.
  }
}

Approximate NLO \EW\ corrections are incorporated by 
replacing the usual NLO \QCD\ $\Bbar_n$ function of Eq.\ (\ref{eq:mepsnlo}) 
with \cite{Kallweit:2015dum,Gutschow:2018tuk}
\begin{equation}\label{eq:mepsnlo_qcdpew}
  \begin{split}
    \overline{\mr{B}}_n^\QCDpEW(\Phi_n)
    =&\;\overline{\mr{B}}_n^\QCD(\Phi_n)
        +\mr{B}_n(\Phi_n)\,\deltaEWapprox(\Phi_n)
        +\mr{B}_n^\text{sub}(\Phi_n)
  \end{split}
\end{equation}
in an additive combination of \QCD\ and \EW\ corrections or 
\begin{equation}\label{eq:mepsnlo_qcdtew}
  \begin{split}
    \overline{\mr{B}}_n^\QCDtEW(\Phi_n)
    =&\;\overline{\mr{B}}_n^\text{QCD}(\Phi_n)\left(1+\deltaEWapprox(\Phi_n)\right)
        +\mr{B}_n^\text{sub}(\Phi_n)
  \end{split}
\end{equation}
in a multiplicative manner. 
In both cases, the approximate \EW\ correction is defined as 
\begin{equation}\label{eq:deltaEWapprox}
  \begin{split}
    \deltaEWapprox(\Phi_n)
    =&\;\frac{\mr{V}_n^\EW(\Phi_n)+\mr{I}_n^\EW(\Phi_n)}{\mr{B}_n(\Phi_n)}\;.
  \end{split}
\end{equation}
Herein, $\mr{V}_n^\EW$ represents the renormalised virtual \EW\  
corrections and $\mr{I}_n^\EW$ the approximate \EW\ real-emission 
corrections integrated over the real-emission phase 
space.\footnote{In practice, we use the Catani--Seymour $\mr{I}$~operator 
\cite{Schonherr:2017qcj} with $\alpha=1$.}
Finally, $\mr{B}_n^\text{sub}$ are possible Born contributions 
at subleading orders, which are, however, zero 
in the processes under consideration in this article.
We stress that the modified $\Bbar_n$ functions and, thus, the \EW\ 
corrections also enter 
the local $K$~factor in Eq.~\refeq{eq:local_k-factor} applied to the
higher-multiplicity processes. 
They thus receive approximate \EW\ corrections through $k_n$ from 
the $\nmaxnlo$-jet process, guaranteeing their continuity across 
the merging parameter $\Qcut$ and consistency with respect to whether
an additional jet at LO accuracy is merged or not.

It should be noted, however, that \change{both the phase-space-point-wise
definition of the additive and multiplicative combinations of Eq.\ 
\eqref{eq:mepsnlo_qcdpew} and \eqref{eq:mepsnlo_qcdtew} differ from the 
usual definition constructed on histogram level in the case of fixed-order 
calculations. 
Both constructions apply the
approximate \EW\ corrections only to the \QCD\ \Bbar-function (collecting 
the virtual corrections and the soft-collinear limit of the real-emission 
corrections, but not to its hard wide-angle radiation part). 
Both $\overline{\mr{B}}_n^\QCDpEW$ and 
$\overline{\mr{B}}_n^\QCDtEW$ are then dressed with \QCD\ radiation 
through $\widetilde{\mc{F}}$.
Thus, the results will differ from the bin-by-bin additive and 
multiplicative combination of \QCD\ and \EW\ corrections of the
fixed-order calculation, respectively. 
Major differences will occur between the two additive combinations 
if selection criteria acting on additional jet activity (like jet vetoes) 
are present, while the multiplicative combination will differ substantially 
if hard real radiation described through the
$\mr{H}_n$~events forms a large part of the event sample. 
We have checked, however, that the latter} is not the case for any observable
presented in this paper.  In fact, in our set-up for $\PW\PW+{}$ jets
production, the $\mr{H}_n$-terms' contribution never exceeds 20\% (15\%)
for the zero-jet (one-jet) selection for inclusive observables and 5\%
(5\%) in the \EW\ Sudakov regime.
\change{We would like to emphasise that the approximate inclusion of
  EW corrections does not improve predictions for inclusive observables. 
It has an intrinsic uncertainty of a few per cent and primarily catches leading
logarithmic EW corrections relevant in the high-energy limit.}

Finally, the approximation of integrated-out real-emission corrections can 
be problematic for leptonic observables in particular. 
Here, radiative energy loss through photon bremsstrahlung 
can amount to $\order{1}$ effects below the on-shell \PW-pair 
production threshold or the \PZ pole (see for instance
\citere{Biedermann:2016guo}).
If not via explicit real-radiation matrix elements, 
these effects can be included through either QED parton showers 
or a soft-photon resummation, the latter of which will be employed 
in this work. 
It is important to note, however, that both solutions lead to 
a double counting of virtual QED corrections which remains 
unresolved. 
As these corrections are applied to the respective \PW decays 
only, it is ensured that they do not interfere with the \EW\  
logarithms in the Sudakov regime \cite{Kallweit:2015dum,Kallweit:2017khh}. 
The unitarity of these resummations also ensures that inclusive 
cross sections remain unaffected.

\subsection{Validation and technical aspects}

The results presented here have been obtained with the combination
\SherpaRecola~\cite{Biedermann:2017yoi} which has already been used
for several NLO \QCD\ and \EW\ computations
\cite{Schonherr:2018jva,Reyer:2019obz,Baberuxki:2019ifp}.
\Sherpa~\cite{Bothmann:2019yzt,Gleisberg:2008ta} is a multi-purpose
Monte Carlo event generator capable to compute both \QCD\ and \EW\ 
corrections in a general and automated way. It implements the
tree-level matrix-element generators {\sc{Comix}}~\cite{Gleisberg:2008fv}
and {\sc{Amegic}}~\cite{Krauss:2001iv} and employs an implementation
of the Catani--Seymour dipole subtraction method for both \QCD\ and \QED\ 
soft and collinear singularities~\cite{Gleisberg:2007md,Schonherr:2017qcj}.
To simulate \QCD\ parton cascades it employs the dipole-shower algorithm
presented in~\citere{Schumann:2007mg}. For the merging of
parton-shower-evolved hard processes at tree and one-loop level a truncated-shower
approach is employed~\cite{Hoeche:2009rj,Hoche:2010kg,Hoeche:2012yf}.
Higher-order QED corrections are effected through the soft-photon 
resummation of Yennie, Frautschi, and Suura (YFS) \cite{Yennie:1961ad}, as 
implemented in \citere{Schonherr:2008av}. 

\Recola \cite{Actis:2016mpe,Actis:2012qn} is a matrix-element generator
that provides any one-loop amplitude in the Standard Model.
It relies on the \collier library
\cite{Denner:2014gla,Denner:2016kdg} to numerically evaluate the
one-loop scalar
\cite{'tHooft:1978xw,Beenakker:1988jr,Dittmaier:2003bc,Denner:2010tr}
and tensor integrals
\cite{Passarino:1978jh,Denner:2002ii,Denner:2005nn}.  The interface
between \Sherpa and \Recola is fully general and hence enables the
computation of any NLO cross section in the Standard Model.  The interface is
compatible with {\sc Recola2} \cite{Denner:2017wsf} which features a
considerable reduction of the memory needs when computing many
channels at a time.  This is made possible by the use of crossing
symmetries in order to compute the minimum number of processes and
has already been exploited in \citere{Chiesa:2019ulk}.

The computation of NLO \QCD\ corrections with \Sherpa has by now become a
standard.  However, the possibility to compute one-loop \EW\ corrections
in an automated manner is still rather recent~\cite{Schonherr:2017qcj}.
To that end, we have carefully tested the implementation of the \EW\ 
corrections against an independent program, namely the combination
\MoCaNLORecola, that has already been used for a variety of processes
including V+jets production
\cite{Biedermann:2016yds,Biedermann:2017bss,Denner:2019tmn}.  

The \Sherpa framework has already been utilised for NLO \QCD\ and \EW\ 
computations for multi-jet \cite{Reyer:2019obz}, 
V+jet \cite{Kallweit:2014xda,Kallweit:2015dum}, 
di-boson \cite{Kallweit:2017khh,Chiesa:2017gqx}, 
tri-boson \cite{Greiner:2017mft,Schonherr:2018jva},
and $t\bar{t}$+jet production \cite{Gutschow:2018tuk}. 
Di-boson-production processes, in particular, have
been cross-validated among various programs including \SherpaRecola
and \MoCaNLORecola in \citere{Bendavid:2018nar}.

\subsection{Set-up}
\label{se:setup}

\subsubsection*{Numerical inputs}

The predictions presented here are obtained for the LHC operating at
$\sqrt{s} = 13 \TeV$.  For the parton distribution function (PDF), the
\texttt{NNPDF31\_nlo\_as\_0118\_luxqed} set \cite{Bertone:2017bme} is used 
and interfaced through \textsc{Lhapdf} \cite{Buckley:2014ana}.
It is based on \citere{Manohar:2016nzj} for the extraction of the
photon content. The choice of the renormalisation and factorisation
scales follows the one of \citere{Cascioli:2013gfa} and reads
\begin{equation}
 \muR = \muF = \frac12 \left( E_{\rm T, \PW^+} + E_{\rm T, \PW^-} \right)
 \equiv\tfrac{1}{2}\,\ETWmean\; ,
\end{equation}
with $E_{\rm T, \PW} = \sqrt{\MW^2 + \left( \vec{p}_{\rm T,
    \Pl} + \vec{p}_{\rm T, \nu} \right)^2}$. The value of the strong coupling
is chosen consistently with the used PDF set, \emph{i.e.}\
\begin{equation}
  \alphas(M^2_\PZ)= 0.118\;.
\end{equation}
To fix the \EW\ coupling, the $G_\mu$ scheme \cite{Denner:2000bj,Dittmaier:2001ay}
is employed throughout with
\begin{equation}
  \alpha = \frac{\sqrt{2}}{\pi} G_\mu \MW^2 \left( 1 - \frac{\MW^2}{\MZ^2} \right)  \qquad \text{and}  \qquad   \GF    = 1.16637\times 10^{-5}\GeV^{-2}\;.
\end{equation}
We use the following values for the masses and widths:
\begin{alignat}{2}
                  \Mt   &=  173.21\GeV,       & \quad \quad \quad \Mb &= 4.8 \GeV,  \nonumber \\
                \MZOS &=  91.1876\GeV,      & \quad \quad \quad \GZOS &= 2.4952\GeV,  \nonumber \\
                \MWOS &=  80.385\GeV,       & \GWOS &= 2.085\GeV,  \nonumber \\
                M_{\rm H} &=  125.0\GeV,       &  \GH   &=  4.07 \times 10^{-3}\GeV.
\end{alignat}
Both the top- and bottom-quark widths are taken to be zero as these
particles do not appear as  resonances in our computations.  The values for
the Higgs-boson mass and width follow the recommendations of the Higgs
cross section working group \cite{Heinemeyer:2013tqa}.  The pole
masses and widths used for the simulations are obtained from the
measured on-shell (OS) values for the W and
Z~bosons according to~\cite{Bardin:1988xt}
\begin{equation}
        M_{\text{V}} = \frac{\MVOS}{\sqrt{1+(\GVOS/\MVOS)^2}}\;,\qquad
\Gamma_{\text{V}} = \frac{\GVOS}{\sqrt{1+(\GVOS/\MVOS)^2}}\;,
\end{equation}
with ${\text{V}}=\PW, \PZ$.

\subsubsection*{Event selection}

The event-selection criteria are based on \citere{Aaboud:2016mrt} and
are specified for both processes in the following.  To cluster \QCD\ 
jets we use the anti-$k_\rT$ algorithm \cite{Cacciari:2008gp} with a
jet-resolution parameter of $R=0.4$. Photons are recombined with
charged leptons and \QCD\ jets using a standard cone algorithm with a
radius parameter of $R=0.1$.

\begin{itemize}
 \item \underline{$\Pp\Pp \to \mu^+ \nu_\mu \Pe^- \bar \nu_\Pe$}

The charged leptons are required to fulfil
\begin{align}
 \ptsub{\Pl} >  20\GeV\;,\qquad |y_{\Pl}| < 2.5\;,\qquad \Delta R_{\Pl\Pl'} > 0.1\;,
\label{eq:lepton_cuts_1}
\end{align}
as well as
\begin{align}
 m_{\Pl \Pl} >  10\GeV\;,\qquad p_{\mathrm{T},\text{miss}} >  20\GeV\;.
\label{eq:lepton_cuts_2}
\end{align}
In addition, a jet-veto is applied in order to limit the size of the \QCD\ corrections.
In particular, any event with an identified jet such that
\begin{align}
\label{jet:def}
 \ptsub{\Pj} >  25\GeV = \ptsub{\Pj,\text{cut}} 
\qquad {\rm and} \qquad |y_{\Pj}| < 2.5
\end{align}
is rejected.

\item \underline{$\Pp\Pp \to \mu^+ \nu_\mu \Pe^- \bar \nu_\Pe\, \Pj$}

For the production of two off-shell $\PW$ bosons in association with a
jet, the lepton cuts \refeq{eq:lepton_cuts_1} and \refeq{eq:lepton_cuts_2}
are preserved.
In addition, one jet has to fulfil Eq.~\eqref{jet:def} and
\begin{align}
  \Delta R_{\Pl^\pm \Pj} > 0.4\;.
\end{align}
A veto with the same parameters is then applied on the 
occurrence of any additional jet.

\end{itemize}

\noindent
\change{The implementation of the applied jet veto follows the experimental
  analysis in \citere{Aaboud:2016mrt} and is not \change{driven} by theoretical considerations.}
Note that experimentally, a further veto on b-jets is usually applied, both,
for $\PW\PW$ and $\PW\PW\Pj$ production, thereby eliminating contributions
from single-top and top-pair production. 
However, as we exclude events with final-state bottom quarks in our computation,
\emph{i.e.}\ resonant top-quark propagators, there is no need to apply such \change{a} veto here.
\change{The listed event selections and the observable calculations for the results presented in
the following} have been implemented in \textsc{Rivet} \cite{Buckley:2010ar}.

\subsubsection*{\MEPSatNLO calculation}

For the \MEPSatNLO predictions we merge the NLO \QCD\ matrix elements
for $\Pp\Pp \to \mu^+ \nu_\mu \Pe^- \bar \nu_\Pe$ and $\Pp\Pp \to \mu^+
\nu_\mu \Pe^- \bar \nu_\Pe \Pj$ and the tree-level matrix elements for
$\Pp\Pp \to \mu^+ \nu_\mu \Pe^- \bar \nu_\Pe \Pj\Pj$ and $\Pp\Pp \to \mu^+
\nu_\mu \Pe^- \bar \nu_\Pe \Pj\Pj\Pj$, provided by \Comix~\cite{Gleisberg:2008fv}.
The merging scale is set to
\begin{equation}
  \Qcut = 30\;\GeV\,.
\end{equation}
Per default we use the CKKW scale-setting prescription of
Eqs.~\refeq{eq:muRCKKW} and \refeq{eq:mQCKKW} to define the
renormalisation, factorisation, and resummation scales, with the scale of
the inner core process, $\mucore$, given by
\begin{equation}
  \mucore=\tfrac{1}{2}\,\overline{E}_{\mr{T},\PW}\,.
\end{equation}
As discussed in \refse{se:merged_predictions}, this directly 
fixes the renormalisation scale to the CKKW scale, $\muR=\muCKKW$, 
and the factorisation and parton-shower starting scale to the 
core scale $\muF=\muQ=\mucore$. 
We describe the $\PW\PW$ and $\PW\PW\Pj$ production processes at NLO \QCD,
\emph{i.e.}\ $\nmaxnlo=1$
and evaluate approximate NLO \EW\ corrections up to this order. 
LO contributions are taken into account for $\PW\PW\Pj\Pj$ and
$\PW\PW\Pj\Pj\Pj$ production processes, \emph{i.e.}\ $\nmax=3$,
that are subject to local $K$~factors, cf.\ Eq.~\eqref{eq:local_k-factor}.
All Standard Model parameters and event-selection criteria are defined as
detailed above, thus, in compliance with the fixed-order calculations.

\section{Numerical results}
\label{se:results}

In this section, we discuss the numerical results obtained for the processes
$\Pp\Pp \to \mu^+ \nu_\mu \Pe^- \bar \nu_\Pe$ and $\Pp\Pp \to \mu^+
\nu_\mu \Pe^- \bar \nu_\Pe \Pj$.  We present both, fiducial cross sections
and differential distributions.  We provide theoretical predictions at LO,
NLO \QCD\ and \EW, as well as \MEPSatNLO incorporating \EW\ corrections in
the virtual \EW\ approximation. Particular emphasis is put on the combination of
\QCD\ and \EW\ corrections and the impact of the \QCD\ parton shower. 

In this article, the NLO \QCD\ and \EW\ cross sections are defined as
\begin{equation}
\sigma^{\mathrm{NLO}}_{\QCD} = \sigma^{\mathrm{Born}}\change{\left(\vh 1 + \delta^{\mathrm{NLO}}_{\QCD}\right)} \qquad \text{and} \qquad
\sigma^{\mathrm{NLO}}_{\EW} = \sigma^{\mathrm{Born}}\change{\left(\vh 1 + \delta^{\mathrm{NLO}}_{\EW}\right)}\,,
\end{equation}
respectively. The additive prescription to combine \QCD\ and \EW\ corrections reads
\begin{equation}
\sigma^{\mathrm{NLO}}_{\QCDpEW} = \sigma^{\mathrm{Born}}\change{\left(\vh 1 + \delta^{\mathrm{NLO}}_{\QCD} + \delta^{\mathrm{NLO}}_{\EW}\right)}\,,
\end{equation}
while the multiplicative one is defined as
\begin{equation}
  \begin{split}
    \sigma^{\mathrm{NLO}}_{\QCDtEW}
    =&{}\; \sigma^{\mathrm{Born}}
           \change{
           \left(\vh
             1 + \delta^{\mathrm{NLO}}_{\QCD}
           \right)
           \left(\vh
             1 + \delta^{\mathrm{NLO}}_{\EW}
           \right)
           }\\
    =&{}\; \sigma^{\mathrm{NLO}}_{\QCD}
           \change{
           \left(\vh
             1 + \delta^{\mathrm{NLO}}_{\EW}
           \right)
           }
    =      \sigma^{\mathrm{NLO}}_{\EW}
           \change{
           \left(\vh
             1 + \delta^{\mathrm{NLO}}_{\QCD}
           \right)
           }
           \;.
  \end{split}
\end{equation}
The difference between these two prescriptions could be used as an
estimate of the missing \QCDmEW\ mixed corrections.  In this context,
the NLO \QCDtEW\ combination can be understood as an improved
prediction when the typical scales of the \QCD\ and \EW\ corrections are
well separated.  In the following, we argue that this is the case for
the processes at hand.

For the first time, we also present predictions
based on a multiplicative scheme, cf.\ Eq.~(\ref{eq:mepsnlo_qcdtew}), to
implement approximate NLO \EW\ corrections in merged calculations of NLO \QCD\ 
matrix elements for $\PW\PW$ and $\PW\PW\Pj$ production matched to the \Sherpa parton
shower. 
\change{In addition, to validate the virtual EW 
approximation employed in the multijet-merged calculations, 
we study 
\begin{equation}
  \begin{split}
    \sigma^{\mathrm{NLO}}_{\QCDtEWapprox}
    =&{}\; \sigma^{\mathrm{Born}}
    \left(\vh
           1 + \delta^{\mathrm{NLO}}_{\QCD}
             \right)
           \left(\vh
           1 + \deltaEWapprox
             \right)
           \;,
  \end{split}
\end{equation}
with \deltaEWapprox defined in Eq.\ \eqref{eq:deltaEWapprox}.}

To estimate the theoretical uncertainty of our predictions we consider
the usual set of 7-point scale variations, \emph{i.e.}\
$\{(\tfrac{1}{2}\muR,\tfrac{1}{2}\muF)$, $(\tfrac{1}{2}\muR,\muF)$,
$(\muR,\tfrac{1}{2}\muF)$, $(\muR,\muF)$, $(\muR,2\muF)$, $(2\muR,\muF)$,
$(2\muR,2\muF)\}$. The uncertainties quoted for fiducial cross sections
and differential distributions in the following correspond to
the resulting envelope. All systematic variations are evaluated on-the-fly
using the implementation of the algorithm presented in~\citere{Bothmann:2016nao}
in the \Sherpa framework.

\subsection{Fixed-order results}
\label{se:results:fo}

\subsubsection{\texorpdfstring{$\PW\PW$}{WW} production}
\label{se:results:fo:WW}

In Table~\ref{tab:ww}, fiducial cross sections for
$\Pp\Pp \to \mu^+ \nu_\mu \Pe^- \bar\nu_\Pe$ at LO, NLO \QCD, and NLO \EW\ 
accuracy are compiled. Thanks to the jet veto, the \QCD\ corrections
amount to $+0.4\%$ only, while 
the \EW\ corrections reach $-3\%$. The two prescriptions of combining the \QCD\ and \EW\ 
corrections lead to practically identical results. The contribution from
$\gamma\gamma$ initial states contained in the LO cross section amounts
to  $4.006(5)\fb$, \ie $1.3\%$. The \QCD\ scale uncertainty of the LO
prediction is estimated, in the absence of a renormalisation scale dependence, 
by variations of the factorisation scale by factors of $\tfrac{1}{2}$ and $2$.
For the NLO \QCD\ result we consider the full 7-point variations. For the LO prediction
this yields an estimated uncertainty of order $6\%$,
while at NLO \QCD\ it is reduced to the $2\%$ level \change{for our
  set-up. However, this reduction depends strongly on the precise form of the jet veto.}

In addition to the channels considered in Table~\ref{tab:ww}, there exists a
loop-induced contribution from the partonic process
$\Pg\Pg \to \mu^+ \nu_\mu \Pe^- \bar\nu_\Pe$ at order $\mathcal{O} \left(\alphas^2 \alpha^4 \right)$.
In the present set-up, it amounts to $29.38(1)^{+24.6\%}_{-17.9\%} \fb$, \emph{i.e.}\ 
$9.2\%$ of the LO prediction for $\Pq\bar \Pq \to \mu^+ \nu_\mu \Pe^- \bar\nu_\Pe$.

\begin{table}
  \begin{center}
    \begin{tabular}{c|c|c|c|c}\rule[-2ex]{0ex}{3ex}
    $\sigma^{\mathrm{LO}}$  [$\fb$] & 
    $\sigma^{\mathrm{NLO}}_{\QCD}$  [$\fb$] &
    $\sigma^{\mathrm{NLO}}_{\EW}$  [$\fb$] &
    $\sigma^{\mathrm{NLO}}_{\QCDpEW}$  [$\fb$] & 
    $\sigma^{\mathrm{NLO}}_{\QCDtEW}$  [$\fb$] \\
    \hline \rule[-2ex]{0ex}{5ex}%
    319.7(1)$^{+5.2\%}_{-6.3\%}$ &
    321.1(8)$^{+2.1\%}_{-2.2\%}$ &
    310.8(5) & 
    312.2(9) & 
    312.1(9) 
    \end{tabular}
  \end{center}
  \caption{\label{tab:ww}
    Fiducial cross sections for $\Pp\Pp \to \mu^+
    \nu_\mu \Pe^- \bar \nu_\Pe$ at $\sqrt{s}=13\TeV$ at LO,
    NLO \QCD, and NLO \EW. For the combination of NLO \QCD\ and NLO \EW\
    corrections results for the additive and multiplicative
    prescription are quoted. } 
\end{table}

In the following, several differential distributions are presented.
In the upper panels of the plots, absolute predictions at LO, NLO \QCD,
NLO \QCDpEW\change{, and NLO \QCDtEW\ accuracy are shown.
In addition, NLO \QCDtEWapprox\ results are displayed to gauge the 
quality of the approximation to be used when constructing the 
merged results in \refse{sec:results:merged:merged}. 
The} lower panels
contain the corresponding results normalised to the NLO \QCD\ ones. Accordingly,
in what follows we quote corrections/deviations relative to the NLO \QCD\ 
prediction, corresponding to the ratio plots provided. The scale uncertainty of the NLO \QCD\ 
prediction, given by the envelope of the 7-point variations of $\muR$ and $\muF$, is indicated by the green
band.

In Fig.~\ref{fig:transverse_ww}, various transverse-momentum observables
as well as the distribution in the rapidity of the anti-muon are shown.  
\bfig
  \center
  \includegraphics[width=0.45\textwidth]{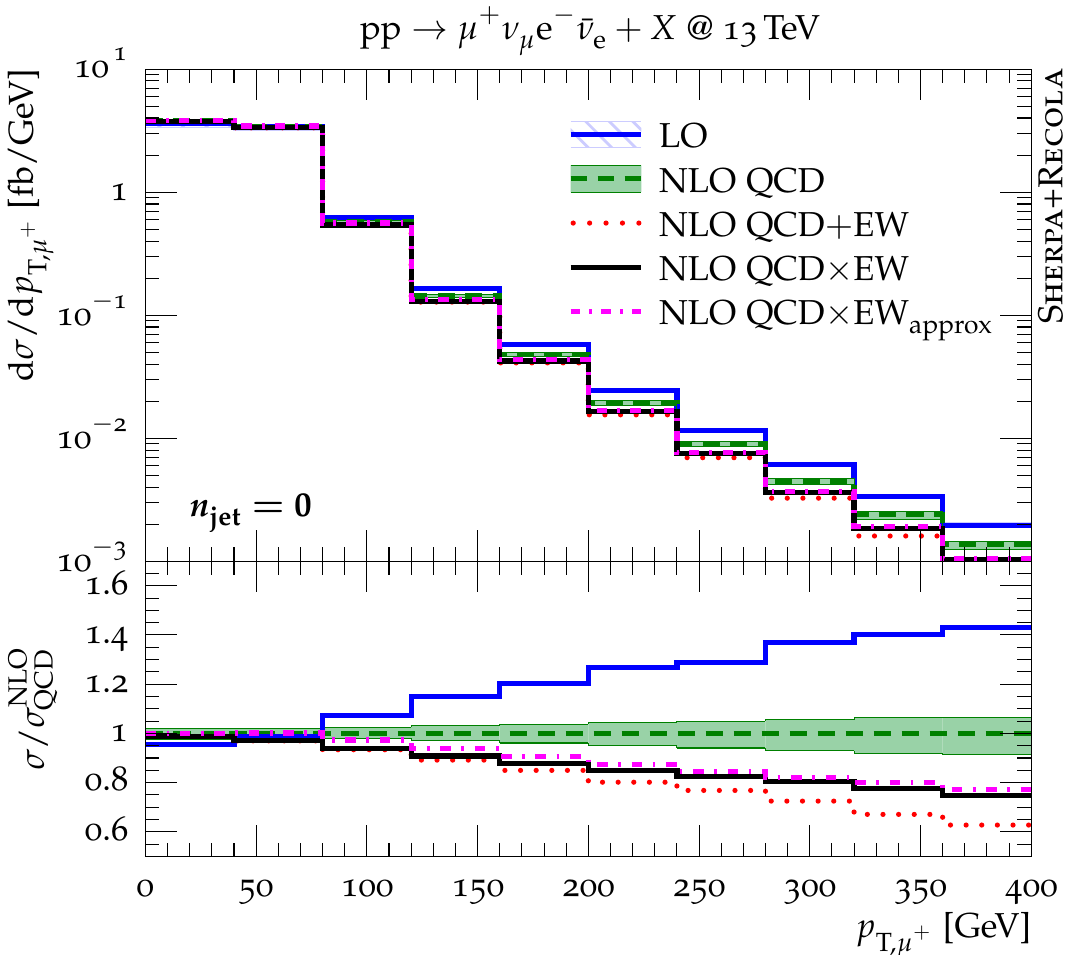}\hfill
  \includegraphics[width=0.45\textwidth]{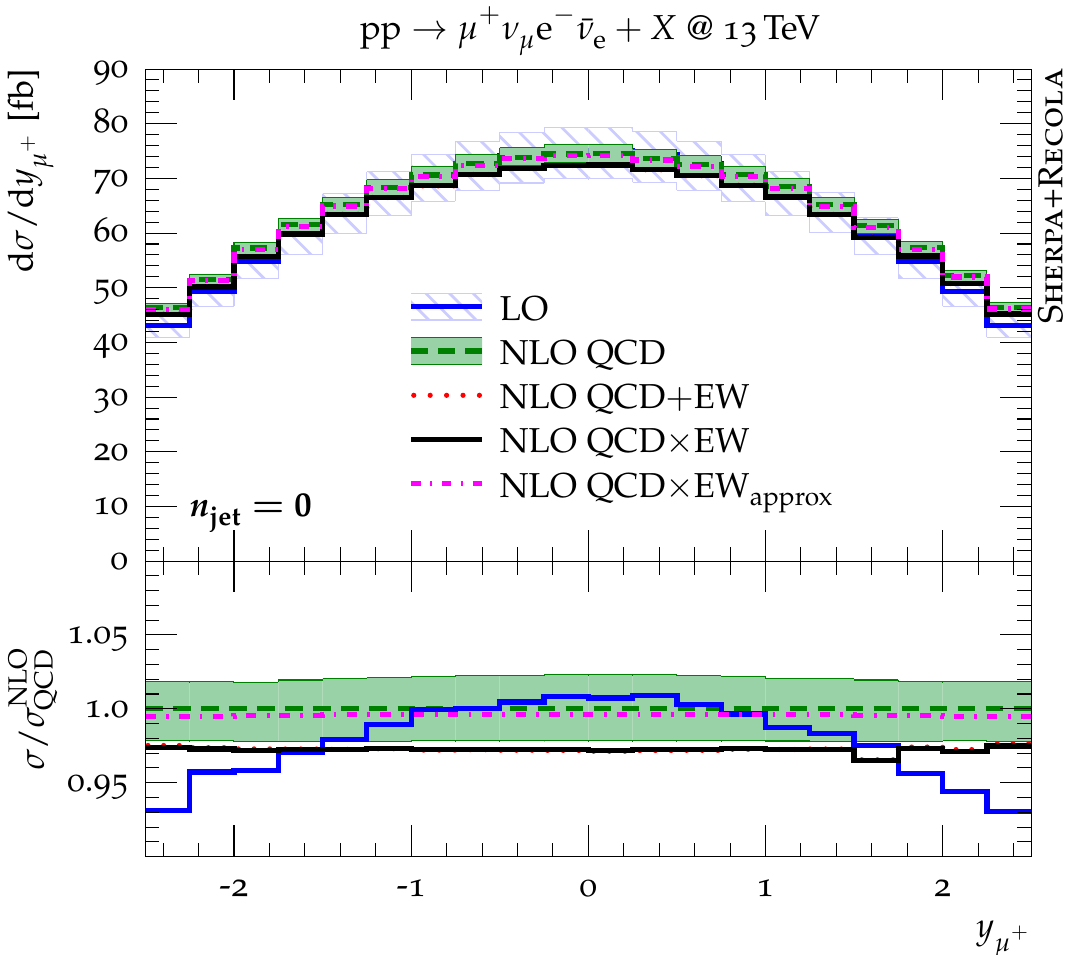}
  \includegraphics[width=0.45\textwidth]{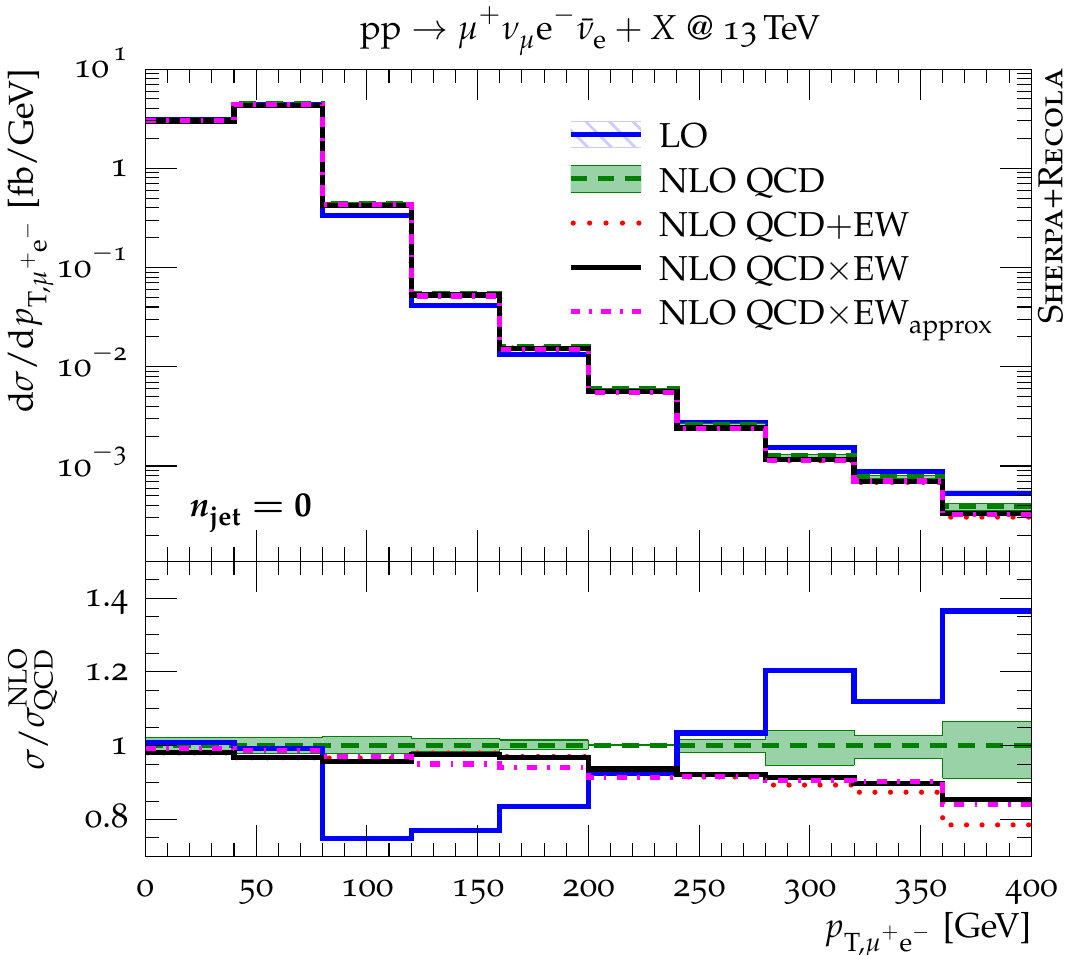}\hfill
  \includegraphics[width=0.45\textwidth]{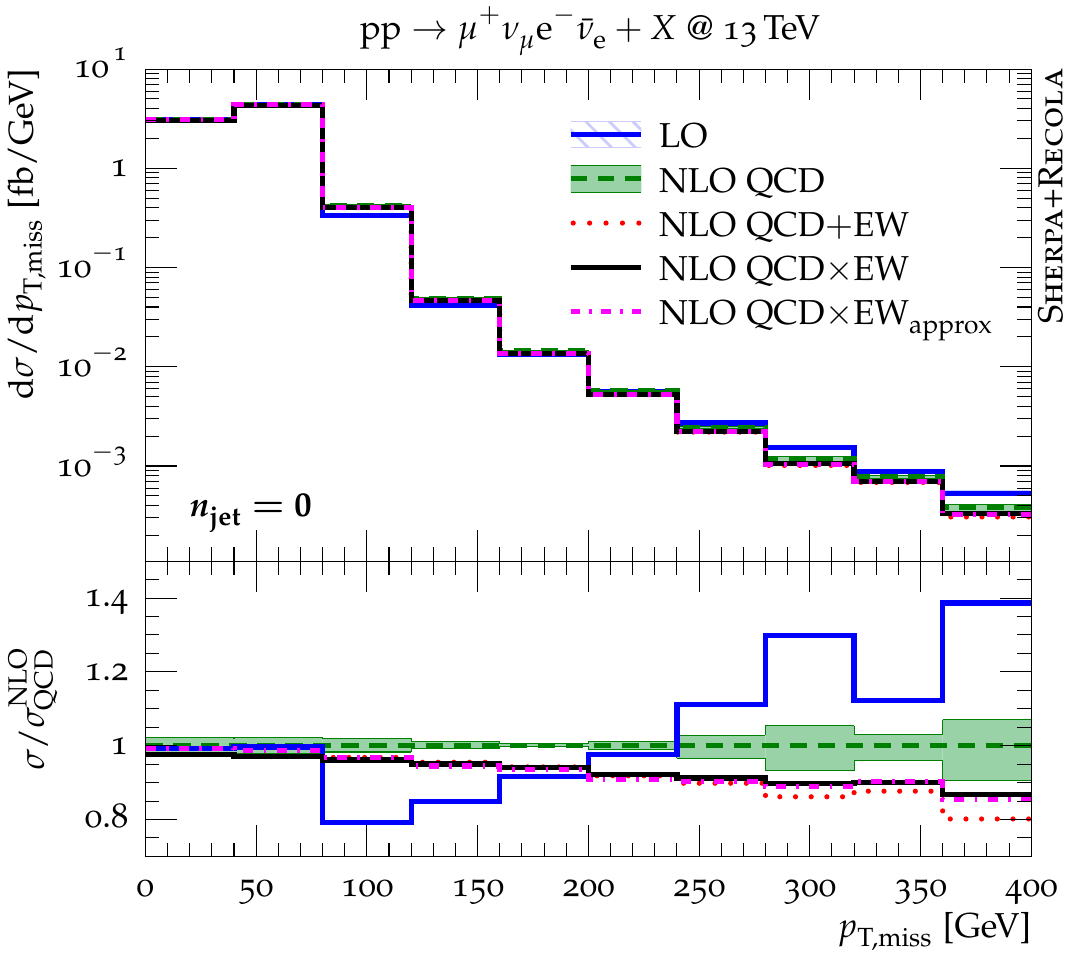}
  \caption{\label{fig:transverse_ww} Differential distributions for
    $\Pp\Pp \to \mu^+ \nu_\mu \Pe^- \bar\nu_\Pe$ at LO, NLO \QCD, NLO
    \QCDpEW, \change{NLO \QCDtEW, and NLO \QCDtEWapprox}: 
  Transverse momentum of the anti-muon (top left),
  rapidity of the anti-muon (top right),
  transverse momentum of the anti-muon--electron system (bottom left), and
  missing transverse momentum (bottom right).
  The upper panels show the absolute predictions, while the lower ones
  display the ratio of the various predictions with respect to the NLO
  \QCD\ predictions. 
 }
\efig%
For the distribution in the transverse momentum of the
anti-muon, the effect of \QCD\ corrections is rather large.  They tend
to lower the predictions for larger transverse momentum and
exceed $-40\%$ at $p_{\mathrm{T},\mu^+}=400\GeV$.  The large negative \QCD\ 
corrections result from the jet-veto cuts.  Owing to the Sudakov
logarithms in the virtual corrections, the \EW\ corrections follow the
same trend and exceed $-25\%$ at $400\GeV$ with respect
to the NLO \QCD\ prediction. For the rapidity distribution of the
anti-muon, the \QCD\ corrections are moderate throughout, being about
$-1\%$ in the central region, while becoming positive in the
peripheral region at a level of $+5\%$.  On the other hand, the \EW\ 
corrections exceed the estimated NLO \QCD\ scale uncertainty but do not
feature a sizeable shape distortion.  The distributions in the
transverse momentum of the anti-muon--electron system and the missing
transverse energy display very similar behaviour both qualitatively
and quantitatively.  This is explained by the fact that the missing
momentum is defined as the sum of the two neutrino momenta.  This
observable has thus a very similar kinematics as the transverse
momentum of the two charged leptons.  In both cases, the NLO \QCD\ 
corrections reach about $-40\%$ at $400\GeV$, while the
\EW\ ones are of order $-15\%$ for the same transverse
momentum.  Around $100\GeV$ the NLO \QCD\ prediction suddenly exceeds
the LO one at a level of $20\%$.  The corrections then turn negative
towards high transverse momentum.  This can be understood as follows.
At LO, contributions with two
resonant W~bosons require these bosons to be back-to-back and therefore
cannot contribute to events with transverse momenta $p_{\mathrm{T},\mu^+\Pe^-}$
or $p_{\mathrm{T},\text{miss}}$ larger than about $\MW$
\cite{Biedermann:2016guo,Kallweit:2017khh}. Thus, at LO such events
can only result from contributions with at most one resonant W boson
and are therefore suppressed. 
At NLO, the momentum of the extra jet can balance the momenta of
the two resonant W~bosons allowing for large $p_{\mathrm{T},\mu^+\Pe^-}$
and/or $p_{\mathrm{T},\text{miss}}$ also in the presence of two
resonant W~bosons.
Going towards higher transverse momenta, such configurations are then
suppressed by the jet veto that forbids hard jets that would balance the $\PW\PW$ system.
\change{The fluctuations in the tails of the
  $p_{\mathrm{T},\mu^+\Pe^-}$ and $p_{\mathrm{T},\text{miss}}$
  distributions are of statistical origin.} 

\change{The NLO \QCDtEWapprox\ results follow closely the exact 
NLO \QCDtEW\ ones, staying within $3\%$ of them in the high-$p_\mathrm{T}$ 
regions, despite the presence of a jet veto which is not accounted for 
in the integrated-out approximate real-emission corrections. 
For inclusive quantities, like the muon rapidity distribution or 
the first two bins of the various transverse-momentum distributions, 
they reproduce the exact NLO EW corrections also within $3\%$. 
The behaviour of the approximation thus meets the expectations based 
on its construction.
}

In Fig.~\ref{fig:invariant_ww}, invariant-mass distributions and
angular distributions are displayed.  
\bfig
  \center
  \includegraphics[width=0.45\textwidth]{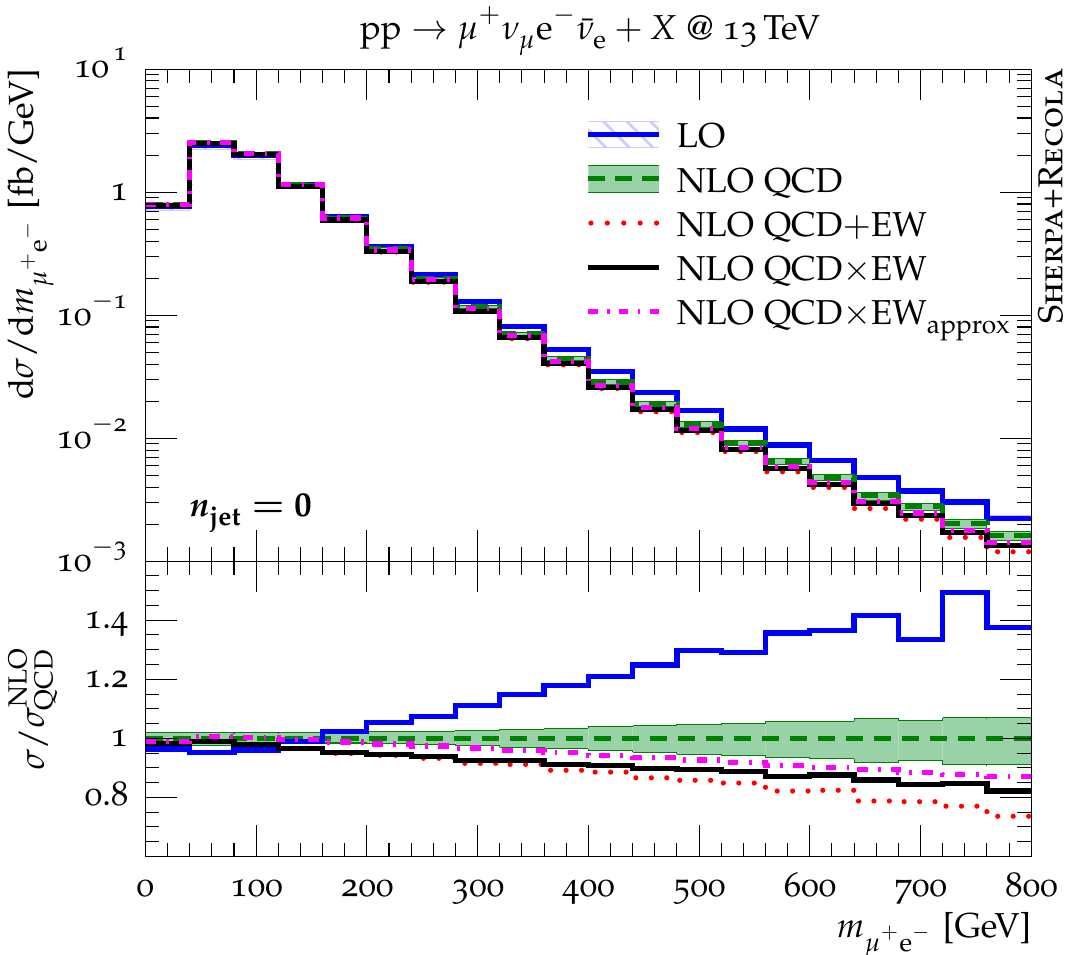}\hfill
  \includegraphics[width=0.45\textwidth]{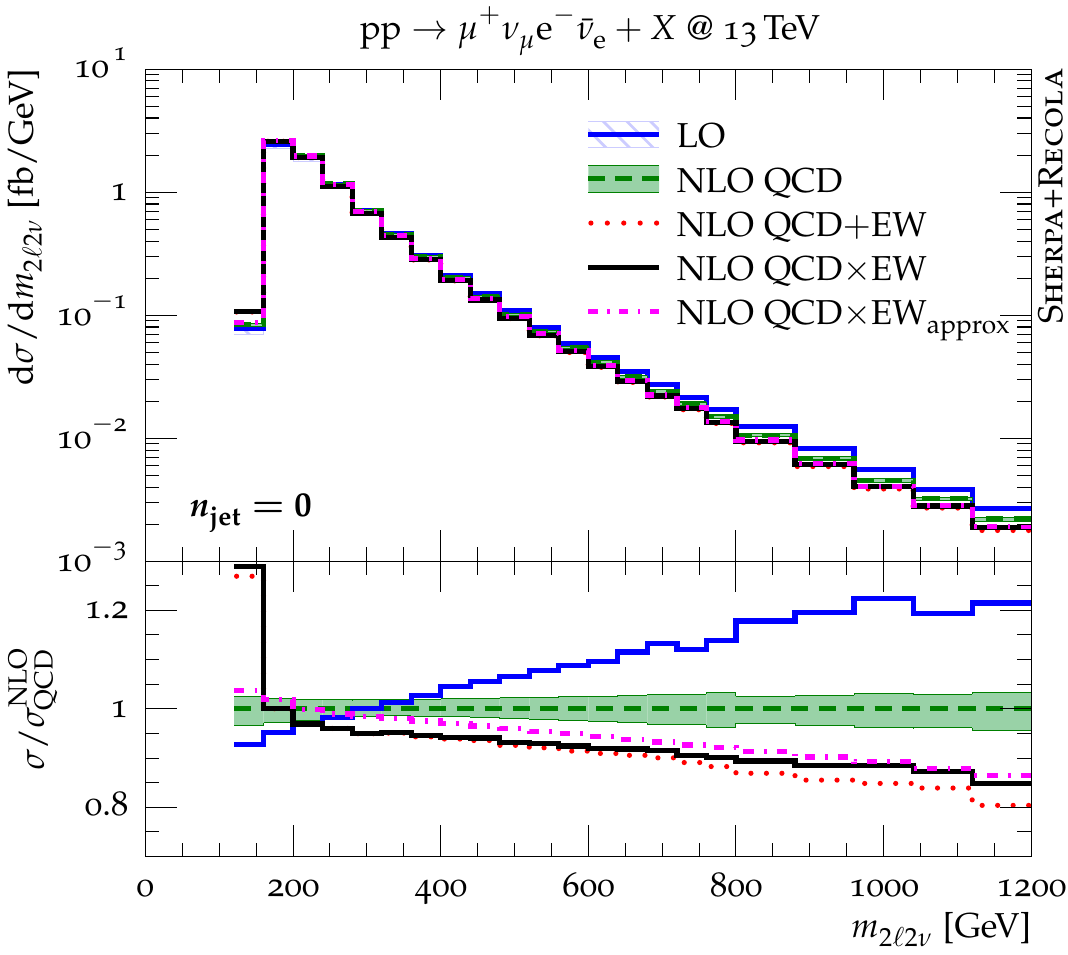}
  \includegraphics[width=0.45\textwidth]{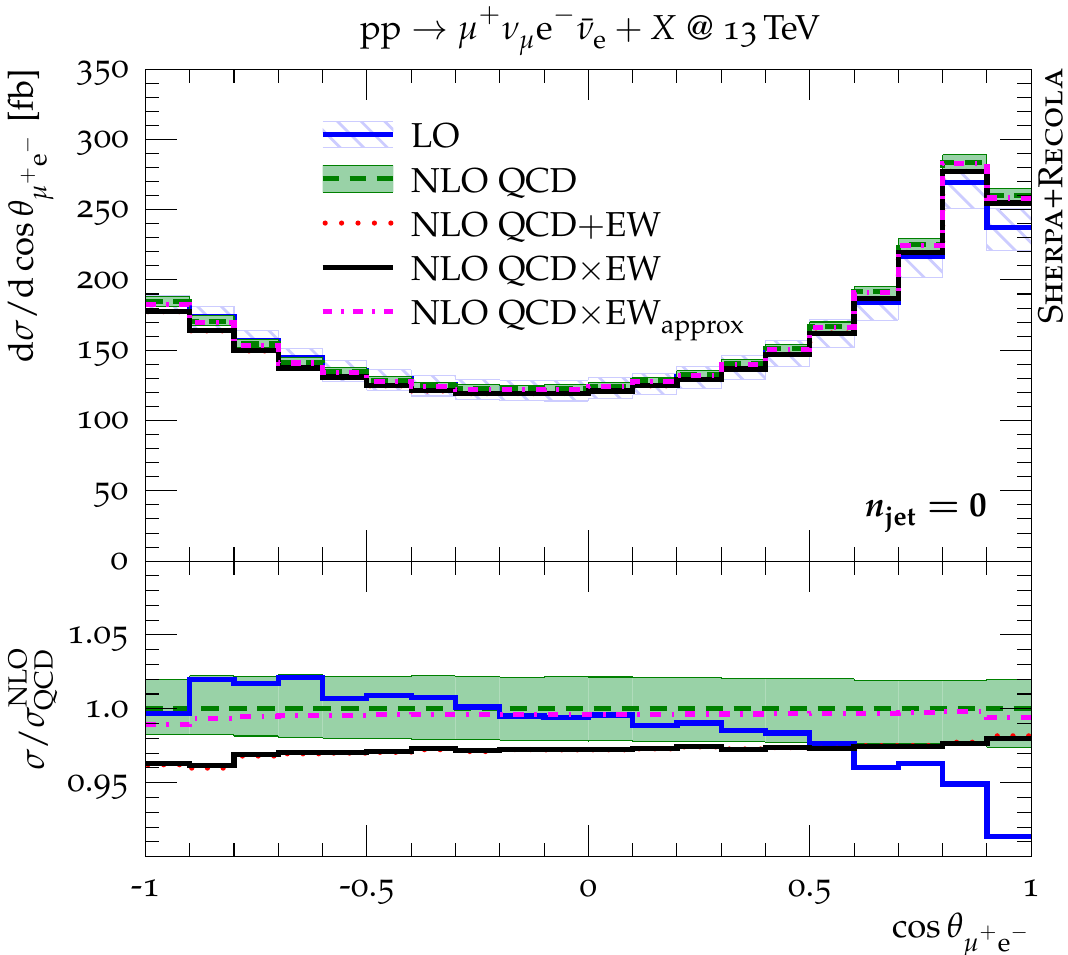}\hfill
  \includegraphics[width=0.45\textwidth]{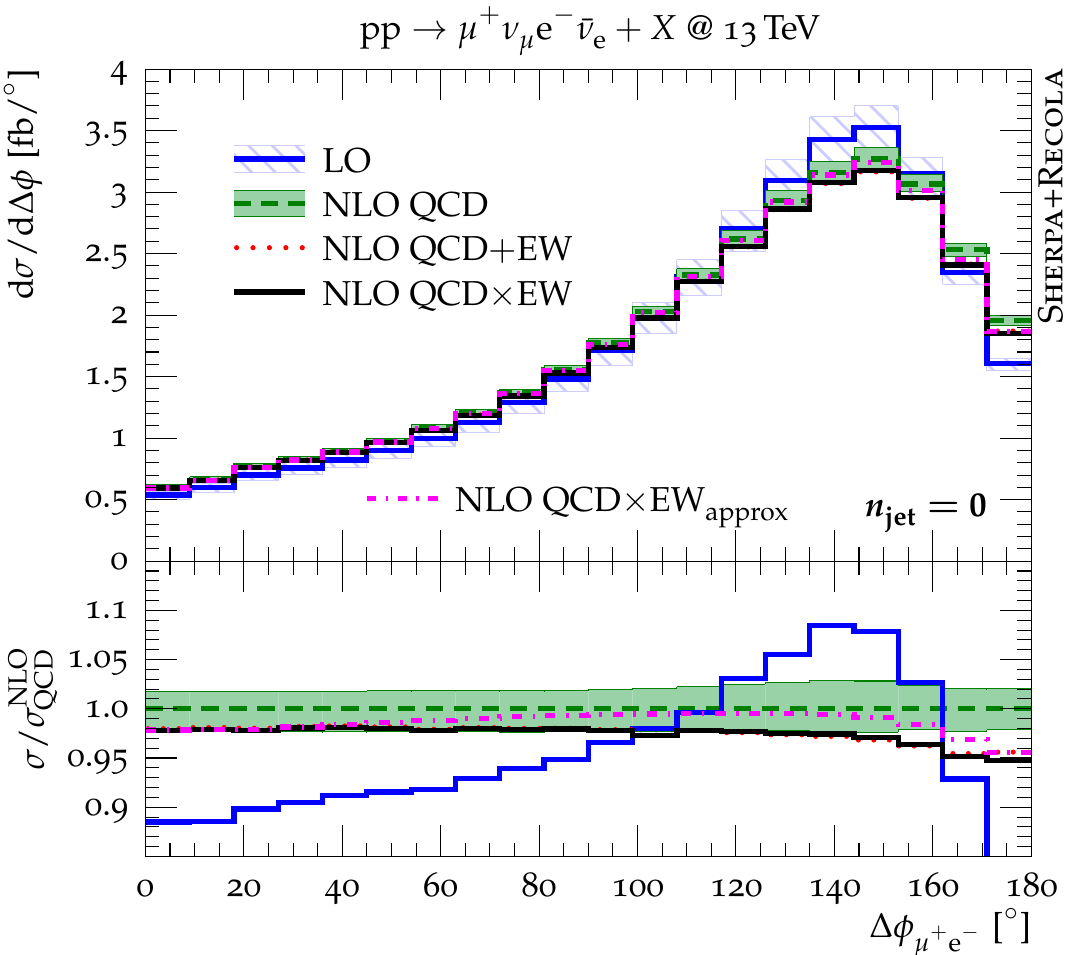}
  \caption{\label{fig:invariant_ww} Differential distributions for $\Pp\Pp \to \mu^+ \nu_\mu \Pe^- \bar\nu_\Pe$ at LO, NLO \QCD, NLO \QCDpEW,\change{ NLO \QCDtEW, and NLO \QCDpEWapprox}:
  Invariant mass of the anti-muon and electron (top left),
  invariant mass of the four leptons (top right),
  cosine of the angle between the anti-muon and the electron (bottom left), and
  azimuthal-angle distance between the anti-muon and the electron (bottom right).
  The upper panels show the absolute predictions, while the lower ones
  display the ratio of the various predictions with respect to the NLO
  \QCD\ predictions. 
 }
\efig
For the distribution
in the invariant mass of the two charged leptons, the \QCD\ corrections
are largely negative, increasing towards high invariant mass and reach
$-50\%$ above $0.8\TeV$.  On the other hand, the \EW\ corrections
steadily approach $-20\%$ at $0.8\TeV$ thanks to the effect of enhanced
\EW\ logarithms.  Despite not being a physical observable, the
distribution in the invariant mass of the four leptons is interesting
to study as it serves as a proxy in studies of physics beyond the Standard Model. 
The behaviour is qualitatively similar to the one of the di-lepton
invariant-mass distribution.  The distribution in the cosine of the
angle between the two charged leptons displays \QCD\ corrections 
\change{smaller than $1\%$} near $\theta=\pi$  reaching \change{$+9\%$} near $\theta=0$.
Thus, when the two leptons are back-to-back the \QCD\ 
corrections are negative, while when they are aligned they turn
positive.  The latter kinematic situation is the most probable and
corresponds to a central production of the two gauge bosons.
The \EW\ corrections are rather smooth and vary by less than $3\%$
between the two extreme kinematic configurations.  Finally, the
distribution in the azimuthal distance between the two charged leptons
displays rather moderate \QCD\ corrections.  They reach $-10\%$  around
$\Delta \phi_{\mu^+\Pe^-} \simeq 140^{\circ}$, \ie when the
two charged leptons are almost in a back-to-back configuration.
In both angular distributions \EW\ corrections are at the level of a few
per cent,
however, exceeding the NLO \QCD\ scale-uncertainty estimate.

\change{
The approximate NLO EW corrections reproduce the invariant-mass spectra 
well in the Sudakov regions, though generally slightly worse than the 
transverse momentum spectra of Fig.~\ref{fig:transverse_ww}. 
Again, the exact NLO EW corrections for inclusive distributions like 
$\cos\theta_{\mu^+\Pe^-}$ and $\Delta\phi_{\mu^+\Pe^-}$ are also
reproduced within about $3\%$. 
}

The difference between the additive and multiplicative prescriptions
for combining \QCD\ and \EW\ corrections is in general small. However, in
regions where both \QCD\ and \EW\ corrections become large, such as for
high transverse momenta or invariant masses, the difference can amount
to ten per cent or more.

\subsubsection{\texorpdfstring{$\PW\PW\Pj$}{WWj} production}
\label{se:results:fo:WWj}

In the same way as for di-boson production, fiducial cross sections at LO, NLO \QCD,
and NLO \EW\ accuracy are given for $\Pp\Pp \to \mu^+ \nu_\mu \Pe^- \bar \nu_\Pe \Pj$ in Table~\ref{tab:wwj}.
Notably, \QCD\ scale uncertainties are almost a factor two larger than
for $\PW\PW$ production 
owing to the additional power in the strong coupling already at LO. 
As before, the inclusion of NLO \QCD\ corrections reduces the scale uncertainties 
observed at LO.
\begin{table}
  \begin{center}
    \begin{tabular}{c|c|c|c|c}\rule[-2ex]{0ex}{3ex}%
      $\sigma^{\mathrm{LO}}$ [$\fb$] & 
      $\sigma^{\mathrm{NLO}}_{\QCD}$ [$\fb$] &
      $\sigma^{\mathrm{NLO}}_{\EW}$  [$\fb$] &
      $\sigma^{\mathrm{NLO}}_{\QCDpEW}$ [$\fb$] & 
      $\sigma^{\mathrm{NLO}}_{\QCDtEW}$ [$\fb$]\\
      \hline \rule{0ex}{3ex}%
      162.5(1)$^{+11.2\%}_{-9.1\%}$ & 
      129.5(5)$^{+5.1\%}_{-8.9\%}$ & 
      155.5(1) & 
      122.5(5) & 
      123.9(5) \\
    \end{tabular}
  \end{center}
  \caption{\label{tab:wwj}
    Fiducial cross sections for $\Pp\Pp \to \mu^+ \nu_\mu \Pe^- \bar \nu_\Pe \Pj$ 
    at $\sqrt{s}=13\TeV$ at LO, NLO \QCD, and NLO \EW.
    Furthermore, results for the additive and multiplicative combination of
    NLO \QCD\ and NLO \EW\ corrections are given.}
\end{table}
In our calculational set-up, the numerical value of the NLO \QCD\ corrections differ
significantly from those for the $\PW\PW$ channel and amount to
$-20\%$ owing to the strong jet veto. The NLO \EW\ corrections
amount to $-4.3\%$, very similar to the case of $\PW\PW$ production.
This could be expected, since the additional gluon does not take part
in the \EW\ interaction.  As a consequence of the sizeable \QCD\
corrections, the additive and multiplicative combination of \QCD\ and \EW\ corrections
differ by about $1\%$. 

As for $\PW\PW$, there exists a loop-induced contribution
at order $\mathcal{O} \left(\alphas^3 \alpha^4 \right)$ from the
partonic process 
$\Pg\Pg \to \mu^+ \nu_\mu \Pe^- \bar\nu_\Pe \Pg$, which is not
included in 
Table~\ref{tab:wwj}. Its fiducial cross section amounts
to $11.941(3)^{+41.3\%}_{-27.5\%} \fb$, \emph{i.e.}\ $7.3\%$ of the tree-level
prediction, which is slightly smaller in comparison to the corresponding contribution
to the $\PW\PW$ process. However, the scale uncertainty on this channel is
particularly large. 

In the following, the same set of distributions is shown as for the
case of $\PW\PW$ production.  In addition, we include the distributions
in the transverse momentum and rapidity of the hardest jet (ordered in
transverse momentum), which are displayed first in
Fig.~\ref{fig:transverse_wwj}. 
\bfig
  \center
  \includegraphics[width=0.45\textwidth]{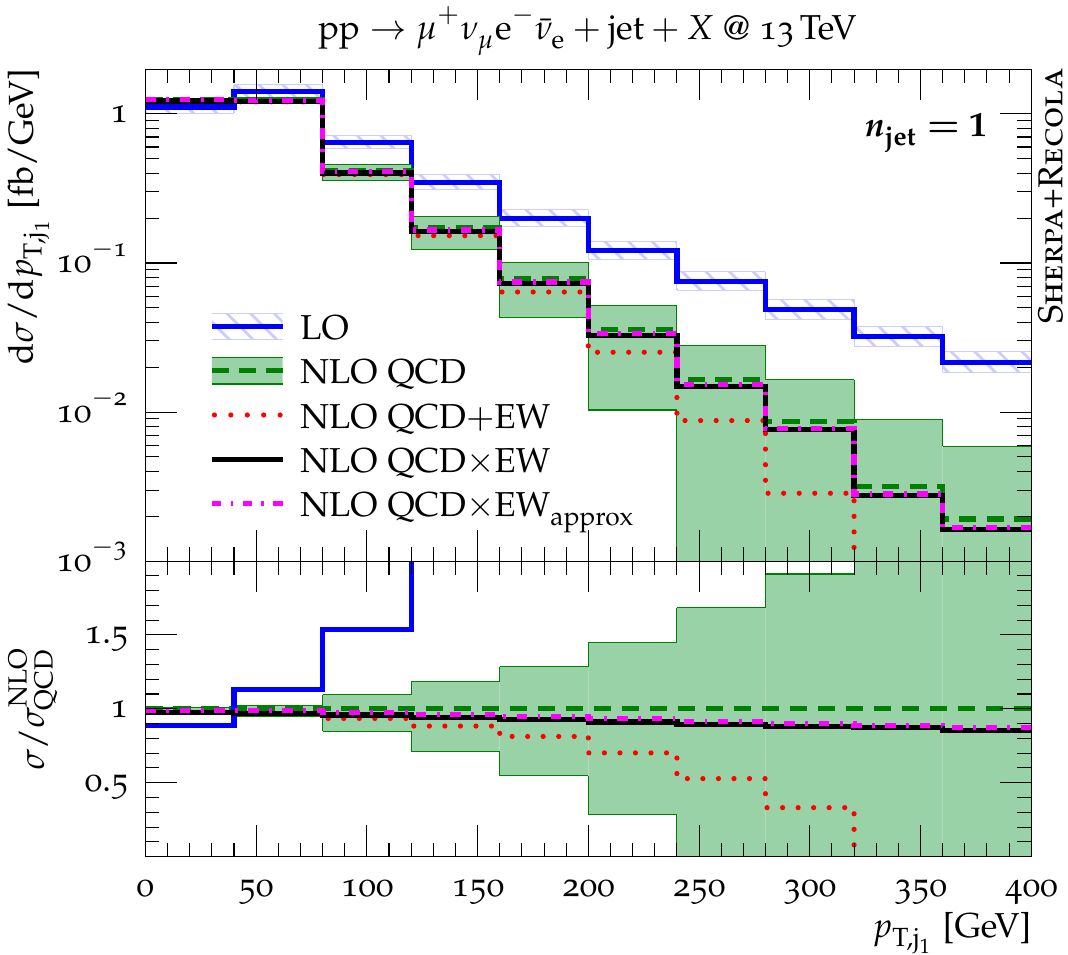}\hfill
  \includegraphics[width=0.45\textwidth]{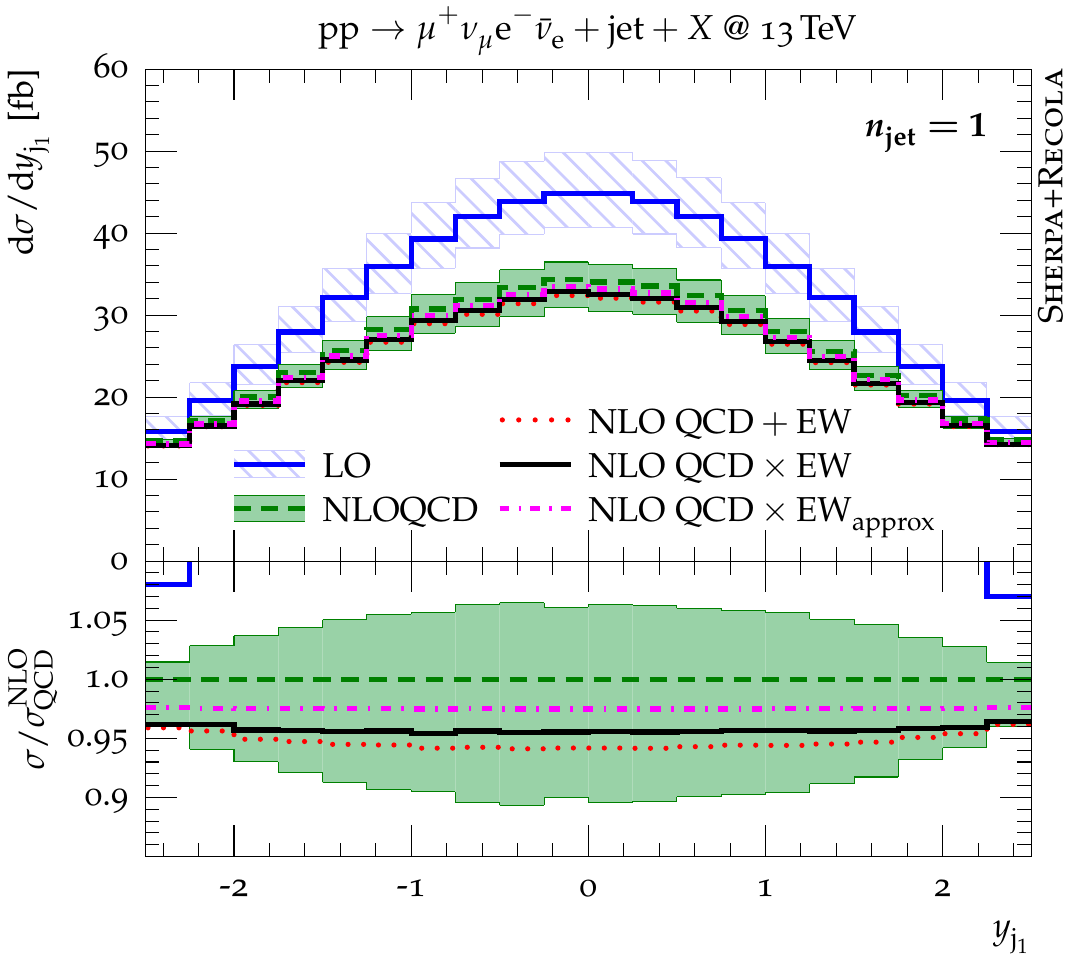}
  \includegraphics[width=0.45\textwidth]{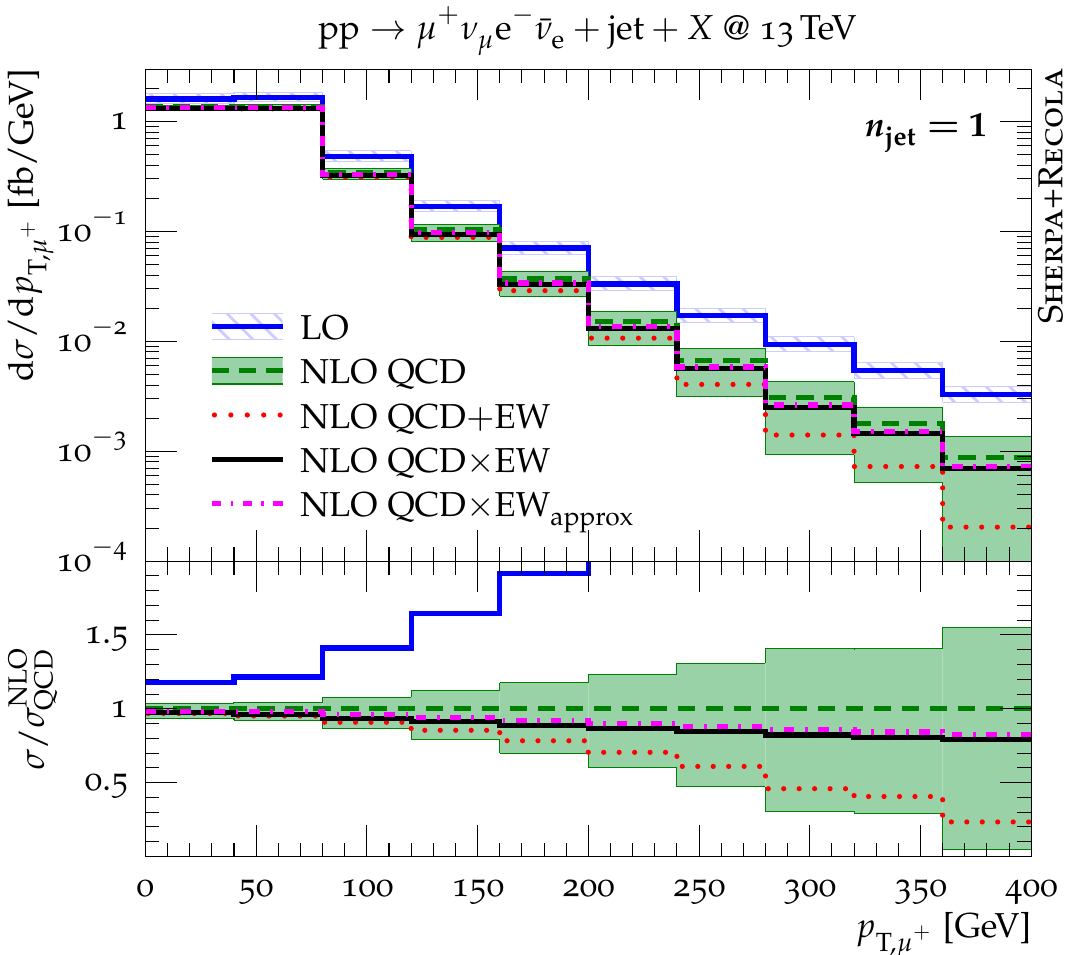}\hfill
  \includegraphics[width=0.45\textwidth]{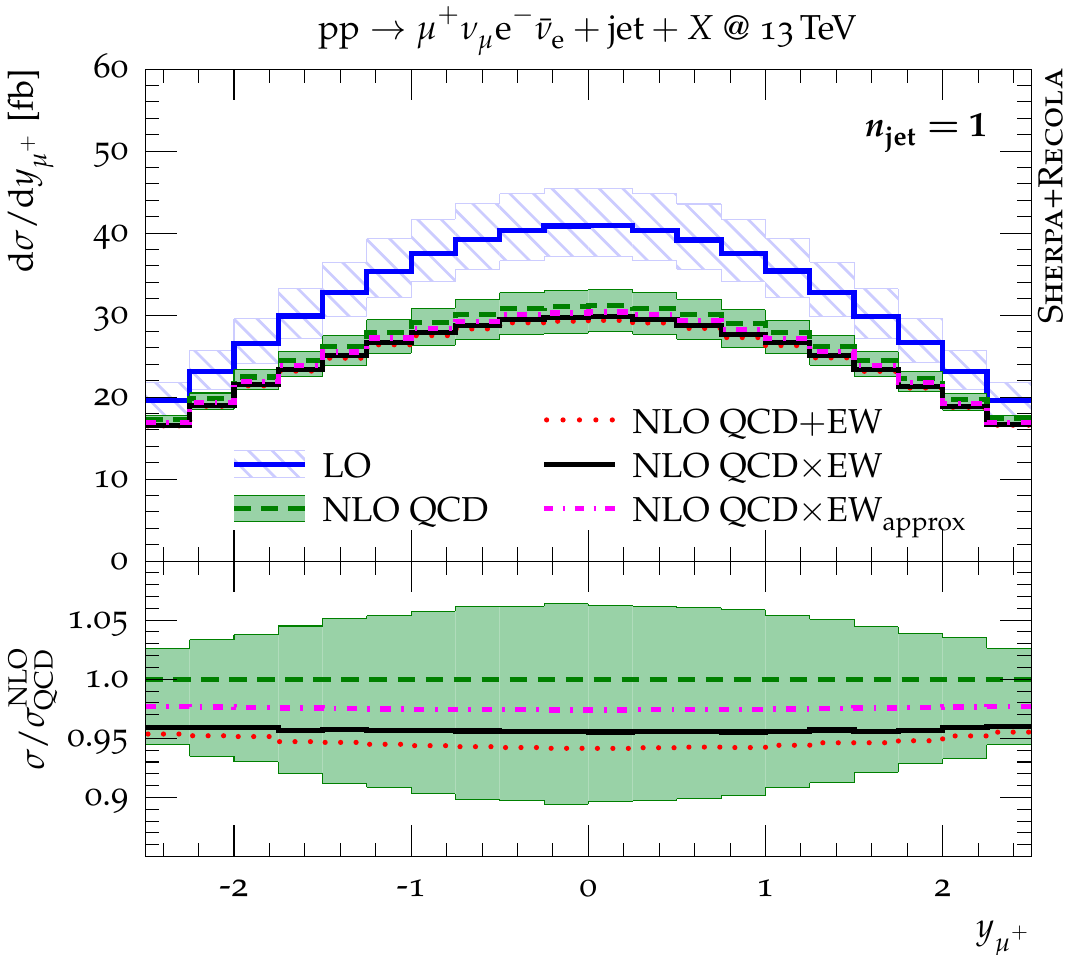}
  \includegraphics[width=0.45\textwidth]{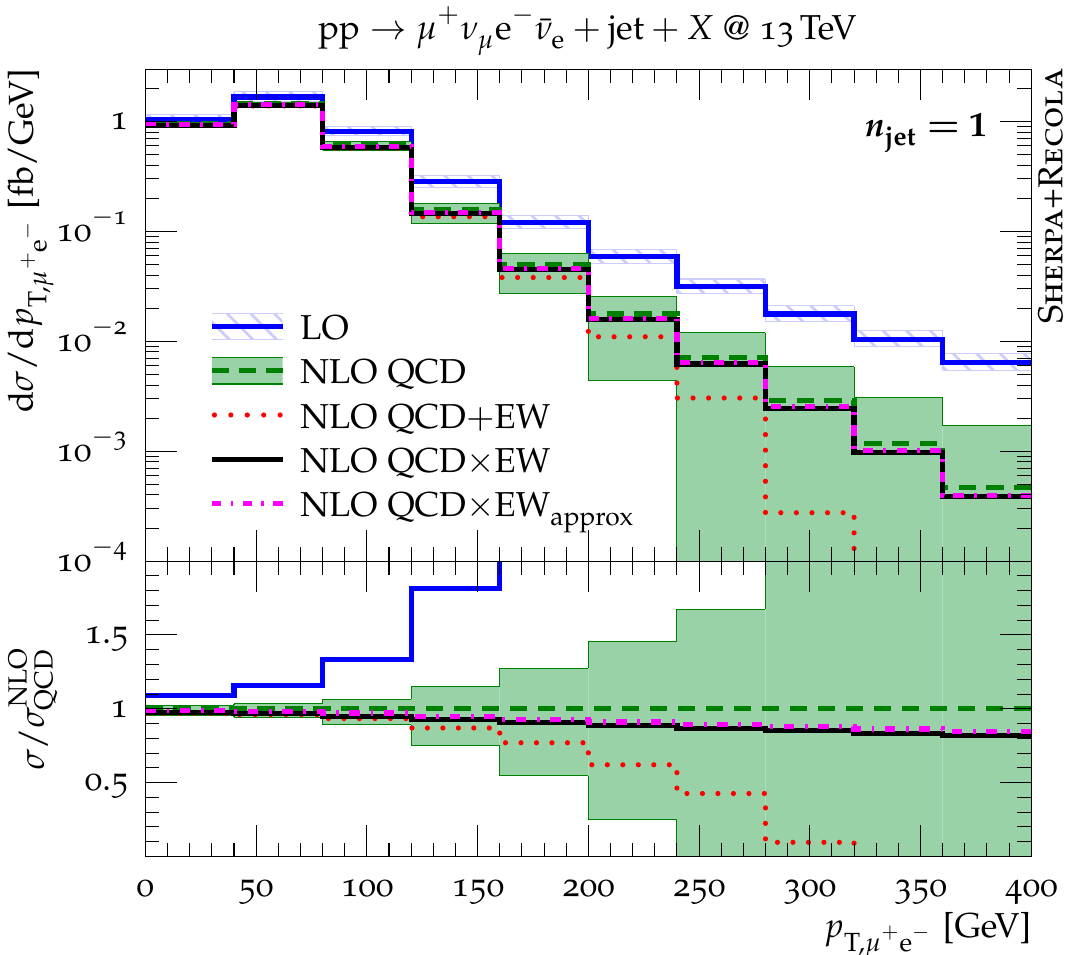}\hfill
  \includegraphics[width=0.45\textwidth]{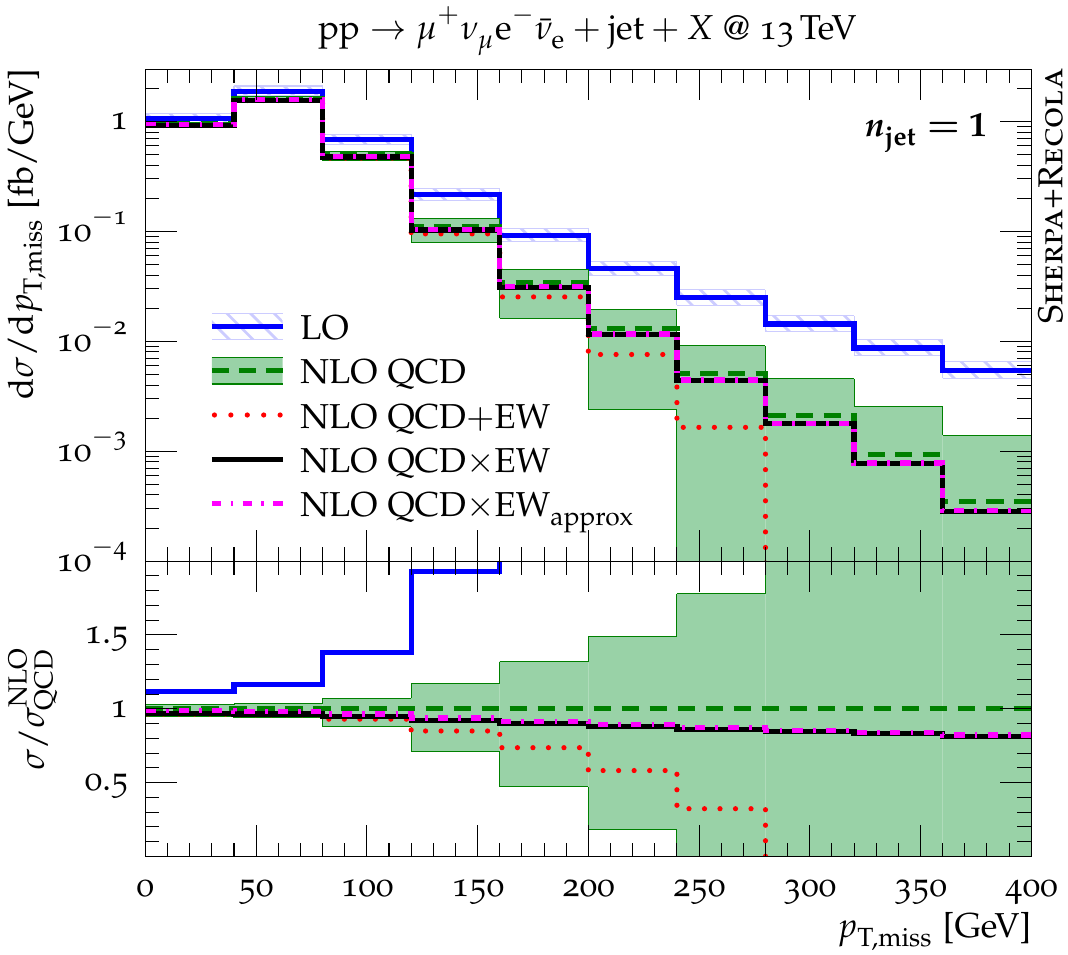}
  \caption{\label{fig:transverse_wwj} Differential distributions for $\Pp\Pp \to \mu^+ \nu_\mu \Pe^- \bar\nu_\Pe\Pj$ at LO, NLO \QCD, NLO \QCDpEW, \change{NLO \QCDtEW, and NLO \QCDtEWapprox}:
  Transverse momentum of the jet (top left),
  rapidity of the jet (top right),  
  transverse momentum of the anti-muon (middle left),
  rapidity of the anti-muon (middle right),
  transverse momentum of the anti-muon--electron system (bottom left), and
  missing transverse momentum (bottom right).
  The upper panels show the absolute predictions, while the lower ones
  display the ratio of the various predictions with respect to the NLO
  \QCD\ predictions. 
 }
\efig
\change{In the tail of the jet transverse-momentum distribution, very large NLO \QCD\ corrections as well
as sizeable NLO \EW\ corrections appear.}  For the
distribution in the rapidity of the jet, the \QCD\ corrections (not
visible on the lower panel) can be as large as $-40\%$.  On the other
hand, the \EW\ corrections are rather stable over the whole kinematic
range.  The qualitative behaviour is similar for the distributions in
the transverse momentum and the rapidity of the anti-muon.
Quantitatively the \QCD\ corrections are smaller for the distribution in
the transverse momentum for the anti-muon than in the one for the
leading jet,
but are at the same level for the rapidity distributions of the
anti-muon and leading jet.
Finally, the \QCD\ corrections to the distributions in
the transverse momentum of the anti-muon--electron system and in the
missing transverse momentum are rather different from the case of
$\PW\PW$ production, being much larger at large transverse momenta.
The different behaviour just above $\MW$ is due to the fact that for
$\PW\PW\Pj$ production configurations 
with two resonant $\PW$~bosons contribute in this phase-space region,
while these are excluded for $\PW\PW$ production at LO.
On the other hand, the \QCD\ corrections to the distributions in $p_{\mathrm{T},\mu^+\Pe^-}$
or $p_{\mathrm{T},\text{miss}}$  are similar to those for the distribution in
the transverse momentum of the jet owing to the recoil of the jet
against the $\PW\PW$ system.  The \QCD\ corrections exceed $-100\%$
above $150\GeV$ (when normalised to LO, the \QCD\ corrections stay
below $100\%$ in the considered range).
This is a \change{consequence} of the \change{applied} jet veto, which reduces the
cross section even stronger in the presence of high-$p_{\mathrm{T}}$ jets.
\change{We note that such large jet-veto logarithms can be avoided by adopting a dynamic definition of the jet veto \cite{Kallweit:2017khh,Kallweit:2019zez}.
These can also be efficiently handled by the parton shower and/or merging procedures as shown below.}
The scale uncertainty grows very large towards high transverse
momentum owing to the large NLO \QCD\ contribution and the cancellations
between LO and NLO \QCD. 
As a consequence, for all observables considered in
Fig.~\ref{fig:transverse_wwj} the NLO \EW\ corrections stay 
within the NLO \QCD\ scale uncertainty bands.

In Fig.~\ref{fig:invariant_wwj} we present the di-lepton and
four-lepton invariant-mass distribution as well as the distributions
in the polar and azimuthal separation of the charged leptons.  
\bfig
  \center
  \includegraphics[width=0.45\textwidth]{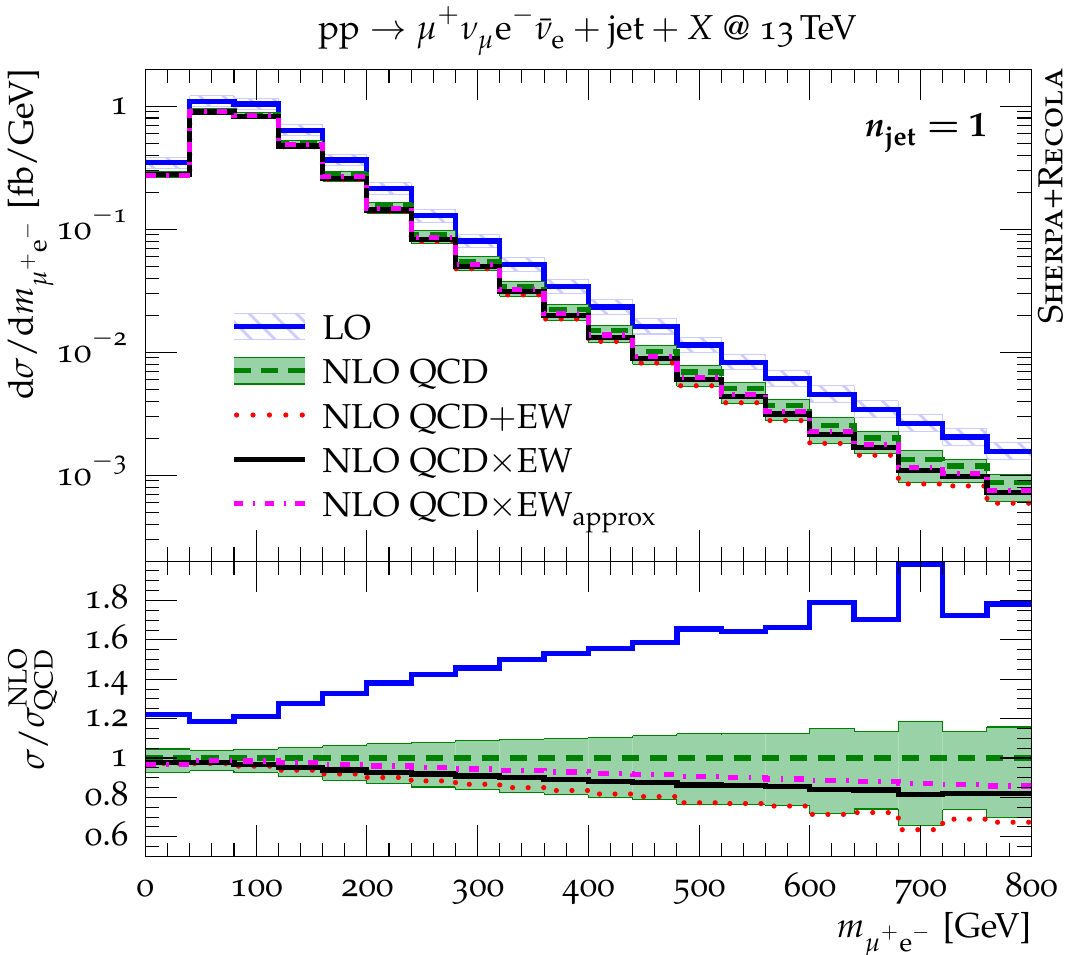}\hfill
  \includegraphics[width=0.45\textwidth]{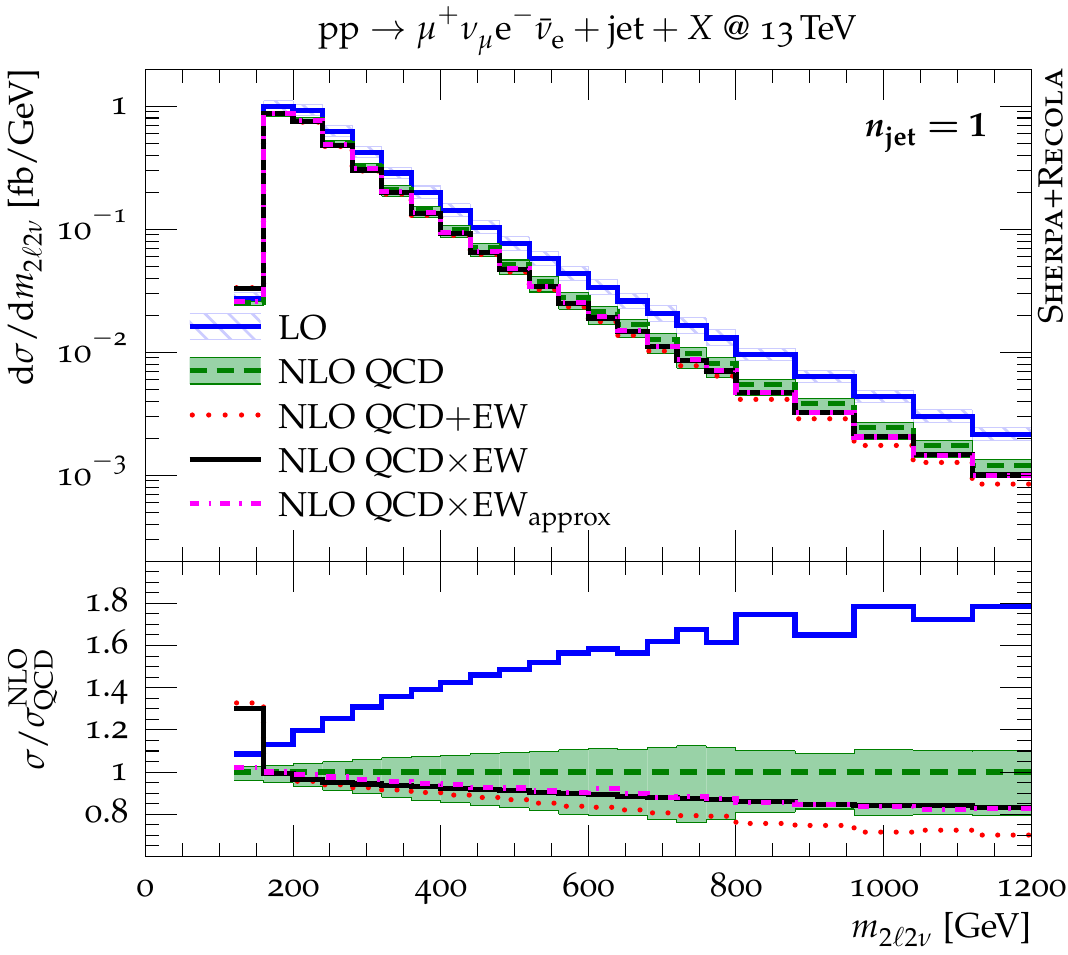}
  \includegraphics[width=0.45\textwidth]{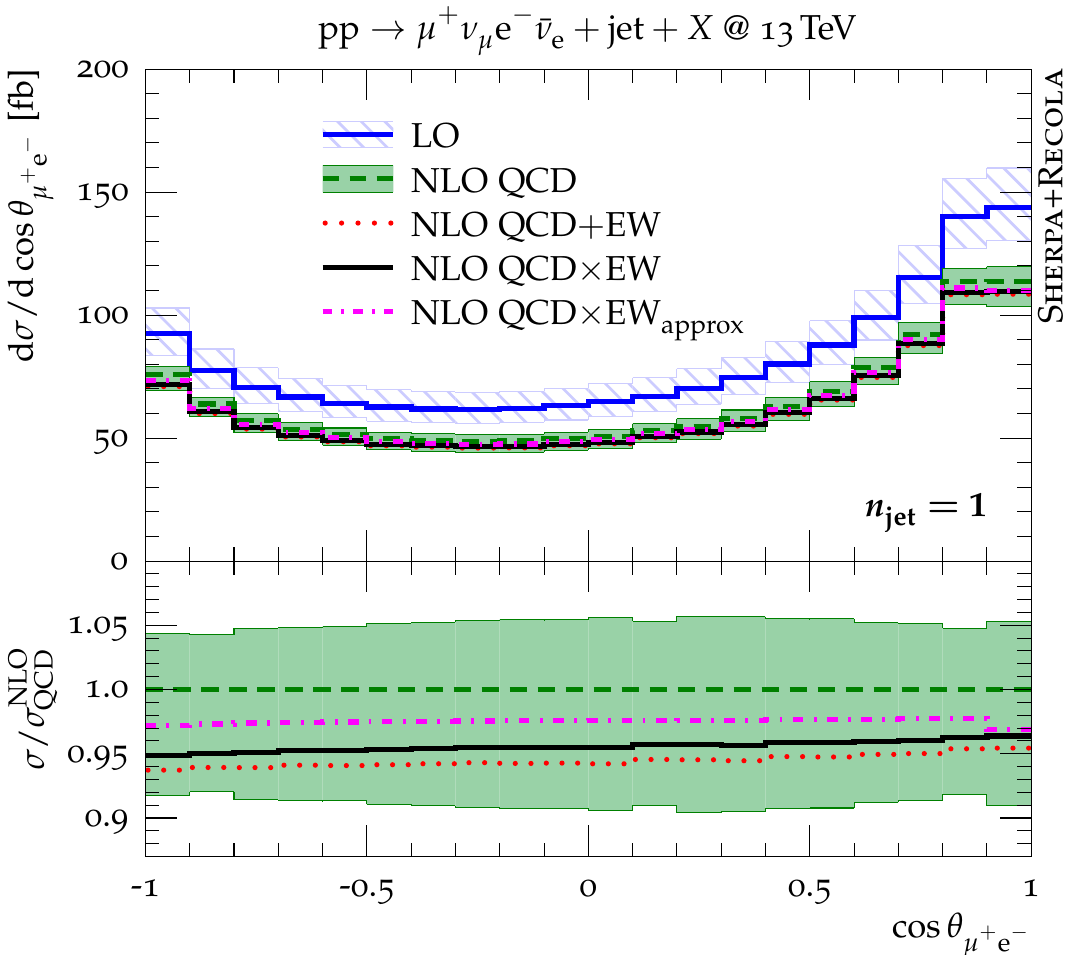}\hfill
  \includegraphics[width=0.45\textwidth]{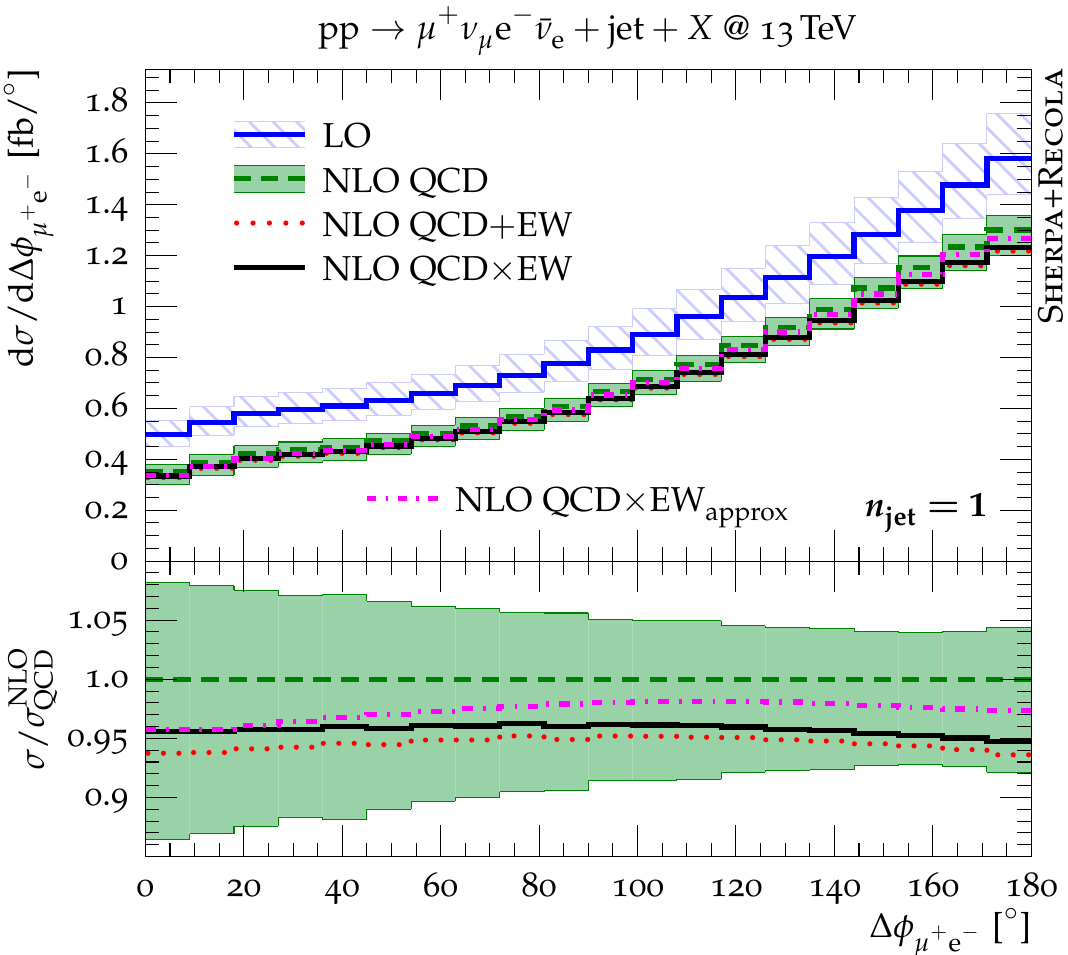}
  \caption{\label{fig:invariant_wwj} Differential distributions for $\Pp\Pp \to \mu^+ \nu_\mu \Pe^- \bar\nu_\Pe \Pj$ at LO, NLO \QCD, NLO \QCDpEW, \change{NLO \QCDtEW, and NLO \QCDtEWapprox}:
  Invariant mass of the anti-muon and electron (top left),
  invariant mass of the four leptons (top right),
  cosine of the angle between the anti-muon and the electron (bottom left), and
  azimuthal-angle distance between the anti-muon and the electron (bottom right).
  The upper panels show the absolute predictions, while the lower ones
  display the ratio of the various predictions with respect to the NLO
  \QCD\ predictions. 
 }
\efig
Both invariant-mass distributions receive \QCD\ corrections
of about $-80\%$ for large invariant masses. The \EW\ corrections are more moderate
and reach $-20\%$ in the considered kinematic range, thereby almost exceeding the
NLO \QCD\ scale uncertainties. As in the case of $\PW\PW$ production, both angular
observables do not exhibit enhanced corrections.  The \QCD\ corrections
(not visible in the lower panels) are
essentially flat, and the \EW\ corrections vary by a few per cent only.

\change{
The inclusion of approximate NLO EW corrections in the NLO \QCDtEWapprox\ 
results again reproduces the exact NLO EW corrections in the same 
manner as previously observed for the zero-jet case, \ie within a few per cent.}

As a consequence of the very large \QCD\ corrections and sizeable \EW\ 
corrections for the invariant-mass and, in particular,
transverse-momentum distributions, the two prescriptions to combine
\QCD\ and \EW\ corrections give rather different predictions at high
invariant masses and transverse momenta. The cross section becomes
even negative in the additive combination for large transverse
momenta.

\subsubsection{Ratios of \texorpdfstring{$\PW\PW$}{WW} and \texorpdfstring{$\PW\PW\Pj$}{WWj}}
\label{se:results:fo:ratios}

In this section, ratios of fiducial cross sections and differential
distributions between $\Pp\Pp \to \mu^+ \nu_\mu \Pe^- \bar \nu_\Pe$
and $\Pp\Pp \to \mu^+ \nu_\mu \Pe^- \bar \nu_\Pe \Pj$ are studied.  Motivated
by the closely related final states of the two processes, the level of universality of
the higher-order \QCD\ and \EW\ corrections can be probed. Furthermore, this cross-section ratio
has been measured by the ATLAS collaboration in \citere{Aaboud:2016mrt} \change{and is expected to
have reduced experimental systematic uncertainties}. Ratios of fiducial
cross sections at LO, NLO \QCD\ and \EW\, as well as their additive and multiplicative combination
are compiled in Table~\ref{tab:ratio}.
\begin{table}
\begin{center}
\begin{tabular}{c|c|c|c|c}
LO & NLO \QCD & NLO \EW & NLO \QCDpEW & NLO \QCDtEW\\
\hline \rule{0ex}{2.4ex}%
0.508$^{+17.5\%}_{-13.5\%}$ & 0.403$^{+2.9\%}_{-6.9\%}$ & 0.500 & 0.392 & 0.397 \\
\end{tabular}
\end{center}
\caption{\label{tab:ratio} Ratios of fiducial cross sections between
  $\Pp\Pp \to \mu^+ \nu_\mu \Pe^- \bar \nu_\Pe \Pj$ and $\Pp\Pp \to
  \mu^+ \nu_\mu \Pe^- \bar \nu_\Pe$  at $\sqrt{s}=13\TeV$ at LO, NLO \QCD, and NLO \EW\ as well as
  for the additive and multiplicative prescription to combine NLO \QCD\ and NLO \EW\ corrections.}
\end{table}
In these ratios as well as in the ratios of distributions below we always treat the
scale uncertainties of $\PW\PW$ and $\PW\PW\Pj$ production as correlated.
While the NLO \QCD\ corrections to the ratio amount to $-20\%$, the \EW\ 
corrections yield $-1.5\%$ only. As a consequence, the additive and
multiplicative prescriptions for the combination of \QCD\ and \EW\ 
corrections agree within $1.3\%$, which is basically
the difference observed for the $\PW\PW\Pj$ cross section. We furthermore note, that the scale
uncertainty on the cross-section ratio (with fully correlated scale
uncertainties for $\PW\PW$ and $\PW\PW\Pj$) significantly reduces when including
the NLO \QCD\ corrections.

Next, we show ratios for those differential distributions that have
already been discussed for the process $\Pp\Pp \to \mu^+ \nu_\mu \Pe^-
\bar\nu_\Pe$ in \refse{se:results:fo:WW}.  In the upper panels, the
ratios
\begin{equation}\label{eq:ratios}
R^{1\Pj}_{0\Pj}(x)= \frac{\frac{\rd\sigma}{\rd x}(\Pp\Pp \to \mu^+ \nu_\mu \Pe^-
\bar\nu_\Pe \Pj)}{\frac{\rd\sigma}{\rd x}(\Pp\Pp \to \mu^+ \nu_\mu \Pe^-
\bar\nu_\Pe)}
\end{equation}
are displayed at LO, NLO \QCD, NLO \QCDpEW, and NLO \QCDtEW\
accuracy. 
\change{As before, the approximate NLO \QCDtEWapprox\ results 
are added to gauge the quality of the approximation before it 
is employed in the construction of the multi-jet merged results 
of \refse{sec:results:merged:ratios}.}
In the lower panels, these ratios are again normalised to the
respective NLO \QCD\ prediction.

In Fig.~\ref{fig:transverse_ratio}, the ratios for the transverse-momentum
and rapidity distribution of the anti-muon are shown.  
\bfig
  \center
  \includegraphics[width=0.45\textwidth]{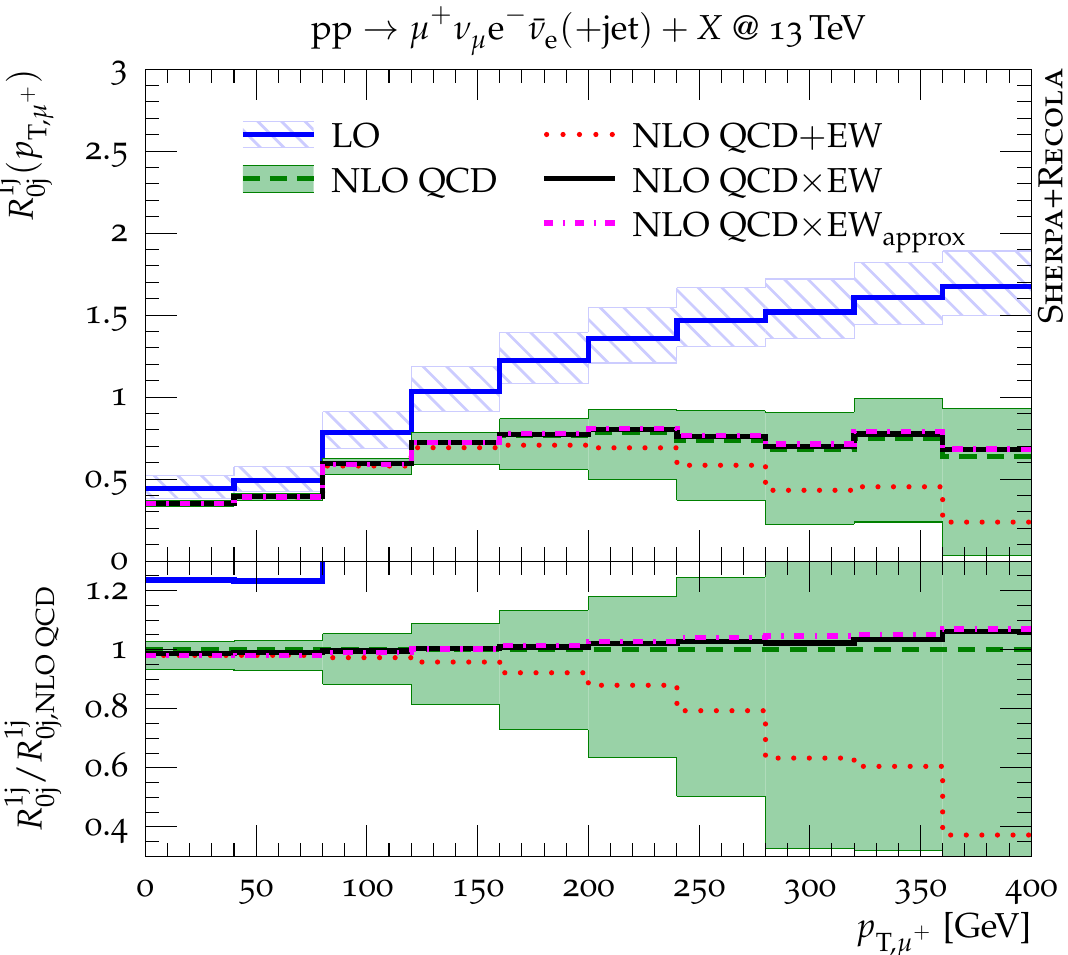}\hfill
  \includegraphics[width=0.45\textwidth]{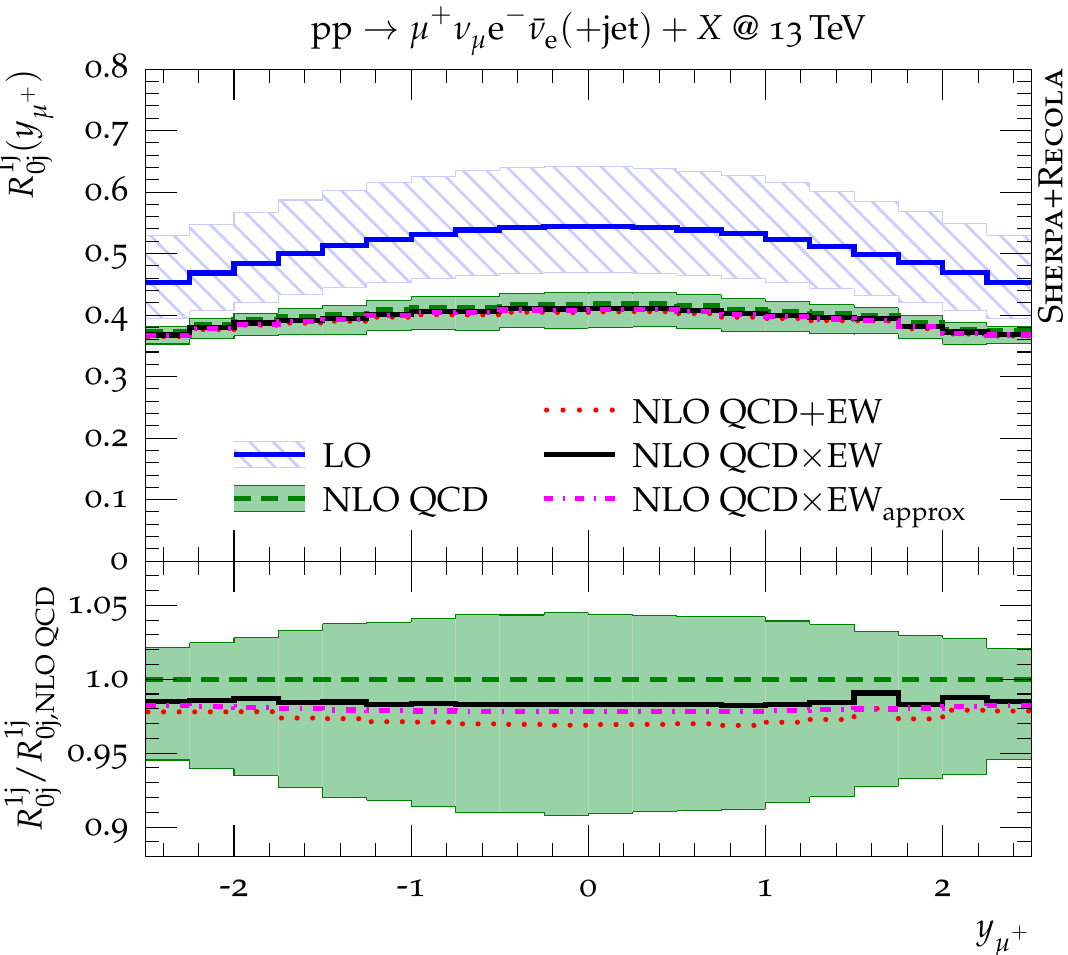}
  \includegraphics[width=0.45\textwidth]{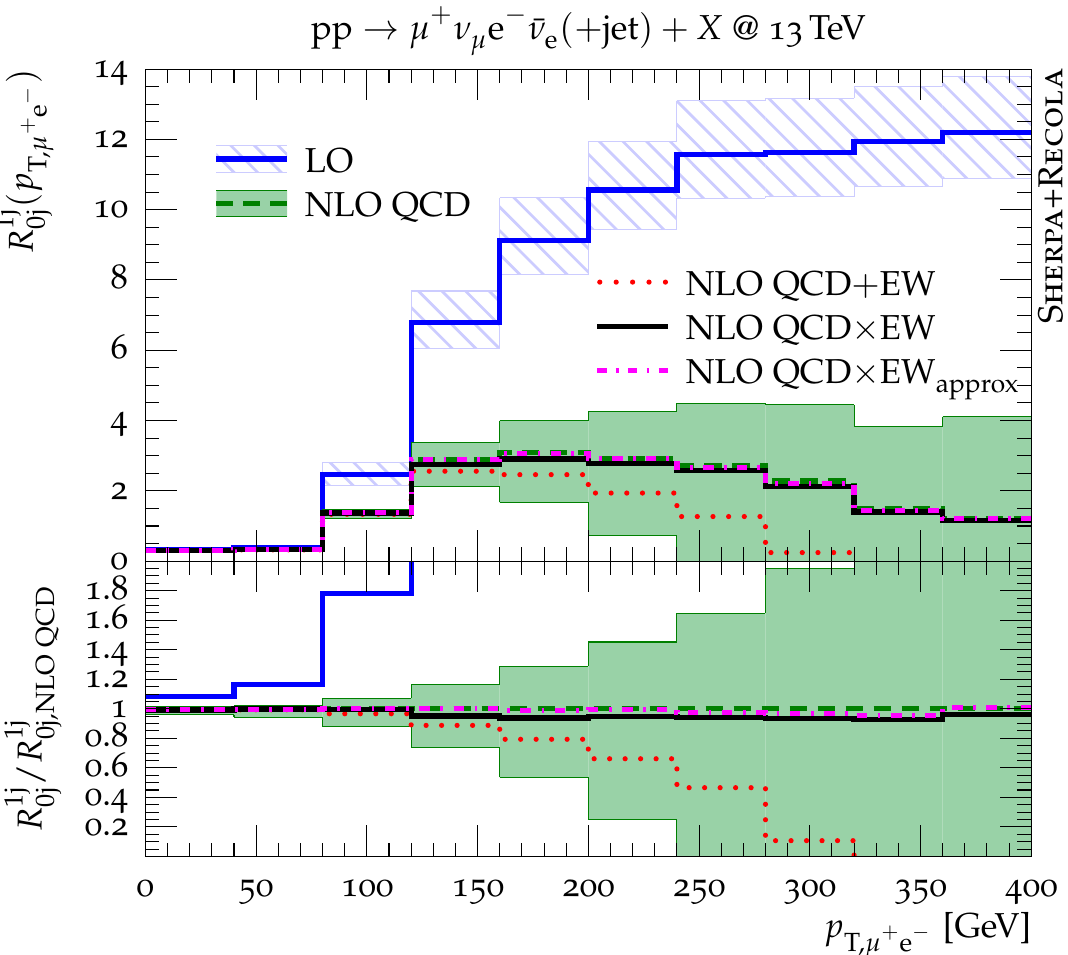}\hfill
  \includegraphics[width=0.45\textwidth]{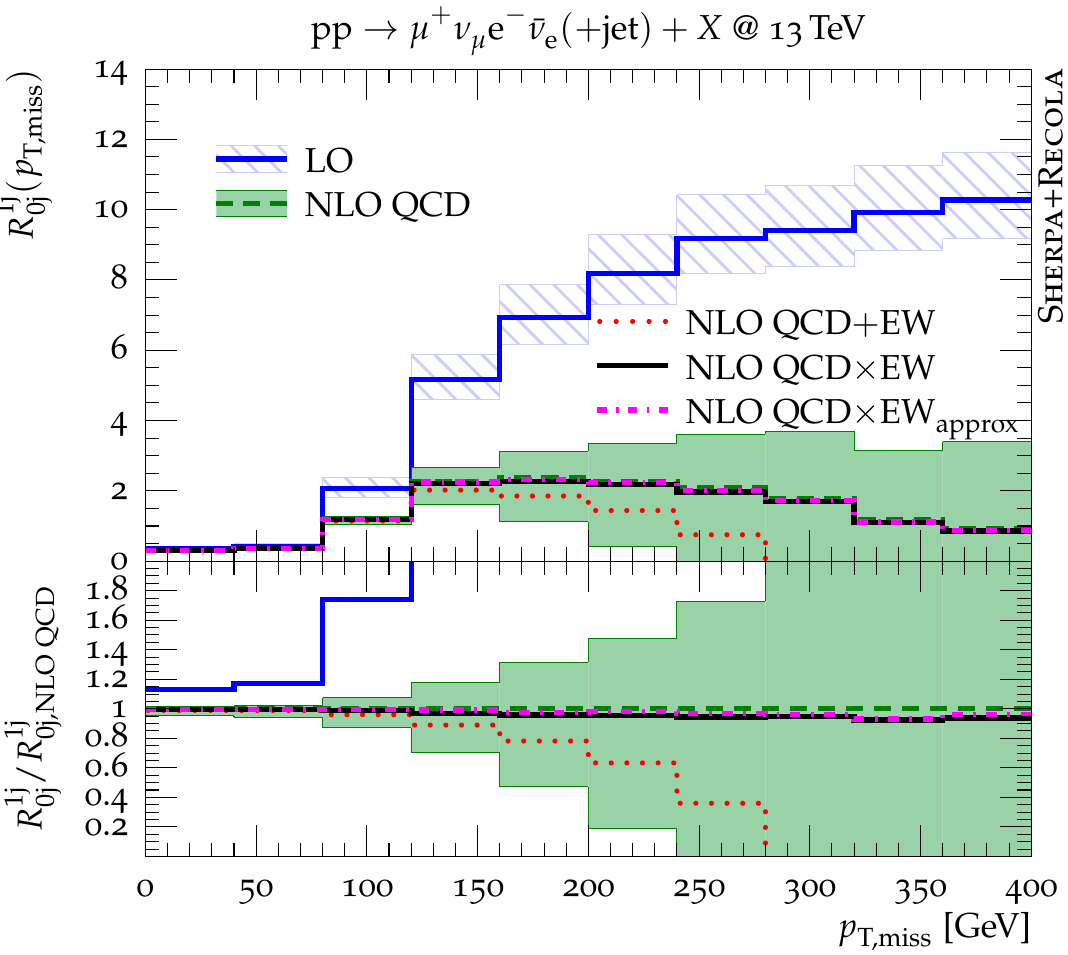}
  \caption{\label{fig:transverse_ratio} Ratios of differential
    distributions between 
    $\Pp\Pp \to \mu^+ \nu_\mu \Pe^- \bar\nu_\Pe \Pj$ and $\Pp\Pp \to \mu^+ \nu_\mu \Pe^- \bar\nu_\Pe$
  at LO, NLO \QCD, NLO \QCDpEW, \change{NLO \QCDtEW, and NLO \QCDtEWapprox}:
  Transverse momentum of the anti-muon (top left),
  rapidity of the anti-muon (top right),
  transverse momentum of the anti-muon--electron system (bottom left), and
  missing transverse momentum (bottom right).
  The upper panels show the absolute predictions, while the lower ones
  display the ratio between the various predictions and the respective NLO
  \QCD\ result. 
 }
\efig
For the transverse-momentum distribution, NLO \QCD\ corrections are very
large and stabilise the ratio towards high transverse momenta. At LO the
jet veto \refeq{jet:def} just affects the $\Pp\Pp \to \mu^+\nu_\mu \Pe^- \bar\nu_\Pe \Pj$
channel, only from NLO on it is active for the $\Pp\Pp \to \mu^+ \nu_\mu \Pe^-\bar\nu_\Pe$
process. The two processes receive NLO corrections of rather different size
providing the observed stabilisation \change{in terms of smaller higher-order corrections} in the ratios. The two prescriptions to
combine \QCD\ and \EW\ corrections behave rather differently.  While the additive
combination differs considerably from the pure \QCD\ result for $p_{\mathrm{T},\mu^+}>200\GeV$,
the multiplicative one stays close to it.  This is in agreement with the observation that
the leading-logarithmic corrections for the two processes  $\PW\PW$
and $\PW\PW\Pj$ are strongly
correlated. In fact, these are related to the \EW\ charges of the external lines,
which are the same for both processes. This can be deduced from the general results
on the leading one-loop \EW\ corrections presented in \citere{Denner:2000jv} and from the results based
on soft-collinear effective theory in \citeres{Chiu:2008vv,Chiu:2009ft}.
This is a strong motivation to prefer the multiplicative prescription.
\change{The difference between the two prescriptions should not be taken as an uncertainty.}
\change{Moreover, it supports} the merging approach presented in
\refse{se:merged_predictions} which rests 
on the assumption that leading \EW\ corrections are rather similar for different
final-state jet multiplicities.
While the anti-muon rapidity distribution does not
exhibit a strong difference between the two combinations over the whole
phase space, the ratio for the multiplicative combination is closer to the
pure \QCD\ result and less dependent on the rapidity.

The ratios of the distributions in the transverse momentum of the two charged leptons and the
missing energy show very large variations. They are particularly sensitive
to the applied jet veto, which is adequately accounted for at NLO \QCD\ only,
in particular for the $\PW\PW$ channel.
Indeed, at LO the $\PW\PW$ and $\PW\PW\Pj$ processes have rather
different kinematics. 
With the inclusion of real radiations the descriptions of both
processes become closer, and the ratios stabilise \change{in terms of
  smaller higher-order corrections}.

The ratios for the invariant-mass and angular distributions shown in
Fig.~\ref{fig:invariant_ratio} confirm the trend seen in the other
distributions. 
\bfig
  \center
  \includegraphics[width=0.45\textwidth]{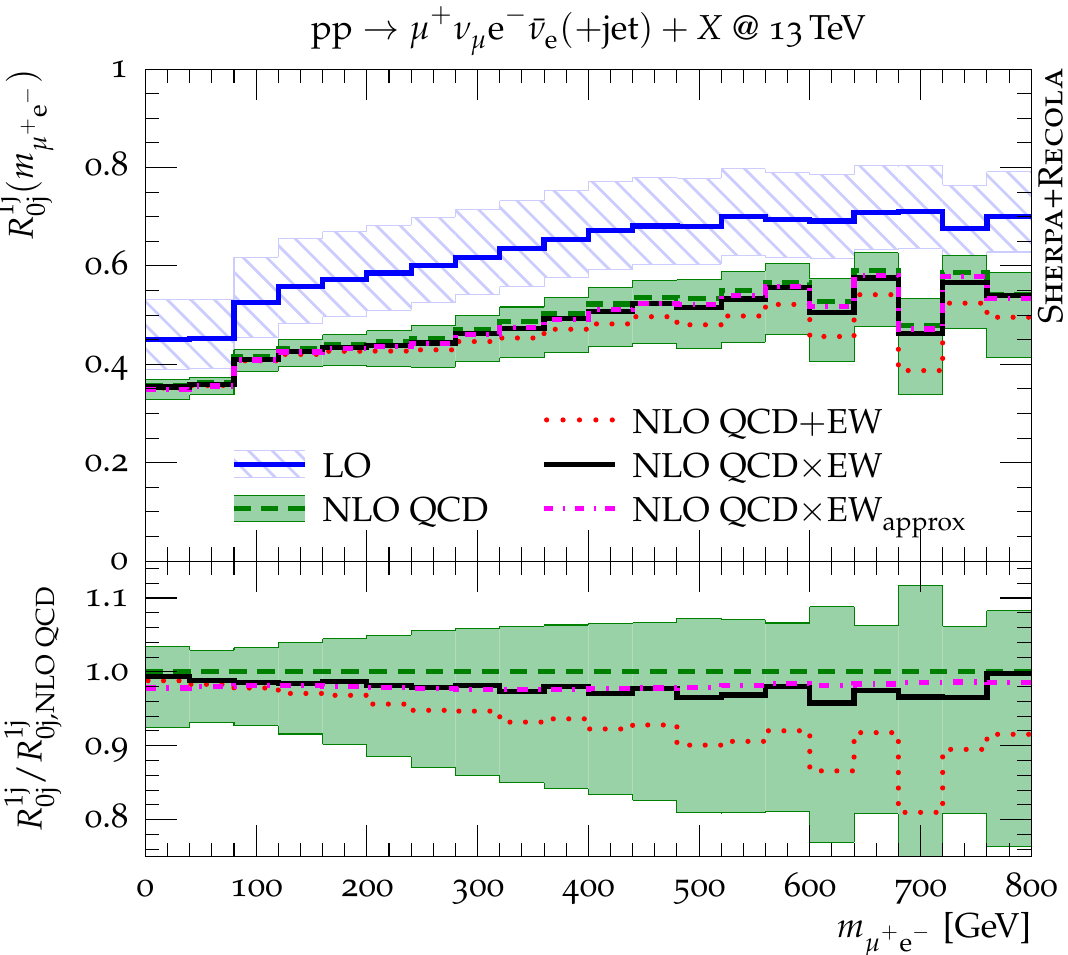}\hfill
  \includegraphics[width=0.45\textwidth]{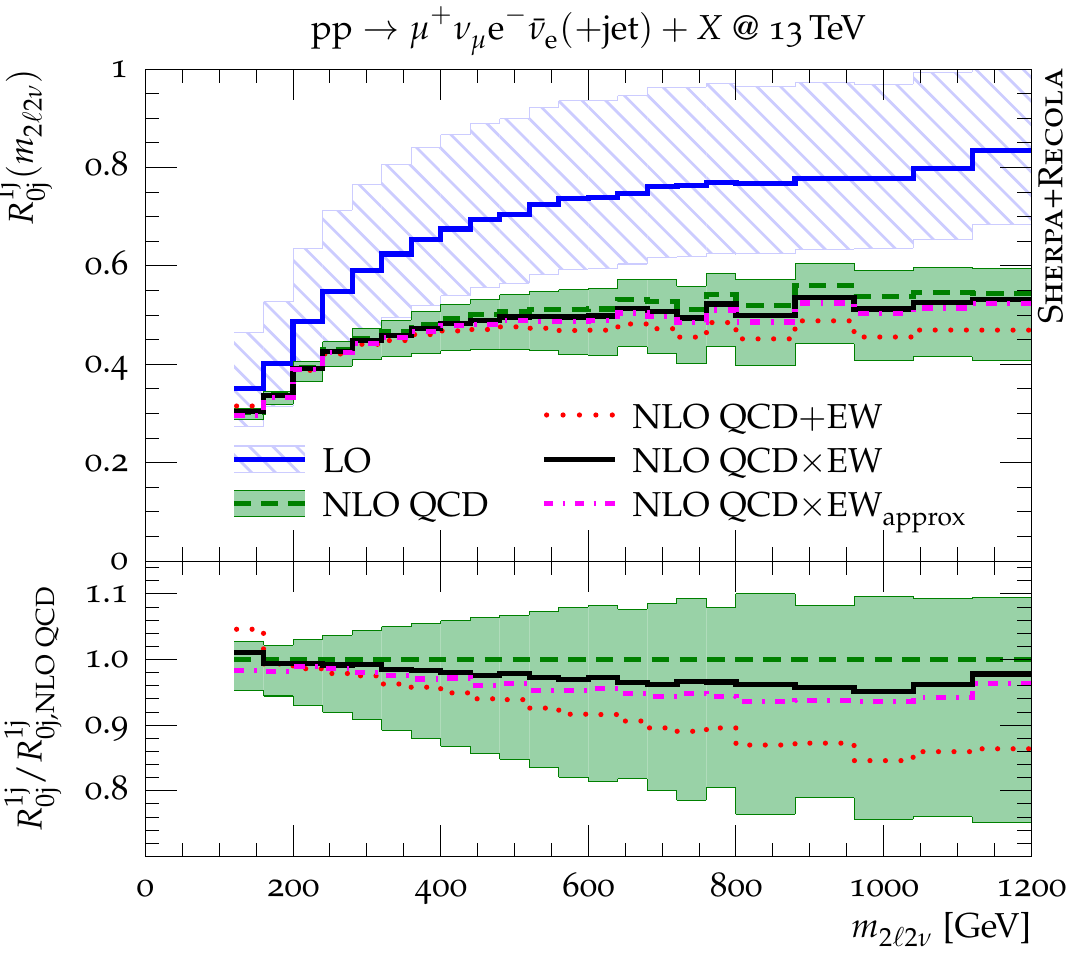}
  \includegraphics[width=0.45\textwidth]{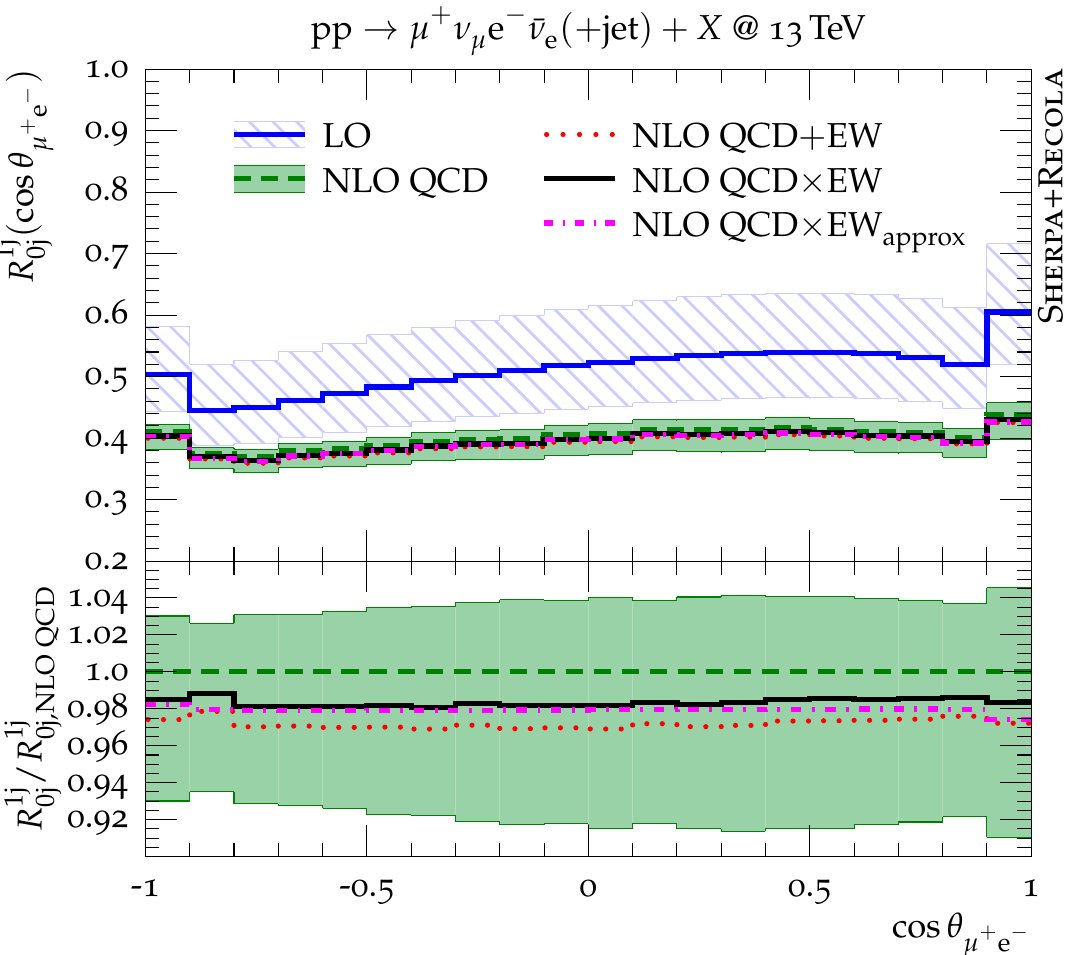}\hfill
  \includegraphics[width=0.45\textwidth]{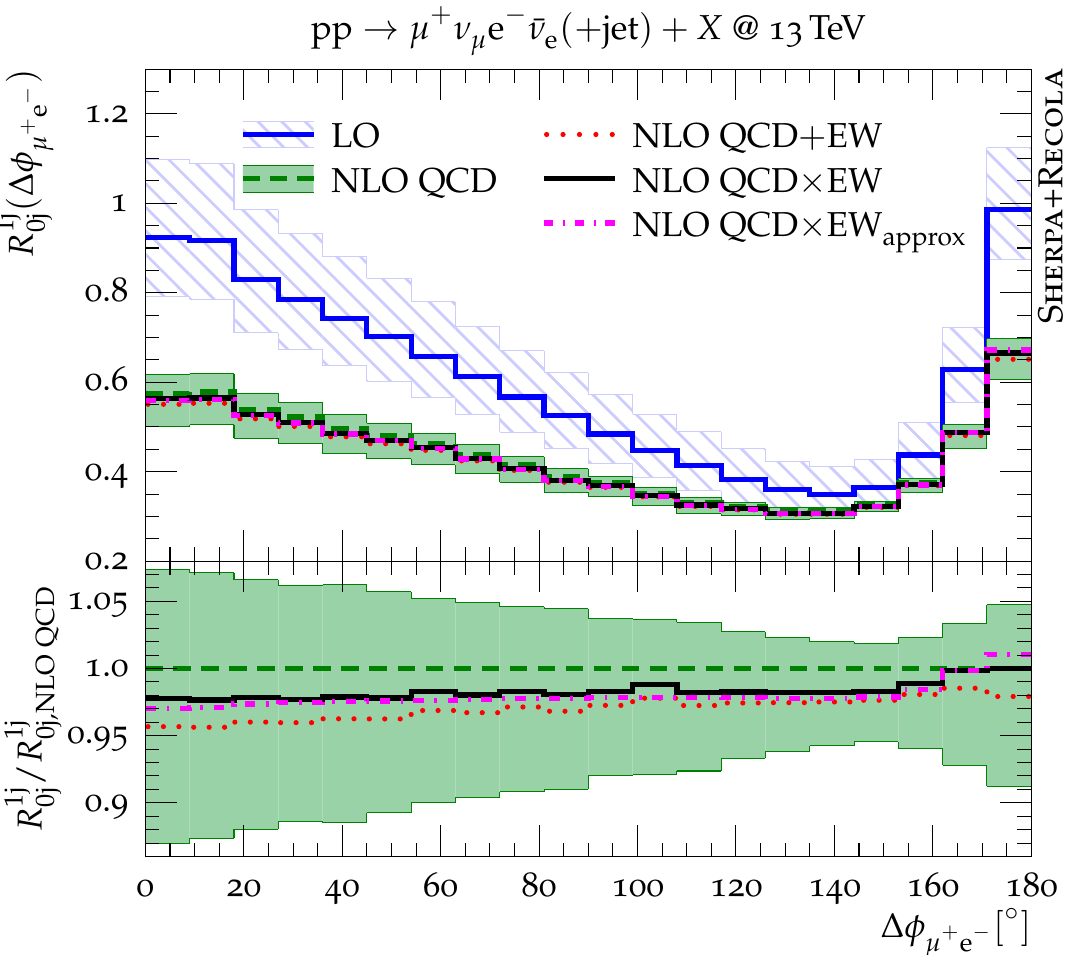}
  \caption{\label{fig:invariant_ratio} Ratios of differential
    distributions 
  between $\Pp\Pp \to \mu^+ \nu_\mu \Pe^- \bar\nu_\Pe \Pj$ and $\Pp\Pp \to \mu^+ \nu_\mu \Pe^- \bar\nu_\Pe$
  at LO, NLO \QCD, NLO \QCDpEW, \change{NLO \QCDtEW, and NLO \QCDtEWapprox}:
  Invariant mass of the anti-muon and electron (top left),
  invariant mass of the four leptons (top right),
  cosine of the angle between the anti-muon and the electron (bottom left), and
  azimuthal-angle distance between the anti-muon and the electron (bottom right).
  The upper panels show the absolute predictions, while the lower ones
  display the ratio between the various predictions and the respective NLO
  \QCD\ result. 
 }
\efig
When using the multiplicative prescription to include
the \EW\ corrections on top of the \QCD\ corrections, the ratios depend
only very weakly on the invariant masses and angles, while for the
additive combination this dependence is more pronounced.  The more or less
decent behaviour of the additive prescription, in particular for the
angular distributions, is due to the smallness of the corresponding
\QCD\ and \EW\ corrections.

\change{
As for the individual cross sections, the approximate NLO EW corrections
reproduce the exact cross-section ratios well in the Sudakov region, for
which they were constructed.
Interestingly, and to some extent accidentally, the exact results for the
ratios of inclusive observables like the muon rapidity,
$\cos\theta_{\mu^+e^-}$ and $\Delta\phi_{\mu^+e^-}$ are also found to be
very well reproduced and in particular better than for the individual
cross sections.
This can be qualitatively understood by the fact that the main feature
of the virtual EW approximation is the crude and inclusive approximation of real
emission corrections via the Catani--Seymour $I$-operator.
Since a large part of the missing effects concerns the final-state
leptons, these real emission corrections are
to a certain degree universal for the different jet multiplicities and thus
cancel to some extent in the ratios.}

\change{The above conclusion that the multiplicative prescription is
  preferred applies for the set-up considered here.
In general, such a statement should be checked on a case-by-case basis
for different phase spaces. 
Nonetheless, as the argument relies on factorisation in the bulk of the cross section, 
the multiplicative approach should also be preferred for set-ups (possibly with a slightly different jet veto) that share a large portion of the phase space.}

\FloatBarrier

\subsection{Multi-jet merged results}
\label{sec:results:merged}

In this section we present predictions based
on the merging of the NLO \QCD\ matrix elements for
$\Pp\Pp \to \mu^+ \nu_\mu \Pe^- \bar\nu_\Pe{}+0,1\,\Pj$ and
tree-level matrix elements for $\Pp\Pp \to \mu^+ \nu_\mu \Pe^- \bar\nu_\Pe{}+2,3\,\Pj$,
each matched to the \Sherpa Catani--Seymour parton shower.
We implement approximate NLO \EW\ corrections in an additive or multiplicative
manner for the zero- and one-jet matrix elements. However, these also enter
the higher-multiplicity tree-level processes through a local
$K$~factor (cf.~\refse{se:merged_predictions}). By merging parton-shower
matched matrix elements of varying final-state
parton multiplicity we arrive at a fully inclusive event
sample for $\PW\PW$ production. This sample can then be analysed for the
zero- and one-jet selection criteria, without
the need to perform dedicated calculations, as it is the case
for the NLO fixed-order predictions. Furthermore, the inclusion of
higher-multiplicity processes and parton-shower
resummation accounts for possible higher jet multiplicities and
in turn provides a more adequate description of the jet-veto
conditions applied. In the following we neglect effects from
the parton-to-hadron transition, as well as underlying-event
contributions appearing in hadron collisions, which allows us
to directly compare fixed-order calculations with perturbative
parton-shower Monte Carlo predictions. 

\subsubsection{Fixed-order vs.\ merged results \change{at NLO \QCD}}
\label{sec:results:merged:fo-vs-merged}

We begin the discussion by comparing \MEPSatNLOQCD predictions from
\Sherpa\ for the zero- and one-jet \change{exclusive} event selections against the fixed-order NLO \QCD\ 
results presented in \refse{se:results:fo}.
\change{Therefore, in the merged predictions for both multiplicities, further jets are vetoed.}
To this end, we present
predictions following the {\emph{default}} CKKW scale-setting prescription
as outlined in \refse{se:merged_predictions}. Accordingly, for each
hard-partonic event configuration a clustering algorithm is
applied to reconstruct the kinematics of the corresponding 
$\Pp\Pp\to \mu^+ \nu_\mu \Pe^- \bar\nu_\Pe$ core process. 
This clustering procedure defines the CKKW scale $\muCKKW$ through 
the reconstructed parton branching scales, cf.\ Eq.\ \eqref{eq:muRCKKW}, 
as well as the core scale $\mucore$ defined on the arrived-at 
core process.
In turn, $\mucore$ then determines $\muF$ and $\muQ$ through 
Eq.\ \eqref{eq:mQCKKW}. 
Through this procedure both the fixed-order accuracies of the 
matrix-element calculations and the resummation accuracy of the 
parton shower are preserved.

Furthermore, we present results based on an {\emph{alternative}} scale-setting
prescription, dubbed \emph{proto-merging}, where for each hard-parton
configuration we use
\begin{equation}
  \muR = \muF =\muQ =\tfrac{1}{2}\,\ETWmean
\label{eq:muproto}
\end{equation}
of the $n$-jet process without clustering any partons first, 
\emph{i.e.}\ we set all three scales equal to the scales used in the 
corresponding fixed-order calculation without reconstructing emission
scales or a core process.
While this respects the fixed-order NLO accuracy of each partonic subsample, 
it spoils the resummation property of the \QCD\ parton shower.\footnote{
  This can be seen by considering two effects:\\[-10pt]
  \begin{enumerate}
    \item The shower starting scale $\muQ$ 
          will not be set to the scale of the reconstructed core process 
          for all processes with at least one parton in the final state.
          This existing parton will then not be correctly embedded in the 
          parton-shower evolution of this core process. 
          Since $\muQ$ will be typically higher than when using a proper 
          merging procedure, the Sudakov vetoes generated will be too large.
    \item The scale of the strong coupling associated with 
          the emission of a parton needs to be set to the relative transverse 
          momentum with respect to its reconstructed emitter parton in order to 
          recover the logarithms produced by the parton shower. 
          Since here a global scale is used which typically is larger 
          than its nodal value, the resulting strong coupling will be too small.
  \end{enumerate}
}
It thus cannot be considered a consistently merged \MEPS description, 
but is included in the following comparison for illustrative purposes.

By invoking the parton shower we include all-order corrections to the
inclusive $\Pp\Pp\to\mu^+ \nu_\mu \Pe^- \bar\nu_\Pe$ production
process and, as a consequence, the  
jet veto affects also the zero- and one-jet selections. As we
focus on \QCD\ corrections, we do not include \QED\ corrections due to
soft-photon emission~\cite{Schonherr:2008av} or \EW\ effects at this stage.

To estimate the dominant theoretical uncertainties we consistently vary the
renormalisation and factorisation scales in the matrix-element and parton-shower 
components~\cite{Bothmann:2016nao,Badger:2016bpw}. As before, we
consider the 7-point variations of the two scales 
$\muR$ and $\muF$ by factors of $\tfrac{1}{2}$ and $2$. 
We do not assess the systematics associated with the choice of the
merging parameter $\Qcut$ as well as the resummation scale $\muQ$ as
these can be expected to be of smaller size \cite{Cascioli:2013gfa,Kallweit:2015dum,
Hoeche:2013mua,Hoeche:2014rya,Hoeche:2014lxa,Hoeche:2014qda}. 

In the following plots we compare the NLO \QCD\ fixed-order results
with \MEPSatNLOQCD predictions for two different scale choices. In the
upper panels we show absolute predictions at NLO \QCD\ accuracy as
well as with \MEPSatNLO parton-shower matching for the default CKKW
scale setting \refeq{eq:muRCKKW} and \refeq{eq:mQCKKW} as well as for
the scale setting \refeq{eq:muproto} corresponding to the fixed-order
results. The lower panels show the corresponding results normalised to
the NLO \QCD\ ones.  Scale uncertainties are indicated by the
envelopes of the bands \change{and should be understood as an order of magnitude estimate of missing higher-order corrections}.

\subsubsection*{\texorpdfstring{$\PW\PW$}{WW} production}

In Figs.~\ref{fig:fo-vs-merged-0j-1} and \ref{fig:fo-vs-merged-0j-2} we
compile the set of exclusive zero-jet observables for the process
$\Pp\Pp \to \mu^+ \nu_\mu \Pe^- \bar\nu_\Pe$ studied in
\refse{se:results:fo:WW} already. 
\bfig
  \center
  \includegraphics[width=0.45\textwidth]{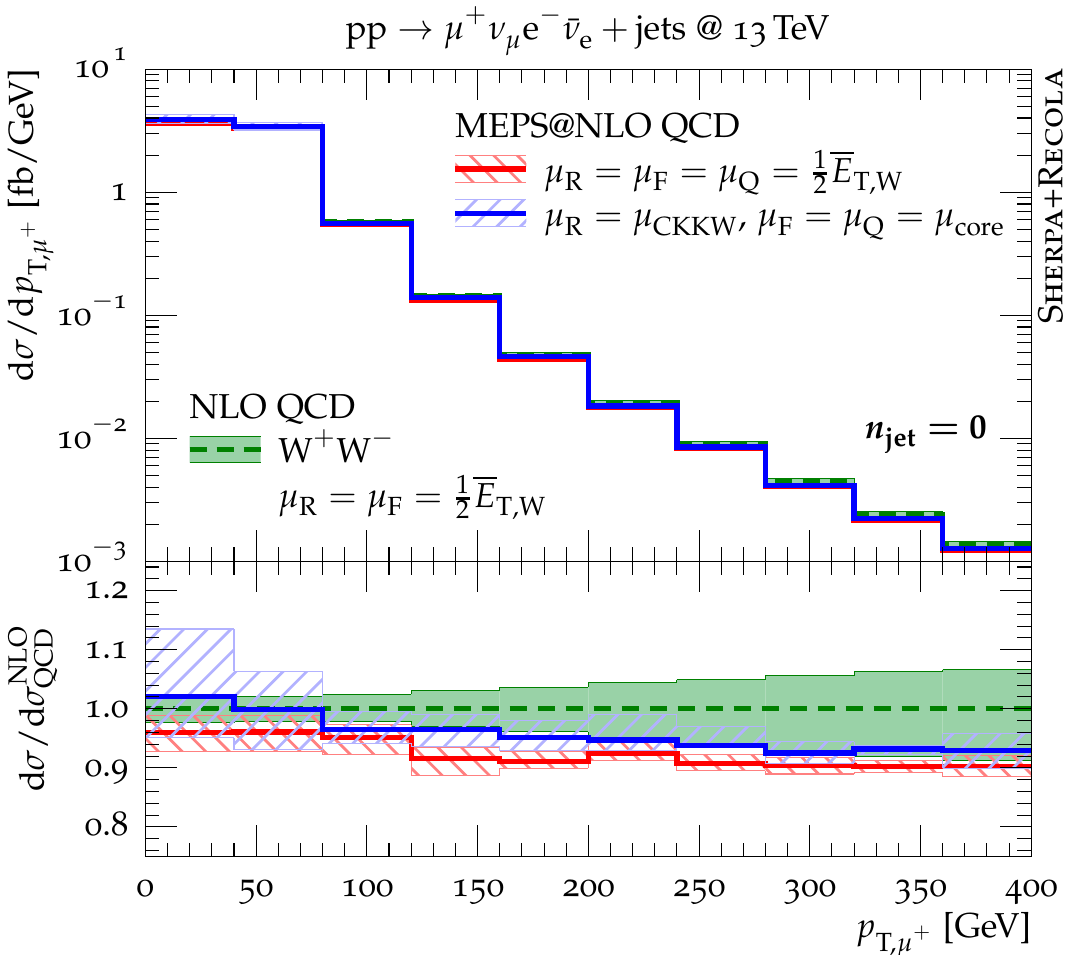}\hfill
  \includegraphics[width=0.45\textwidth]{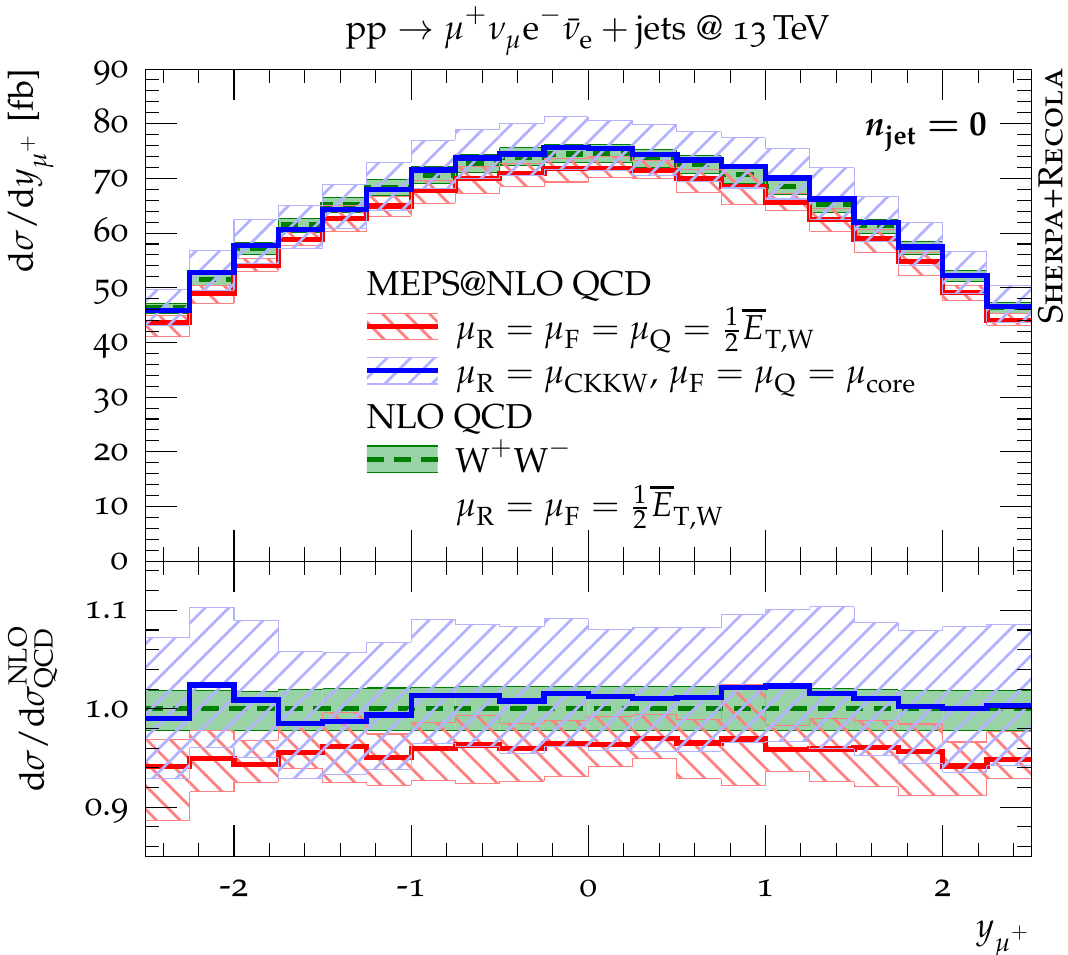}
  \includegraphics[width=0.45\textwidth]{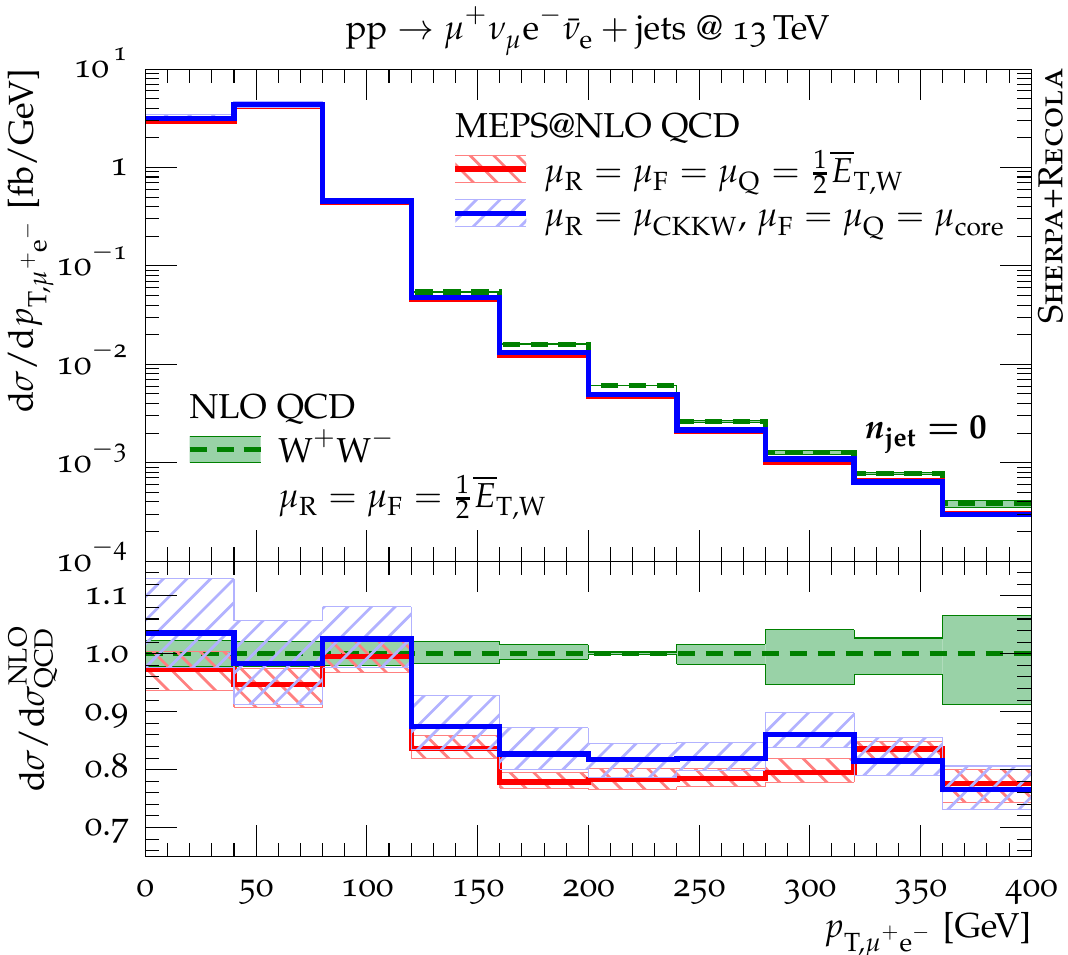}\hfill
  \includegraphics[width=0.45\textwidth]{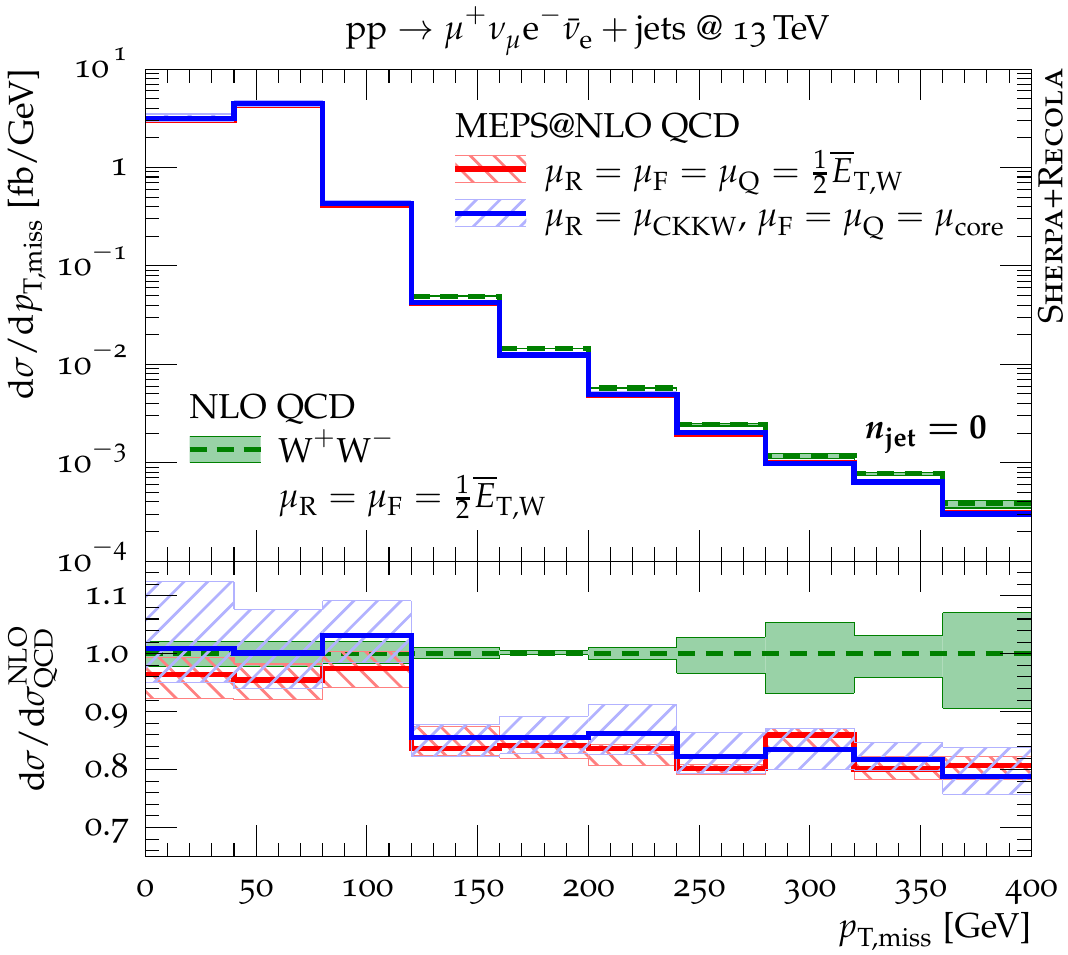}
  \caption{
    Comparison of NLO \QCD\ fixed-order results for the process $\Pp\Pp \to \mu^+ \nu_\mu \Pe^- \bar\nu_\Pe$
    with proto-merged and the fully multi-jet merged \MEPSatNLOQCD predictions in the $\njet=0$ event
    selection:
    Transverse momentum of the anti-muon (top left),
    rapidity of the anti-muon (top right),
    transverse momentum of the anti-muon--electron system (bottom left), and
    missing transverse momentum (bottom right).
    No \QED\ or \EW\ corrections are taken into account here.
    \label{fig:fo-vs-merged-0j-1}
  }
\efig
\bfig
  \center
  \includegraphics[width=0.45\textwidth]{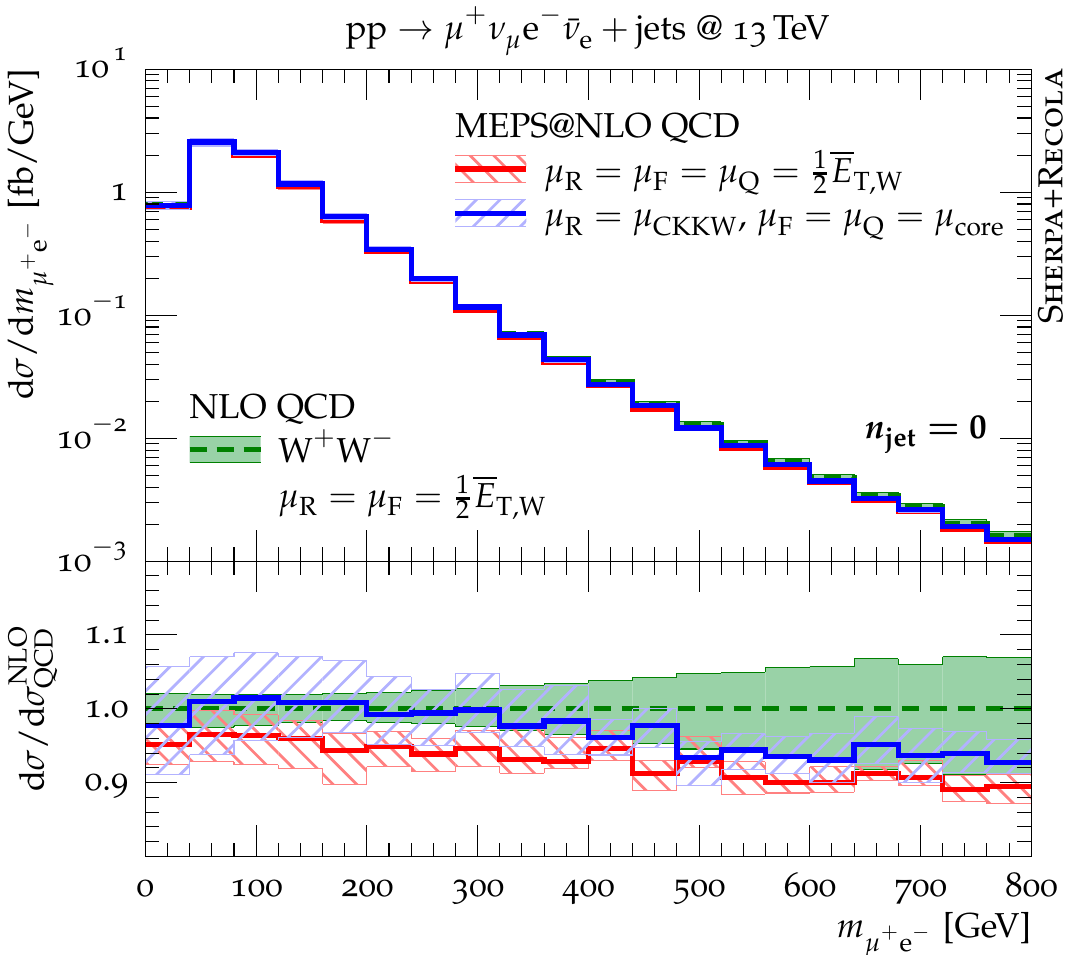}\hfill
  \includegraphics[width=0.45\textwidth]{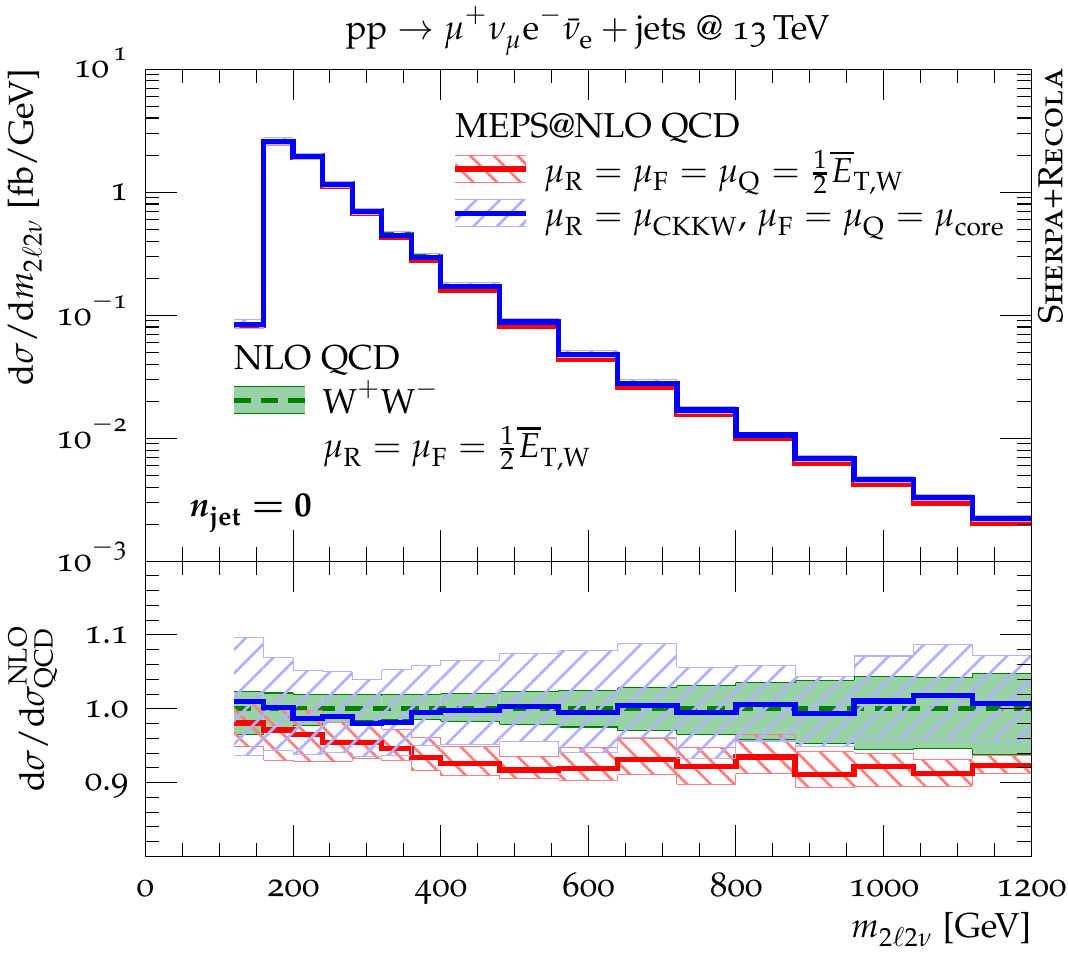}
  \includegraphics[width=0.45\textwidth]{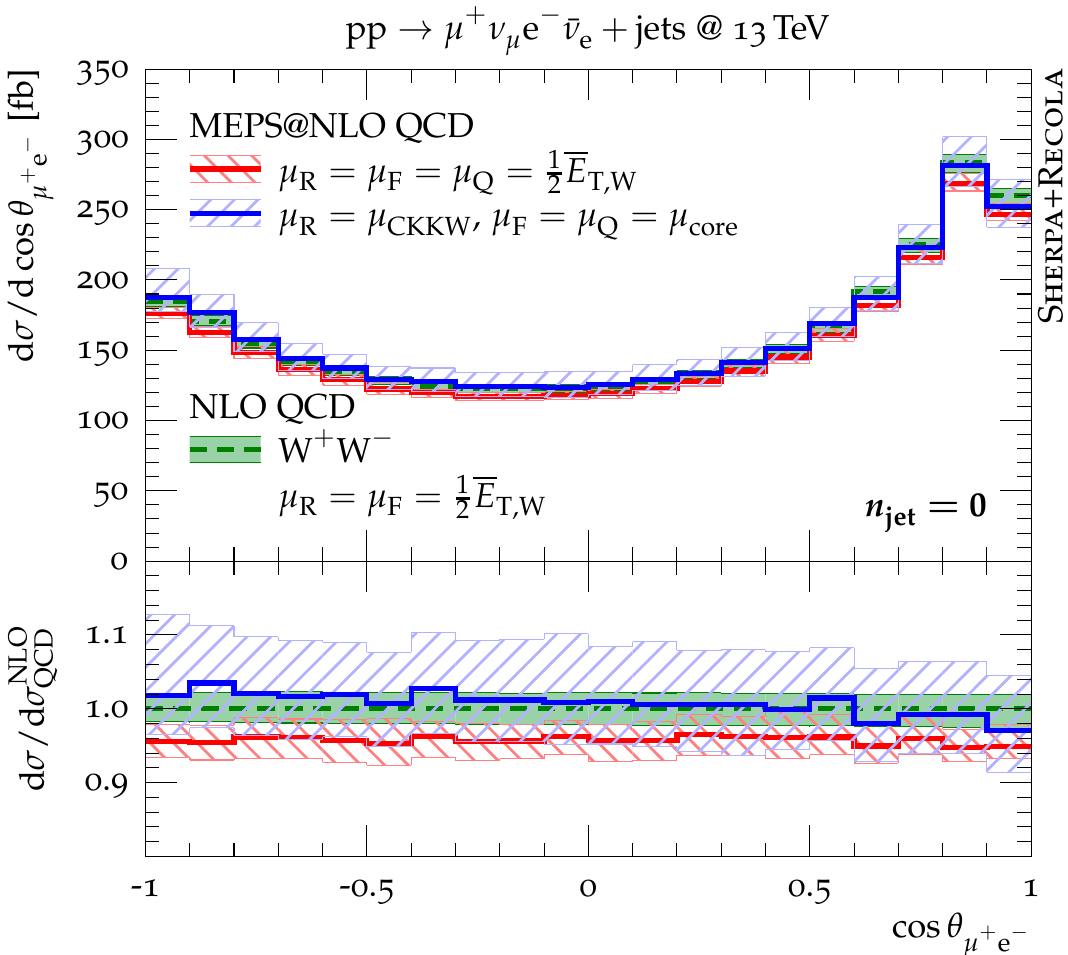}\hfill
  \includegraphics[width=0.45\textwidth]{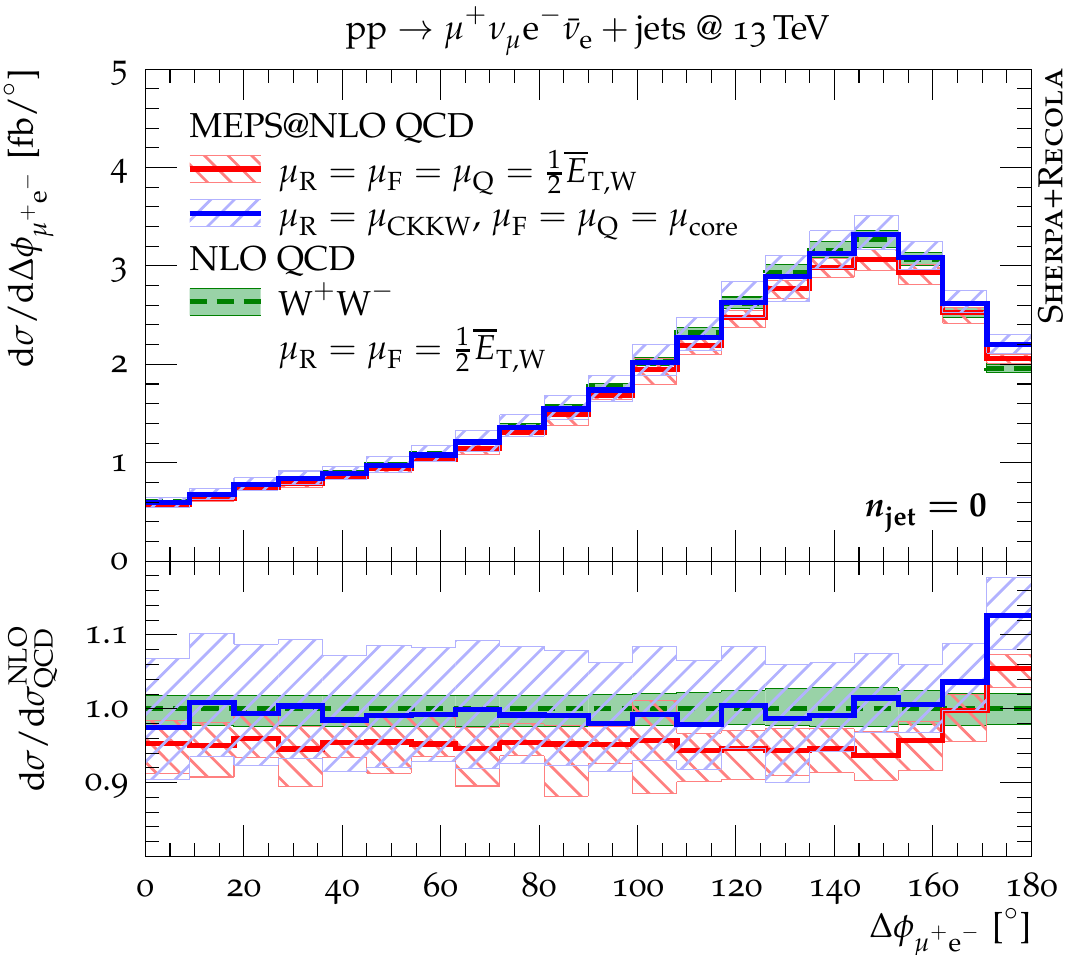}
  \caption{
    Comparison of NLO \QCD\ fixed-order results for the process $\Pp\Pp \to \mu^+ \nu_\mu \Pe^- \bar\nu_\Pe$
    with proto-merged and the fully multi-jet merged \MEPSatNLOQCD predictions in the $\njet=0$ event
    selection: 
    Invariant mass of the anti-muon and electron (top left),
    invariant mass of the four leptons (top right),
    cosine of the angle between the anti-muon and the electron (bottom left), and
    azimuthal-angle distance between the anti-muon and the electron (bottom right).
    No \QED\ or \EW\ corrections are taken into account here.
    \label{fig:fo-vs-merged-0j-2}
  }
\efig
We begin by considering the
anti-muon transverse-momentum distribution in Fig.~\ref{fig:fo-vs-merged-0j-1}.
The two \MEPSatNLOQCD predictions agree \change{rather} well with the NLO \QCD\ result \change{in the bulk of the distribution}.
\change{In the tail of the distribution, which is suppressed by three orders of magnitudes, the two merged predictions are on the edge of the scale-uncertainty band of the fixed-order prediction.}
For low to intermediate transverse momenta
the uncertainty of the proto-merged prediction reproduces well the
fixed-order uncertainty. However, for larger $p_{\mathrm{T},\mu^+}$ it is significantly
reduced, as the \MEPSatNLO method accounts for a proper resummation of
higher-order \QCD\ corrections.
For the default fully multi-jet
merged prediction this effect is also observed. However, at lower transverse
momenta the uncertainty increases, due to the typically smaller value of
the renormalisation scale, determined by the emission scale of the associated
partons. From the anti-muon rapidity distribution one can read off that the
central production rate predicted by the default CKKW scale-setting
prescription is in fact closer to the fixed-order cross section, whereas
it features a larger scale uncertainty of about $\pm 8\%$, compared to
only $\pm 2\%$ of the fixed-order result. While the proto-merged prediction exhibits an uncertainty
more closely resembling the fixed-order estimate, its central cross
section is reduced by about $4\%$.

For the distributions in the transverse momentum of the charged-lepton pair
and the missing transverse momentum the differences in the theoretical
predictions are much more sizeable. Up to about the $\PW$-boson mass the
merged predictions agree well with the fixed-order result, however,
they predict significantly smaller event rates beyond $\MW+\ptsub{\Pj,\text{cut}}$.
This originates from the same reason as in the fixed-order case namely that, due to kinematic constraints, 
the inclusion of real radiations beyond the W-boson mass lifts the cross section.
In addition, above this threshold the jet-veto criterion plays a significant role as in the NLO \QCD\ calculation of
$\Pp\Pp \to \mu^+ \nu_\mu \Pe^- \bar\nu_\Pe$ it is addressed only at
LO accuracy. In the merged calculations, however, the jet veto is
modelled by the NLO \QCD\ $\Pp\Pp \to \mu^+ \nu_\mu \Pe^- \bar\nu_\Pe\Pj$ 
calculation dressed with the parton-shower resummation, 
leading to a more realistic description
of this event-selection criterion. In particular, at high transverse momenta
an increased \QCD\ activity is expected that in consequence triggers the jet
veto thereby reducing the cross section.

For the observables depicted in Fig.~\ref{fig:fo-vs-merged-0j-2} the observed
pattern is further confirmed. The default multi-jet merged predictions agree
nicely with the fixed-order results, with an increased systematic uncertainty
in the bulk. However, in particular in the region of high dilepton invariant 
mass the uncertainty is indeed sizeably reduced with respect to the
one of the fixed-order computation.
For the angular separation between the two charged leptons the region of
$\Delta\phi_{\mu^+\Pe^-}\approx\pi$ is affected by the parton-shower resummation.
This originates from the suppression of the LO for $\PW\PW$ production
in this bin (cf.\ Fig.~\ref{fig:invariant_ww}). As a consequence, the CKKW
prescription differs by more than $10\%$ from the fixed-order prediction.

\FloatBarrier

\subsubsection*{\texorpdfstring{$\PW\PW\Pj$}{WWj} production}

In Figs.~\ref{fig:fo-vs-merged-1j-1} and \ref{fig:fo-vs-merged-1j-2} we
present \MEPSatNLOQCD predictions for the exclusive one-jet selection for the
process $\Pp\Pp \to \mu^+ \nu_\mu \Pe^- \bar\nu_\Pe\Pj$, studied at NLO \QCD\ 
in \refse{se:results:fo:WWj} already.
\bfig
  \center
  \includegraphics[width=0.45\textwidth]{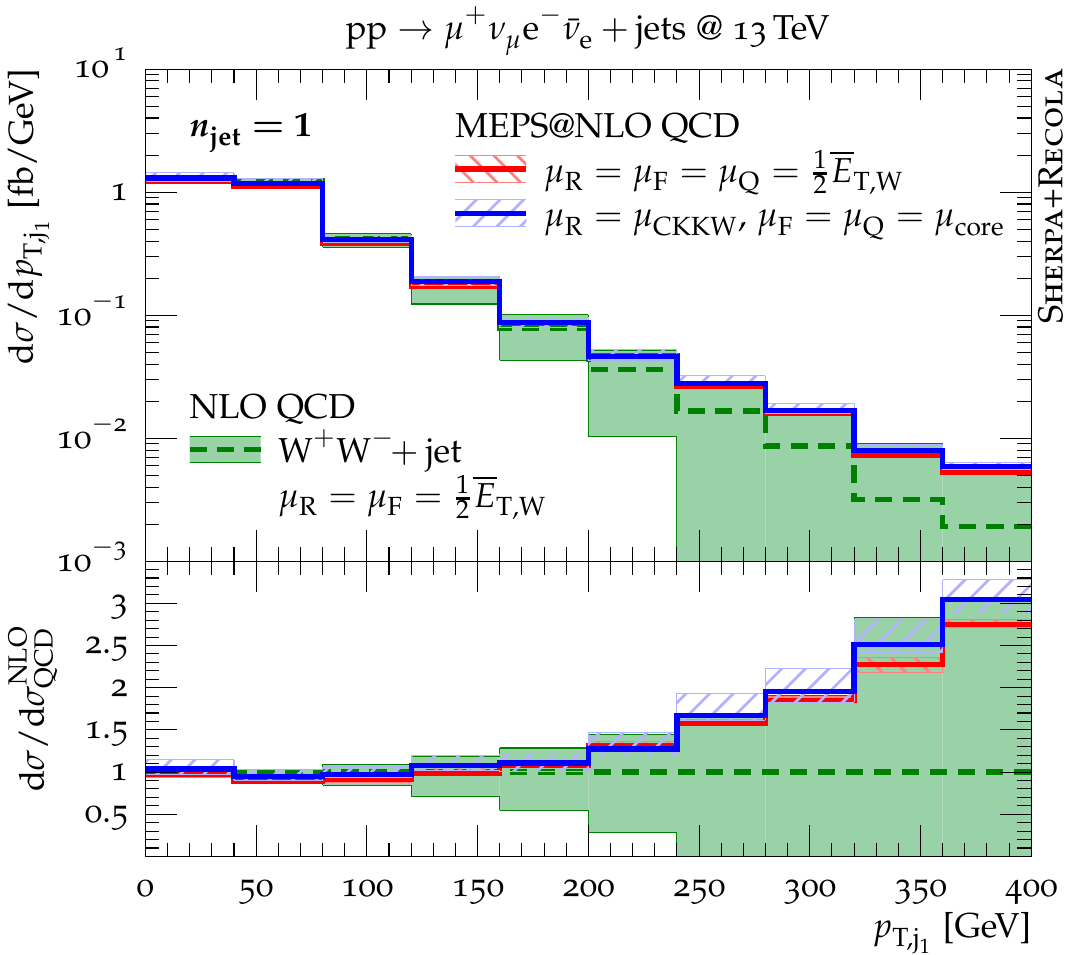}\hfill
  \includegraphics[width=0.45\textwidth]{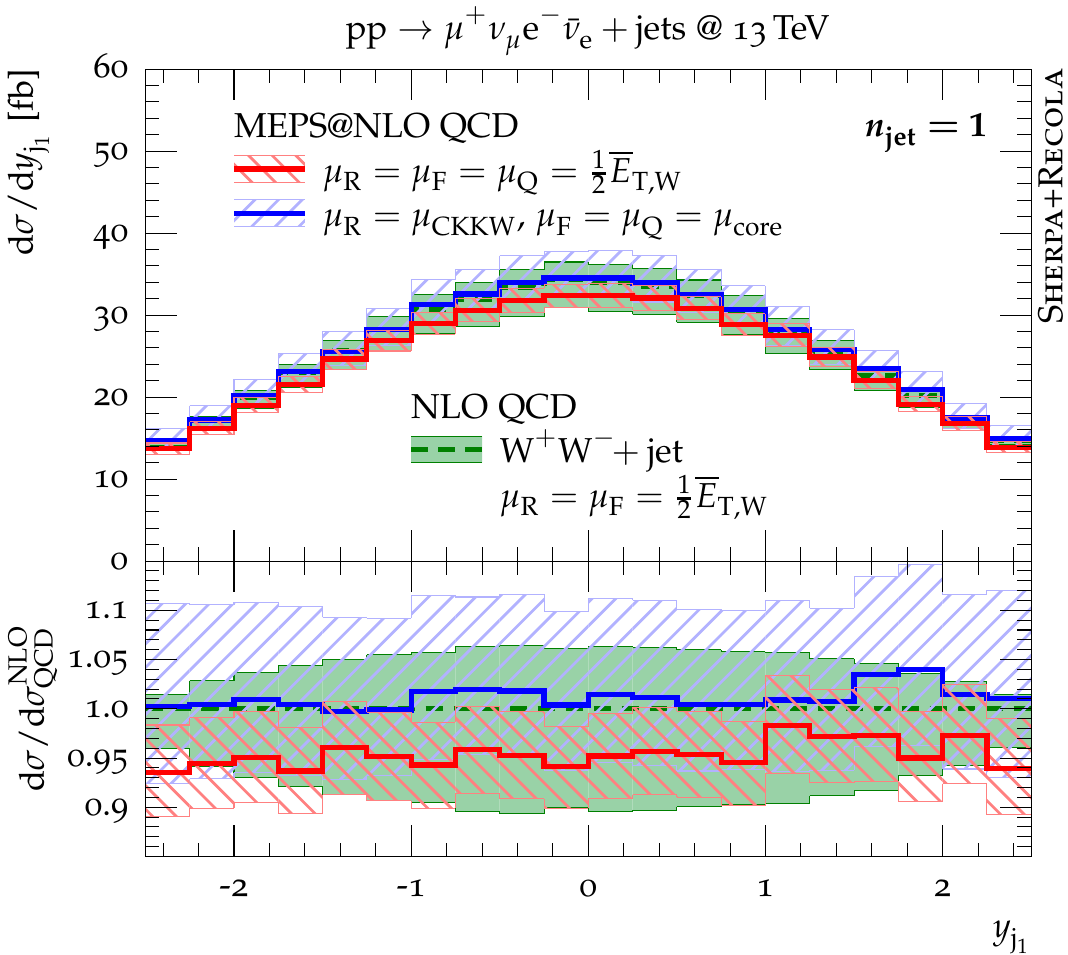}
  \includegraphics[width=0.45\textwidth]{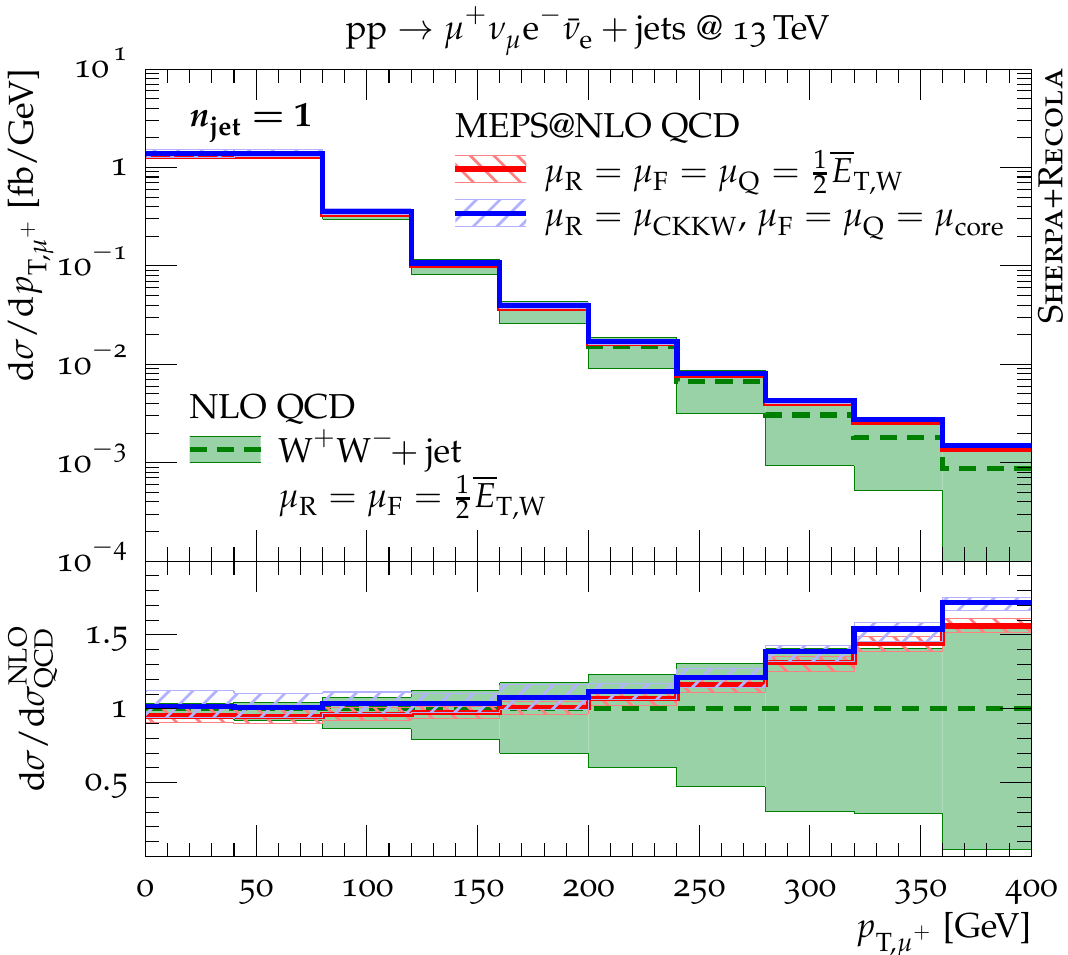}\hfill
  \includegraphics[width=0.45\textwidth]{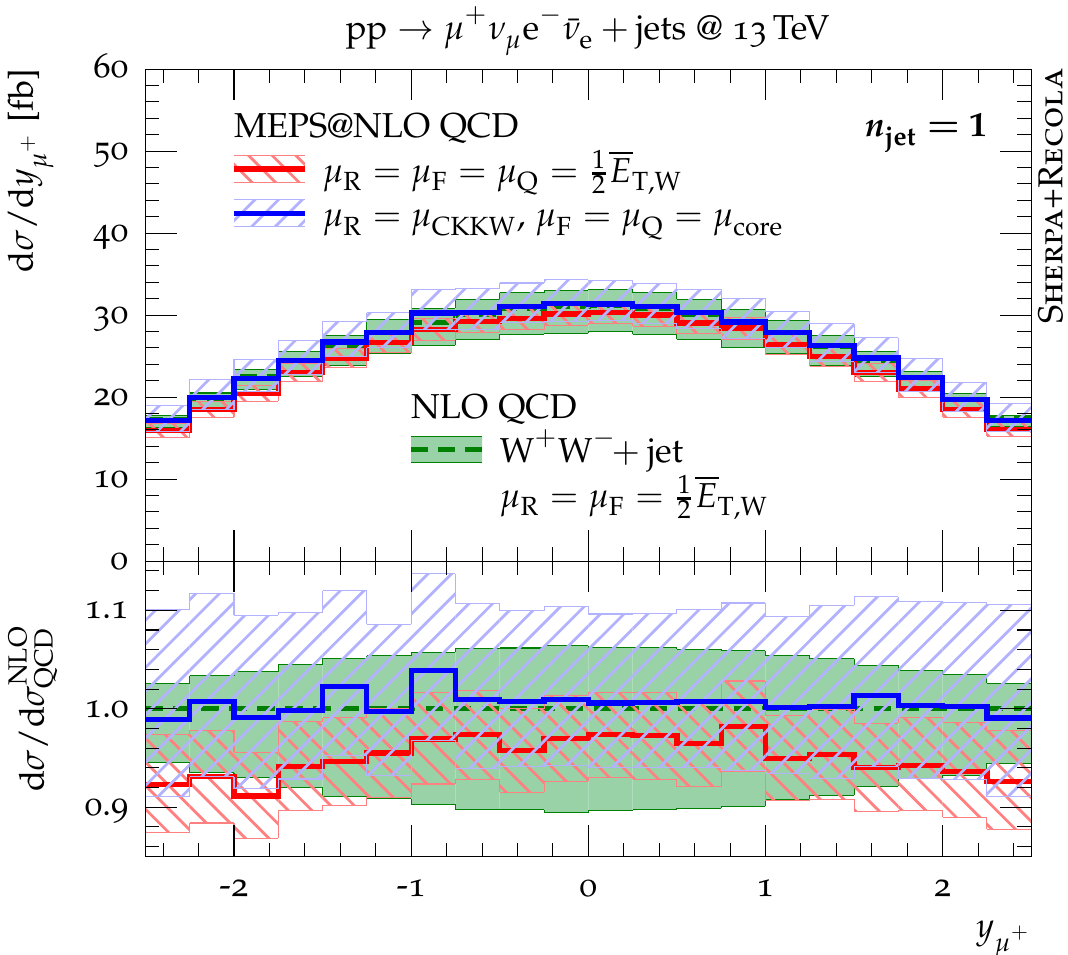}
  \includegraphics[width=0.45\textwidth]{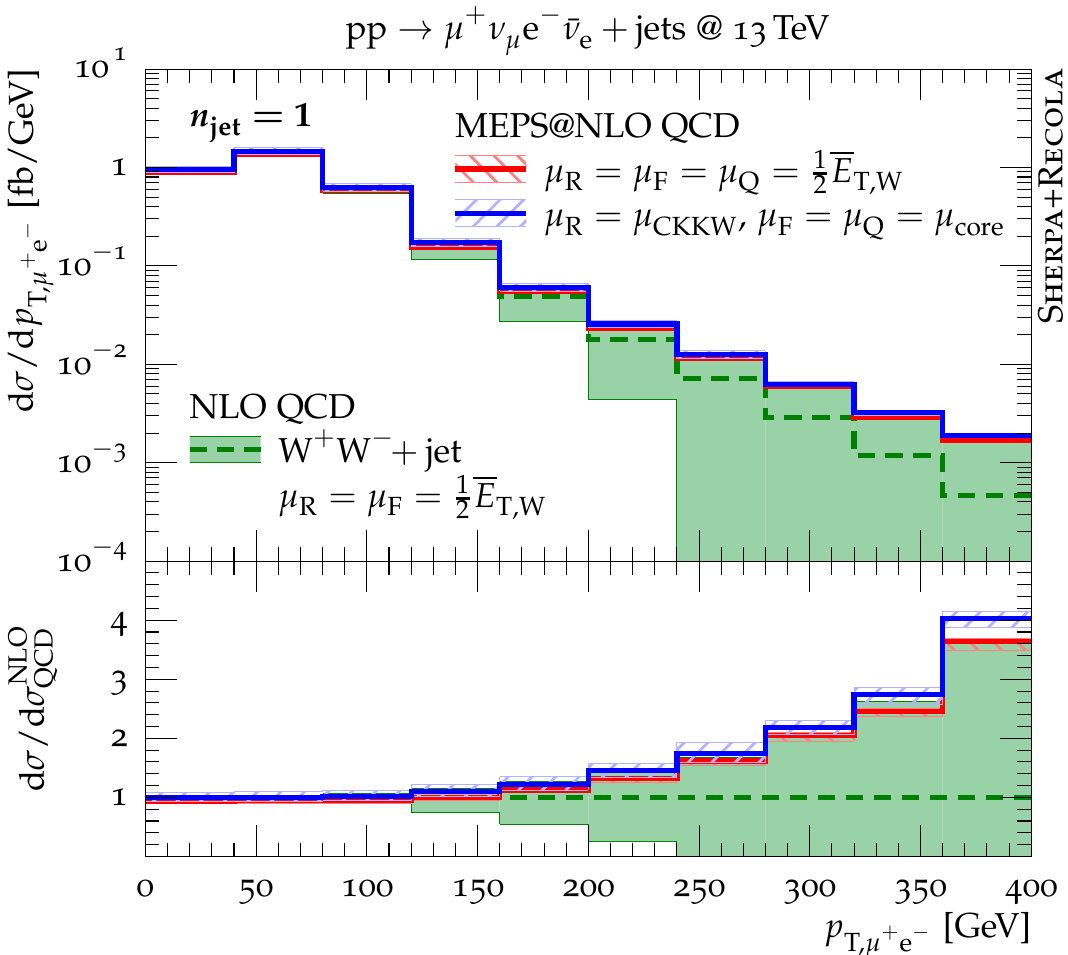}\hfill
  \includegraphics[width=0.45\textwidth]{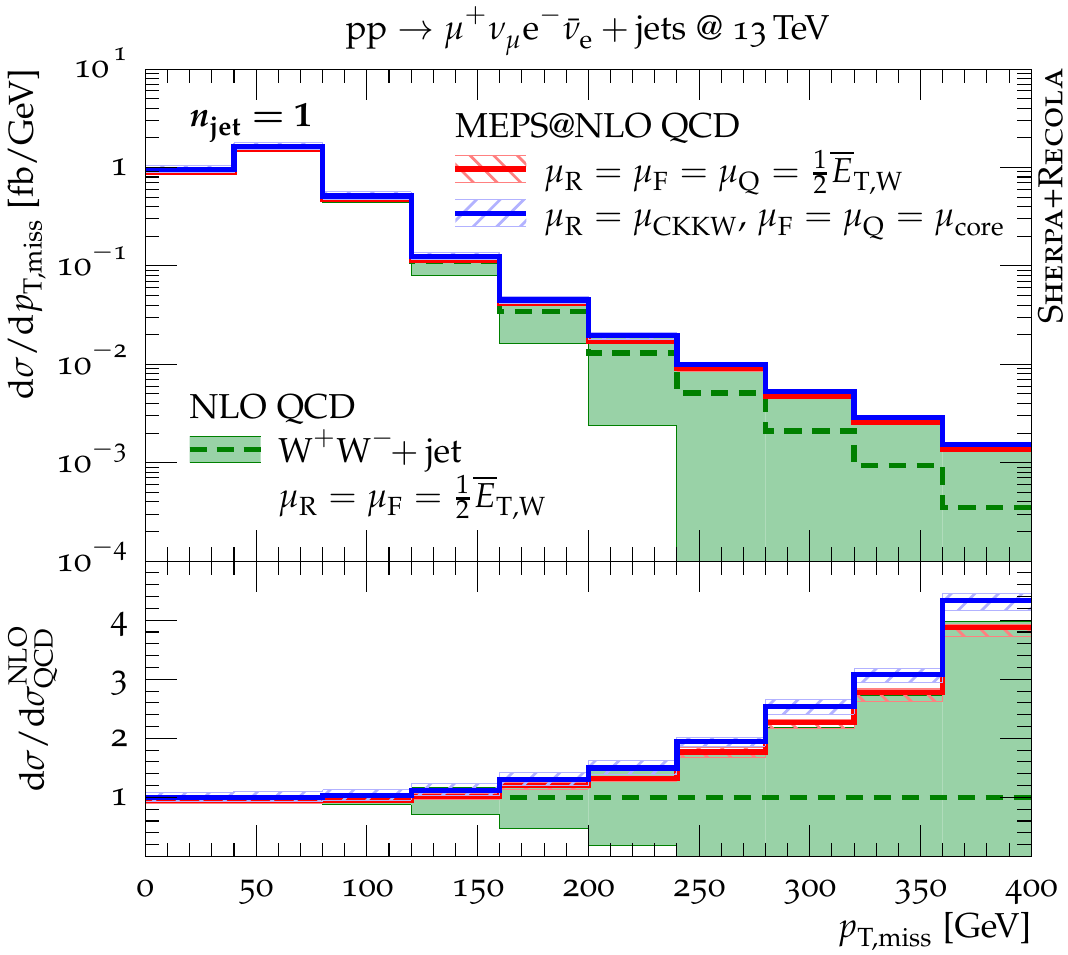}
  \caption{
    Comparison of NLO \QCD\ fixed-order results for $\Pp\Pp \to \mu^+ \nu_\mu \Pe^- \bar\nu_\Pe\Pj$
    with proto-merged and the fully multi-jet merged \MEPSatNLOQCD predictions in the $\njet=1$ event
    selection:
    Transverse momentum of the jet (top left),
    rapidity of the jet (top right),  
    transverse momentum of the anti-muon (middle left),
    rapidity of the anti-muon (middle right),
    transverse momentum of the anti-muon--electron system (bottom left), and
    missing transverse momentum (bottom right).
    No \QED\ or \EW\ corrections are taken into account here.
    \label{fig:fo-vs-merged-1j-1}
  }
\efig
\bfig
  \center
  \includegraphics[width=0.45\textwidth]{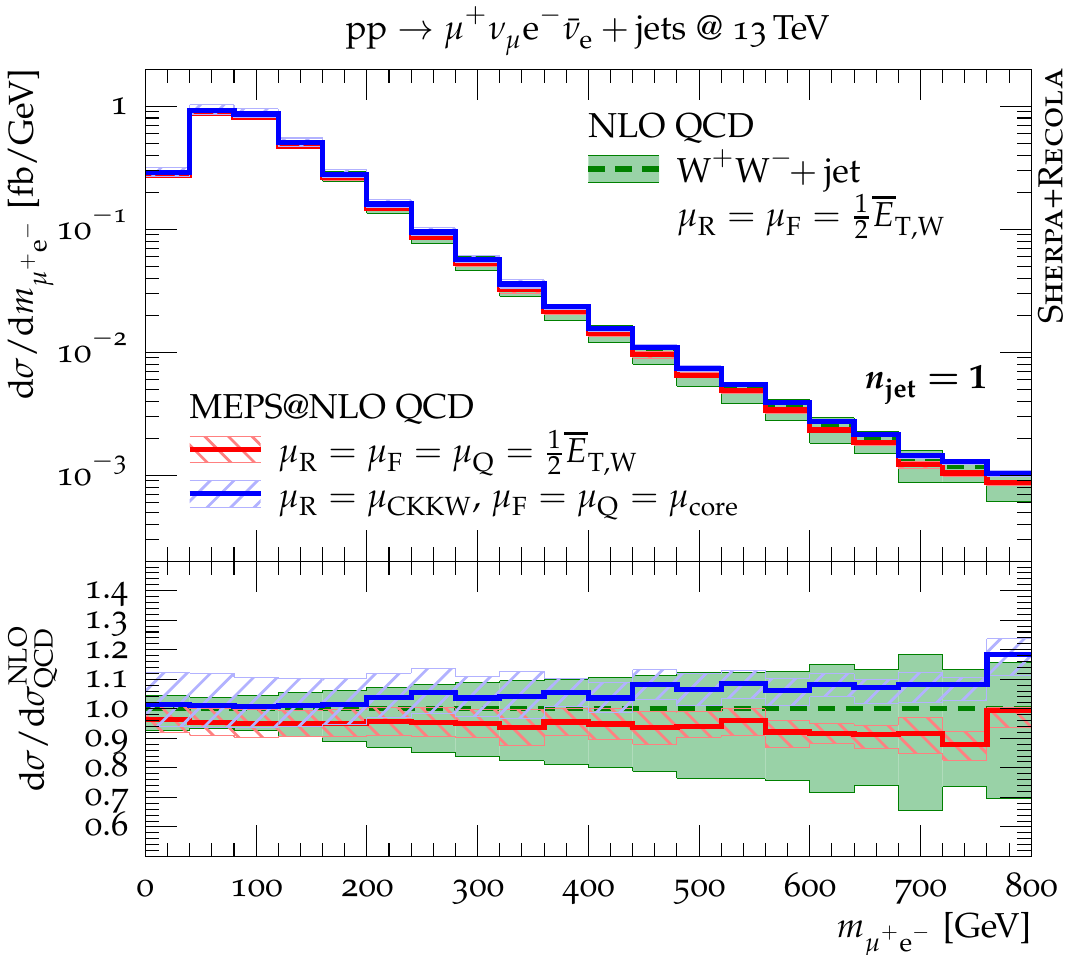}\hfill
  \includegraphics[width=0.45\textwidth]{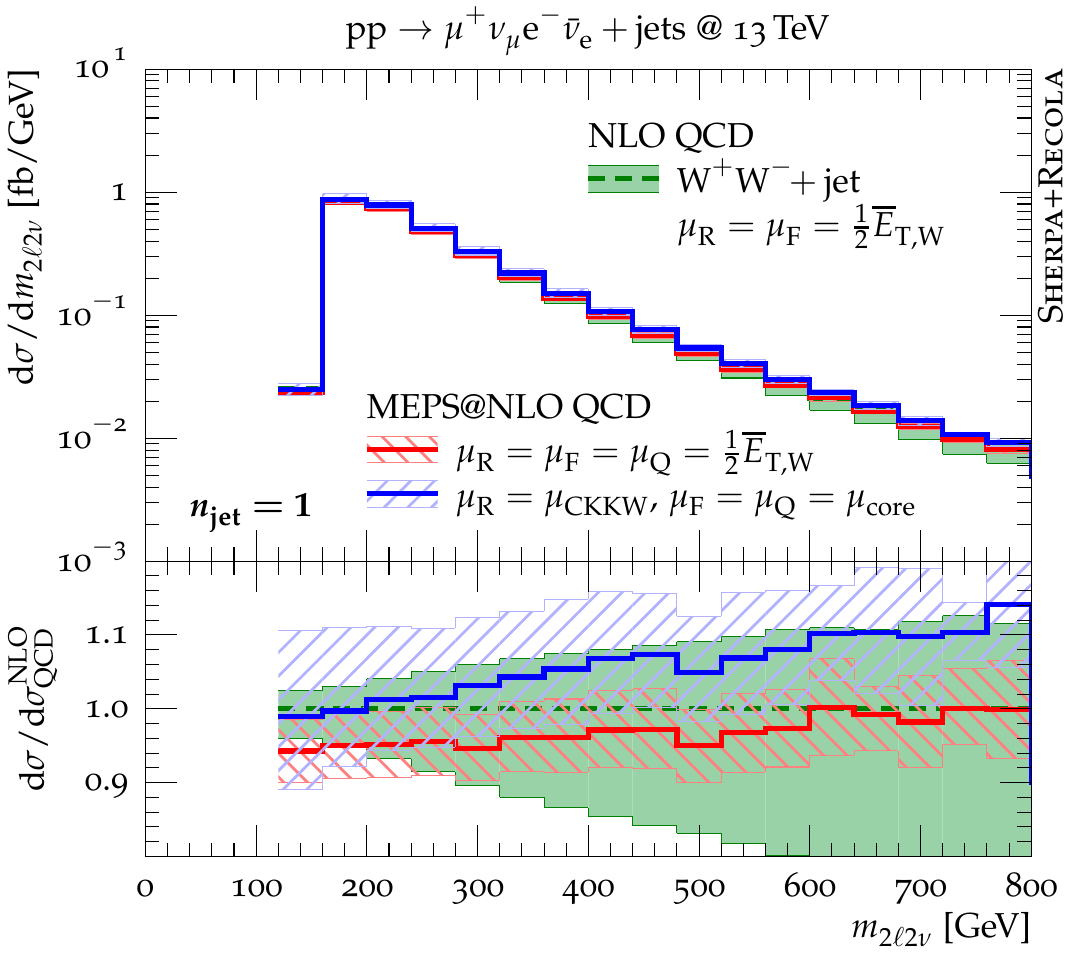}
  \includegraphics[width=0.45\textwidth]{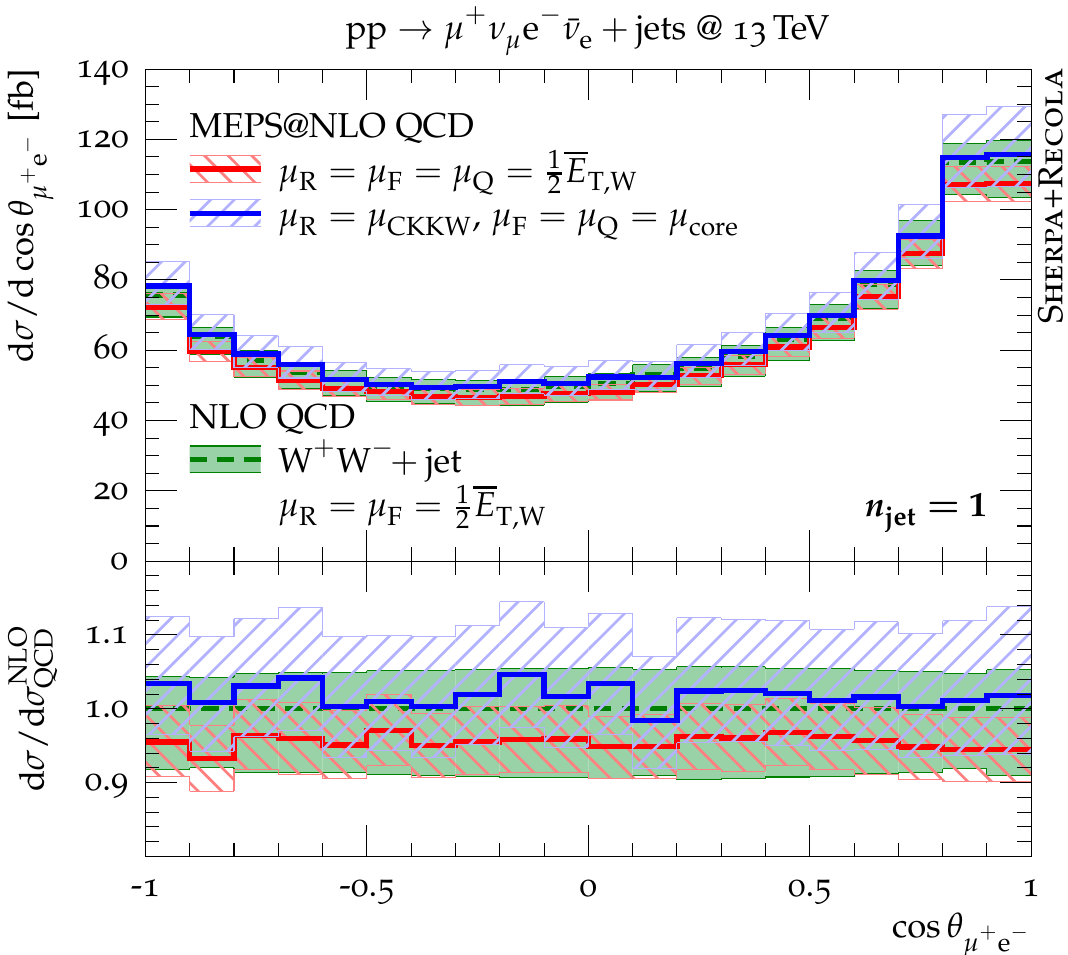}\hfill
  \includegraphics[width=0.45\textwidth]{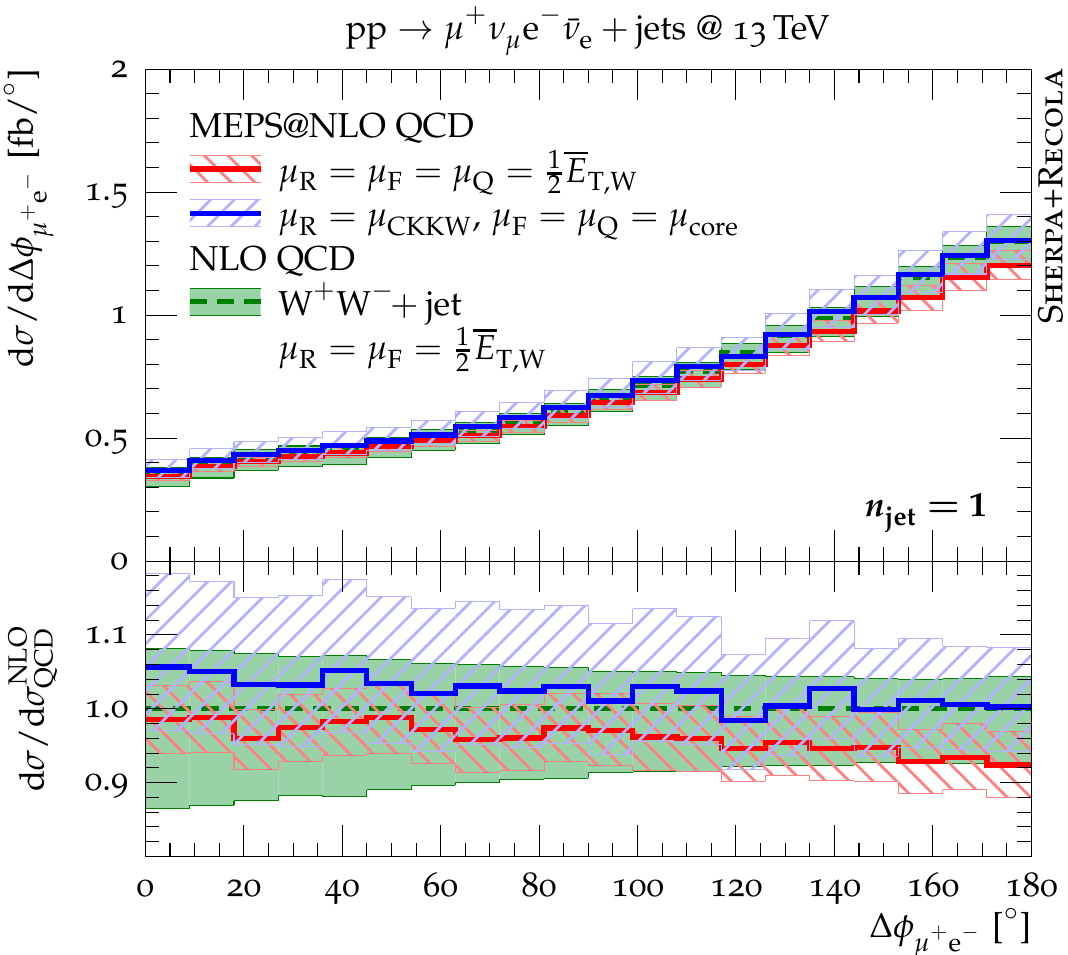}
  \caption{
    Comparison of NLO \QCD\ fixed-order results for the process $\Pp\Pp \to \mu^+ \nu_\mu \Pe^- \bar\nu_\Pe\Pj$
    with proto-merged and the fully multi-jet merged \MEPSatNLOQCD predictions in the $\njet=1$ event
    selection:
    Invariant mass of the anti-muon and electron (top left),
    invariant mass of the four leptons (top right),
    cosine of the angle between the anti-muon and the electron (bottom left), and
    azimuthal-angle distance between the anti-muon and the electron (bottom right).
    No \QED\ or \EW\ corrections are taken into account here.
    \label{fig:fo-vs-merged-1j-2}
  }
\efig
Most notably, a significant reduction of the systematic uncertainty in phase-space
regions affected by the jet-veto criterion is observed as a consequence of
including parton-shower resummation. In particular the various
transverse-momentum distributions, \emph{i.e.}\ $p_{\mathrm{T},\Pj_1}$, 
$p_{\mathrm{T},\mu^+}$, $p_{\mathrm{T},\mu^+\Pe^-}$,
and $p_{\mathrm{T},\text{miss}}$, receive huge corrections from the inclusion of multiple-emission effects 
through the parton shower and the higher-multiplicity matrix elements.
This results in significantly harder $p_{\mathrm{T}}$ spectra, as through multiple 
emissions larger recoil can be achieved without triggering the applied jet veto.
\change{The inclusion of parton-shower effects and merging significantly increases the predictions 
which thus do not feature the large negative corrections seen in the fixed-order case that are associated with the jet veto.}
For the $p_{\mathrm{T},\mu^+\Pe^-}$ and $p_{\mathrm{T},\text{miss}}$ distributions the
difference with respect to the NLO \QCD\ prediction gets as large as a factor of 4.
It is to note, however, that these sizeable differences are still compatible 
with the fixed-order predictions' scale-uncertainty estimates. 
In contrast, for the merged predictions the uncertainties remain at the $10\%$
level.

From the jet and anti-muon rapidity distributions in
Fig.~\ref{fig:fo-vs-merged-1j-1} it is apparent that the total production
rate obtained for the default CKKW scale-setting prescription used in the fully multi-jet
merged \MEPSatNLO calculation is in very good agreement with the fixed-order NLO \QCD\ 
result. In fact, for these rather inclusive observables these two central predictions
almost coincide, while the proto-merged prediction, using the scale
$\muR=\muF=\muQ=\tfrac{1}{2}\,\ETWmean$ throughout
shows a small shape distortion. However, as
observed for the zero-jet process already, the estimated systematic uncertainty of the
fully merged sample is somewhat increased with respect to the proto-merged and the fixed-order
calculation. 

In Fig.~\ref{fig:fo-vs-merged-1j-2} the $\cos\theta_{\mu^+\Pe^-}$ distribution
confirms this pattern. For the charged-leptons' polar-angle separation, however,
a small enhancement towards smaller values of $\Delta\phi_{\mu^+\Pe^-}$ can be observed in
the merged predictions. For the two invariant-mass distributions a significant reduction
of the scale uncertainty in the high-mass regions is observed. The merged sample
tends to populate these phase-space regions somewhat more, though the central predictions
stay \change{mainly} within the uncertainty band of the fixed-order result.

\FloatBarrier

\subsubsection{Including \EW\ corrections via the virtual approximation}
\label{sec:results:merged:merged}

Having compared the \MEPSatNLOQCD predictions against the fixed-order
calculations for the zero- and one-jet selection, we now progress by
considering the inclusion of approximate NLO \EW\ corrections into the
merged calculations. \change{The accuracy of this approximation has been 
examined for the different observables under consideration in this paper 
in \refse{se:results:fo}.}
The central prediction is formed by the fully multi-jet
merged sample based of the NLO \QCD\ matrix elements for
$\Pp\Pp \to \mu^+ \nu_\mu \Pe^- \bar \nu_\Pe$ and $\Pp\Pp \to \mu^+
\nu_\mu \Pe^- \bar \nu_\Pe \Pj$ and the tree-level ones for
$\Pp\Pp \to \mu^+ \nu_\mu \Pe^- \bar \nu_\Pe \Pj\Pj$ and $\Pp\Pp \to \mu^+
\nu_\mu \Pe^- \bar \nu_\Pe \Pj\Pj\Pj$, all matched to the \Sherpa
Catani--Seymour dipole shower.

For all matrix-element multiplicities we now include soft-photon
resummation effects via the YFS approach.
For the zero- and one-jet one-loop matrix elements we furthermore
employ the \EW\ virtual approximation, described in
\refse{se:merged_predictions}, both in the additive and multiplicative manner,
cf.\ Eqs.~(\ref{eq:mepsnlo_qcdpew}) and (\ref{eq:mepsnlo_qcdtew}), respectively.
We would like to remind the reader that an overlap of the QED corrections 
provided by the soft-photon resummation and the approximate \EW\ corrections 
exists. 
It does, however, not impact the accuracy of the method in the targeted 
\EW\ Sudakov regime, cf.\ the discussion at the end of \refse{se:merged_predictions}.

As an additional reference, we furthermore compile predictions from
merging the LO matrix elements for $\Pp\Pp \to \mu^+ \nu_\mu \Pe^- \bar \nu_\Pe+0,1,2,3\,\Pj$
using the \MEPSatLO approach~\cite{Hoeche:2009rj}. 
As for the \MEPSatNLO calculations we use $\Qcut=30\;\GeV$ here. 

In \reftas{tab:xs-j0-merged} and \ref{tab:xs-j1-merged} we compile the
fiducial cross sections for the various theoretical predictions in the
zero- and one-jet selection, respectively. 
\begin{table}
\begin{center}
\begin{tabular}{c|c|c|c}
  \MEPSatLO &\multicolumn{3}{c}{\MEPSatNLO}\\
 \rule[-2ex]{0ex}{3ex}%
  \QCD\ [fb] & \QCD\ [fb] & \QCDpEWapprox\ [fb] & \QCDtEWapprox\ [fb]\\
\hline \rule{0ex}{3ex}%
279.8$^{+7.8\%}_{-8.0\%}$ & 322.8$^{+7.3\%}_{-5.8\%}$ & 318.8 & 318.4 \\
\end{tabular}
\end{center}
\caption{\label{tab:xs-j0-merged} Fiducial cross sections for 
  $\Pp\Pp \to \mu^+ \nu_\mu \Pe^- \bar \nu_\Pe$, \emph{i.e.}\ for the exclusive
  zero-jet event selection, at $\sqrt{s}=13\TeV$ 
  for \MEPSatLO, \MEPSatNLOQCD, \MEPSatNLOQCDpEWapprox and \MEPSatNLOQCDtEWapprox.}
\end{table}
\begin{table}
\begin{center}
\begin{tabular}{c|c|c|c}
  \MEPSatLO &\multicolumn{3}{c}{\MEPSatNLO}\\
 \rule[-2ex]{0ex}{3ex}%
  \QCD\ [fb] & \QCD\ [fb] & \QCDpEWapprox\ [fb] & \QCDtEWapprox\ [fb]\\
\hline  \rule{0ex}{3ex}%
108.7$^{+17.6\%}_{-10.2\%}$ &  131.8$^{+9.6\%}_{-6.9\%}$ & 129.2 & 129.0 \\
\end{tabular}
\end{center}
\caption{\label{tab:xs-j1-merged} Fiducial cross sections for
  $\Pp\Pp \to\mu^+ \nu_\mu \Pe^- \bar \nu_\Pe \Pj$, \emph{i.e.}\ for the exclusive
  one-jet event selection, at $\sqrt{s}=13\TeV$ 
  for \MEPSatLO, \MEPSatNLOQCD, \MEPSatNLOQCDpEWapprox and \MEPSatNLOQCDtEWapprox.}
\end{table}
These can be directly compared to
the respective fixed-order results quoted in \reftas{tab:ww} and \ref{tab:wwj}.
We recognise that the \MEPSatLO cross sections for both event selections are
significantly lower ($12\%$ for $\PW\PW$ and $33\%$ for $\PW\PW\Pj$ production) than at fixed
order, originating from the inclusion of
parton emissions off the respective Born configuration that can trigger the
applied jet veto and thus reduce the na\"ive LO cross section. Both \MEPSatNLOQCD
cross sections are in very good agreement with the respective fixed-order
result (within $2\%$ for both $\PW\PW$ and $\PW\PW\Pj$ production). 
The \MEPS NLO \QCD\ corrections amount to $+15\%$ for $\PW\PW$ and 
$+21\%$ for $\PW\PW\Pj$. While the scale uncertainty is only
marginally reduced by going to NLO \QCD\ for $\PW\PW$, it decreases by
almost a factor of two for  $\PW\PW\Pj$.
The rates for the additive and multiplicative inclusion of approximate
\EW\ NLO corrections come out somewhat larger than at fixed order. In fact, for
both selections these corrections stay below $-1.5\%$.  
Again, it is to note that these \EW\ corrections are tailored to the \EW\ Sudakov 
regime and are not expected to fully reproduce the exact NLO \EW\ corrections 
for inclusive observables. 

In Figs.~\ref{fig:merged-0j-1} and \ref{fig:merged-0j-2} we display
differential distributions for the zero-jet event selection, while results for the
one-jet selection are presented in Figs.~\ref{fig:merged-1j-1} and
\ref{fig:merged-1j-2}.
\bfig
  \center
  \includegraphics[width=0.45\textwidth]{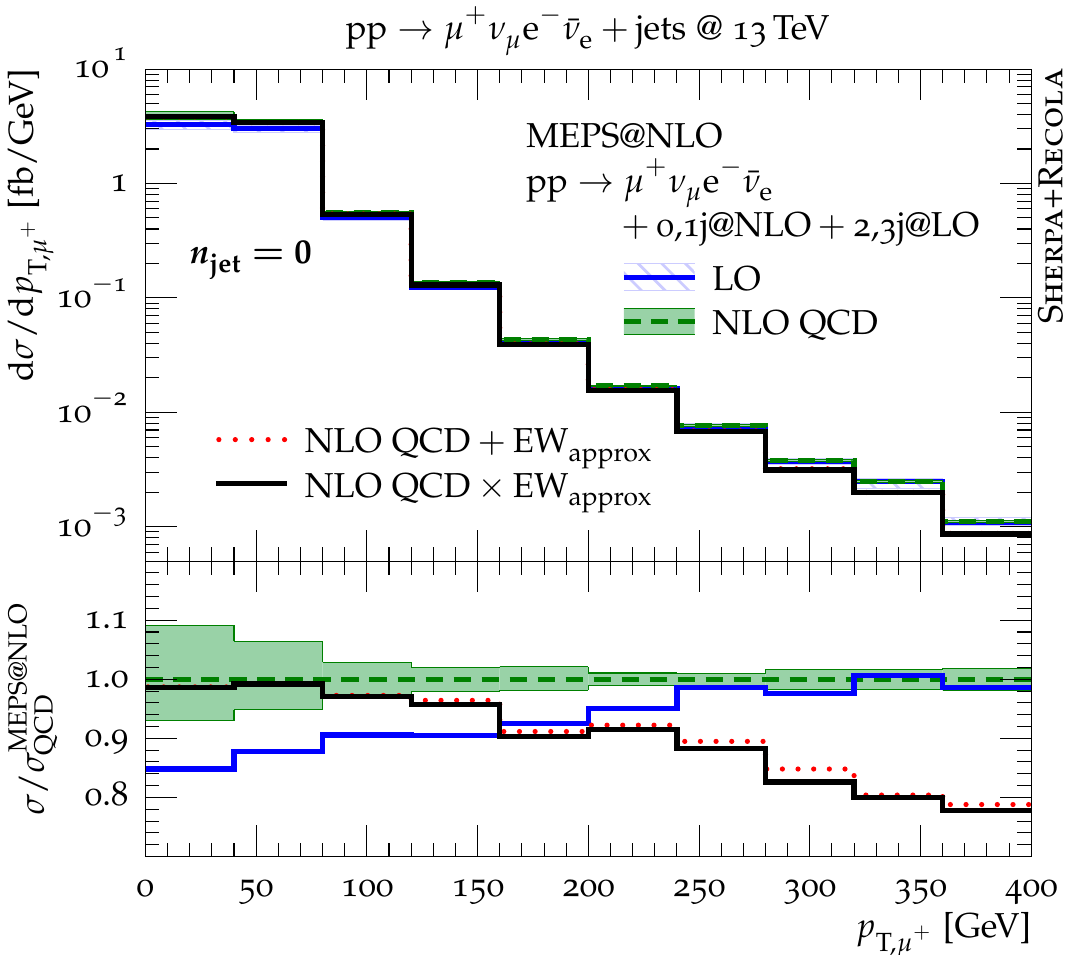}\hfill
  \includegraphics[width=0.45\textwidth]{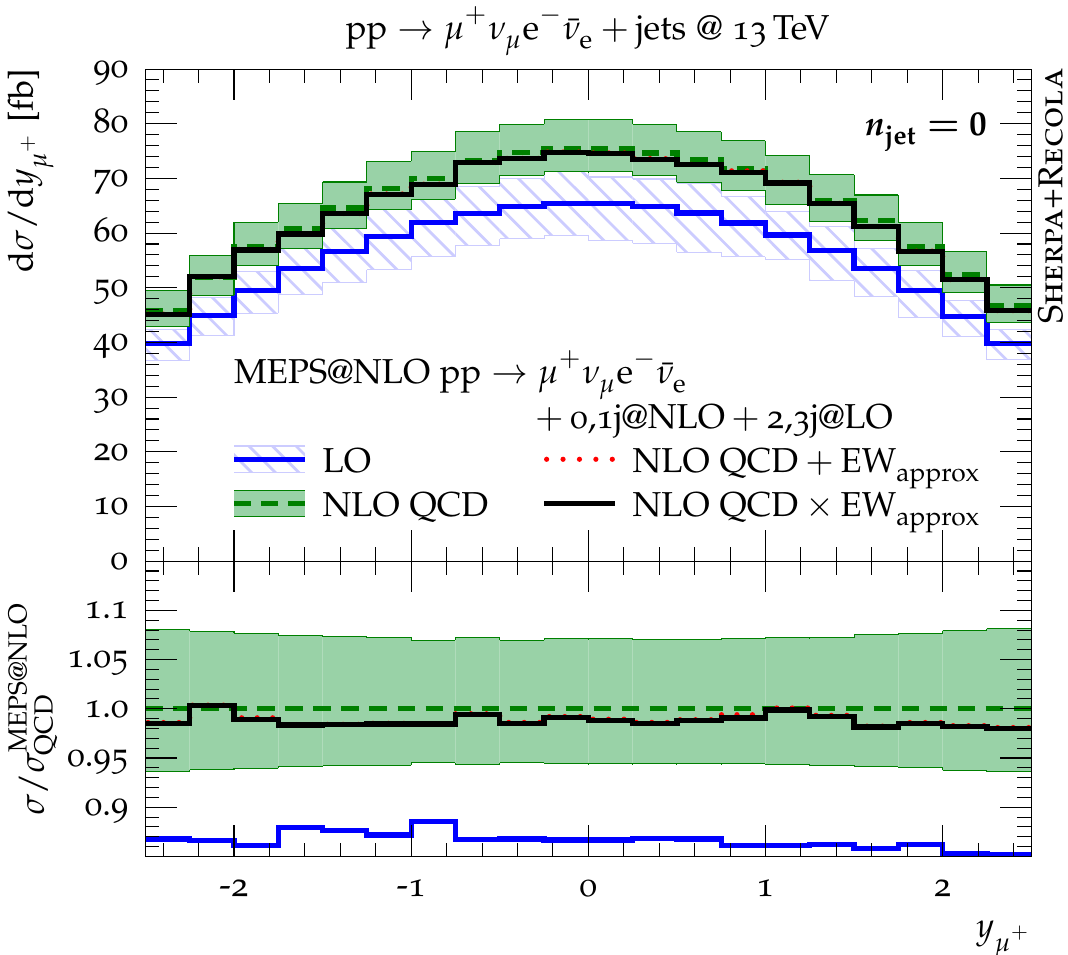}
  \includegraphics[width=0.45\textwidth]{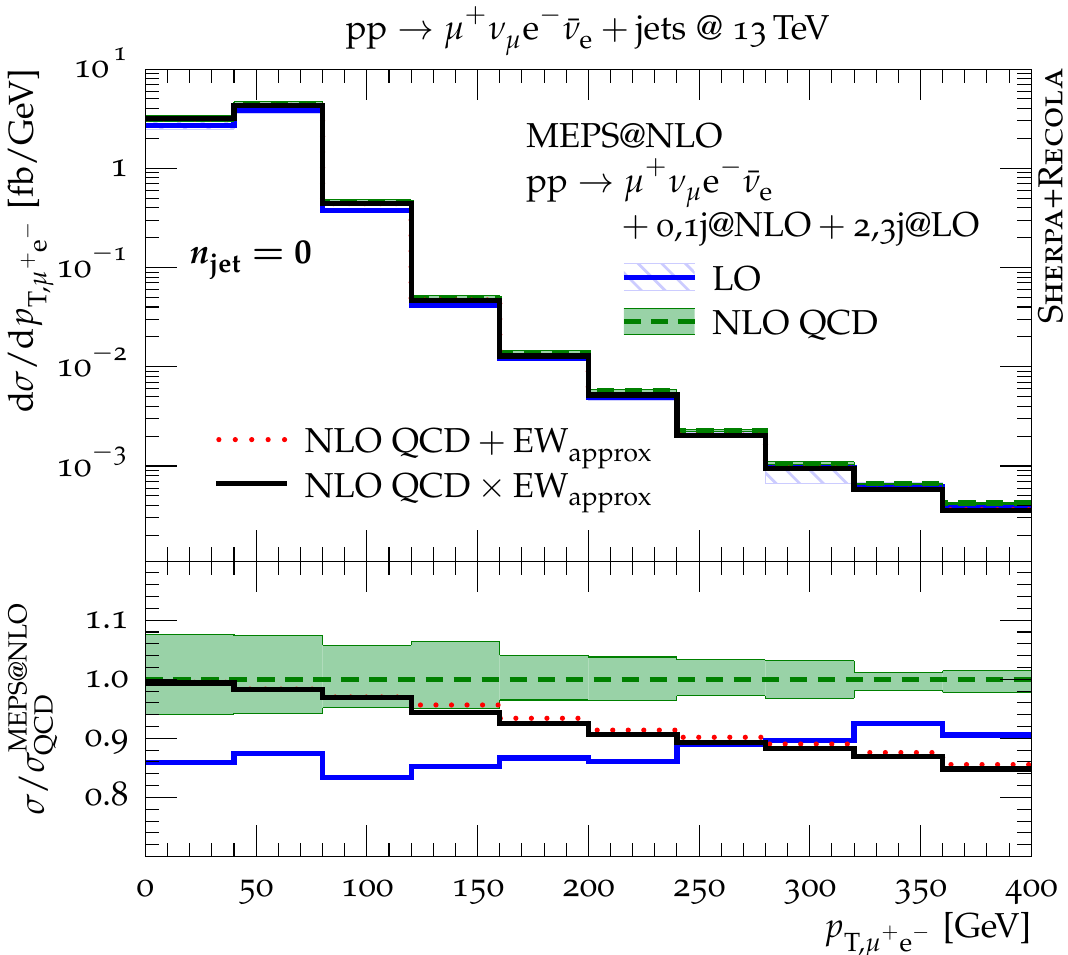}\hfill
  \includegraphics[width=0.45\textwidth]{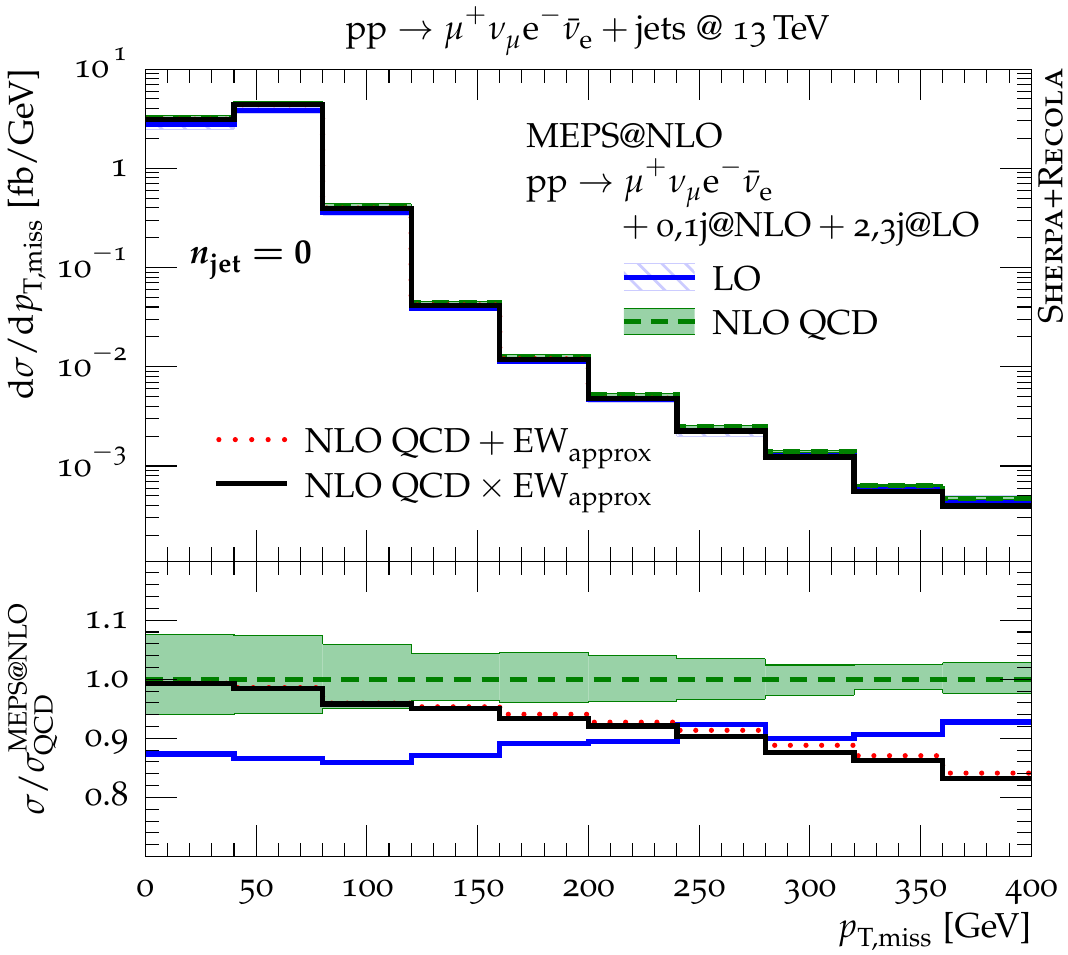}
  \caption{
    \label{fig:merged-0j-1}
    Predictions from multi-jet merged parton-shower simulations for the
    $n_\text{jet}=0$ event selection:
    Transverse momentum of the anti-muon (top left),
    rapidity of the anti-muon (top right),
    transverse momentum of the anti-muon--electron system (bottom left), and
    missing transverse momentum (bottom right).
    All results contain YFS soft-photon resummation.
    For the \MEPSatNLO calculation we present results including
    approximate NLO \EW\ corrections in the additive and multiplicative approach.
  }
\efig
\bfig
  \center
  \includegraphics[width=0.45\textwidth]{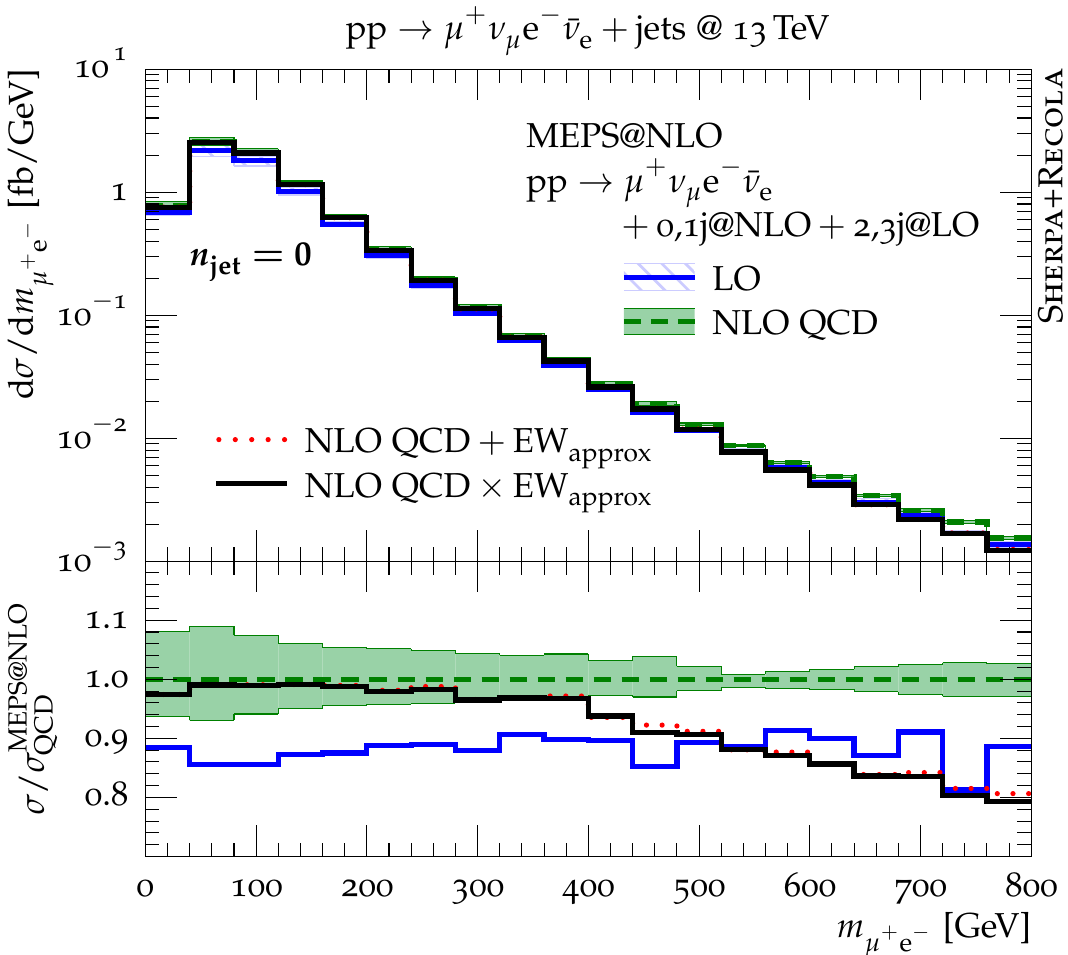}\hfill
  \includegraphics[width=0.45\textwidth]{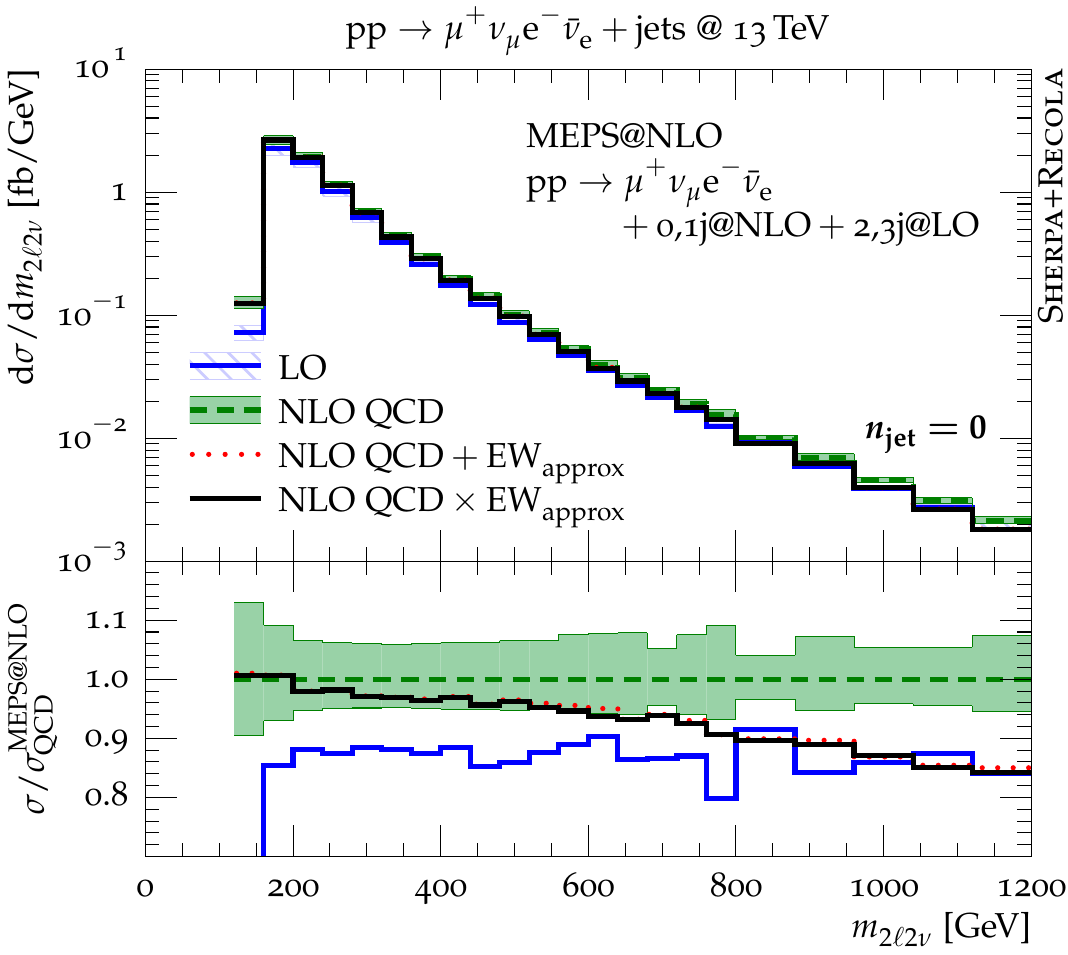}
  \includegraphics[width=0.45\textwidth]{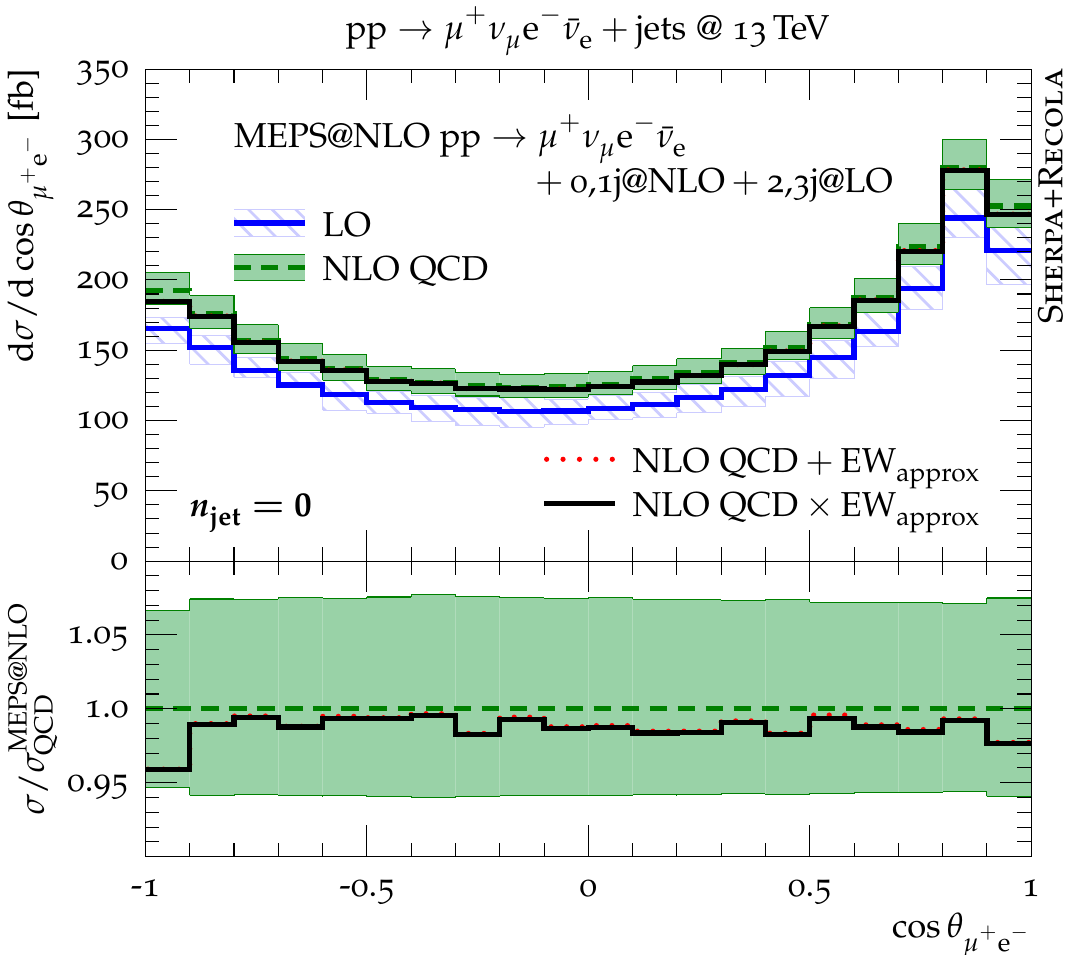}\hfill
  \includegraphics[width=0.45\textwidth]{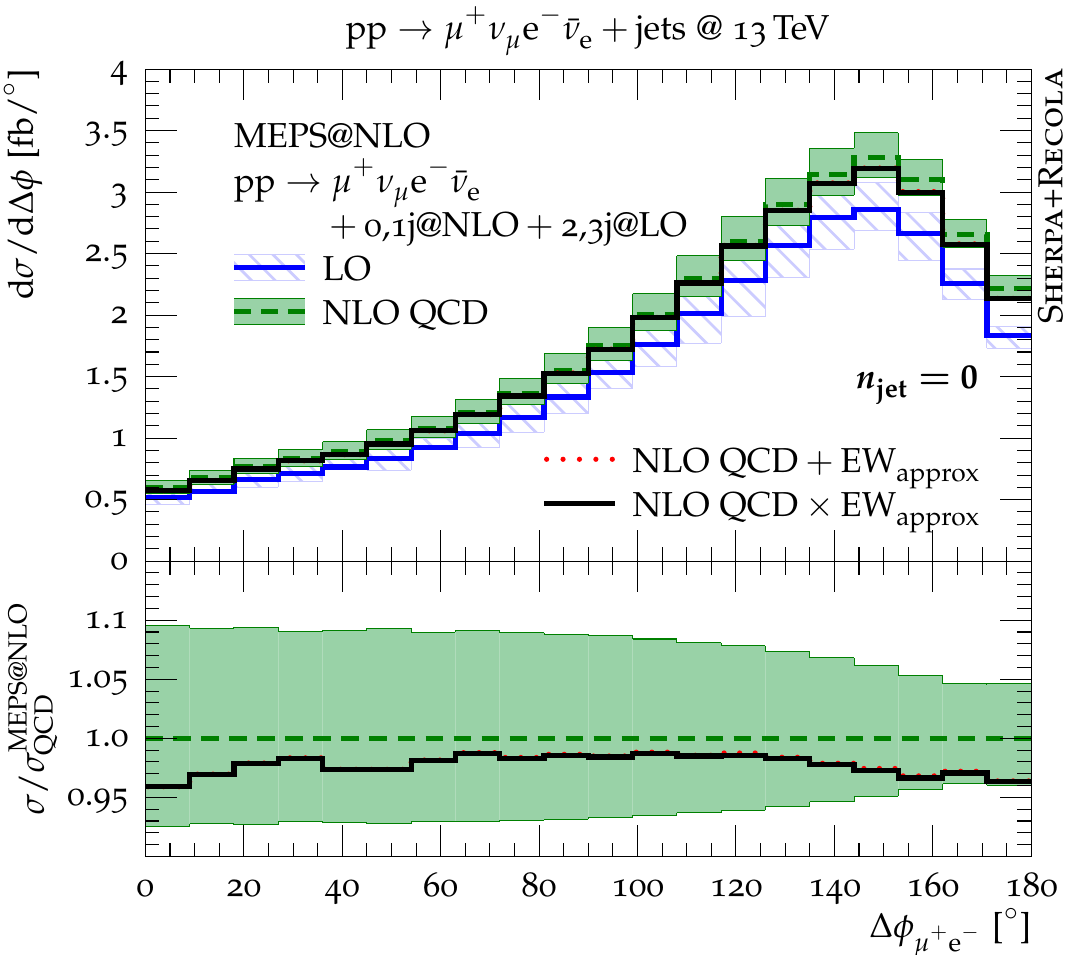}
  \caption{
    \label{fig:merged-0j-2}
    Predictions from multi-jet merged parton-shower simulations for the
    $n_\text{jet}=0$ event selection:
    Invariant mass of the anti-muon and electron (top left),
    invariant mass of the four leptons (top right),
    cosine of the angle between the anti-muon and the electron (bottom left), and
    azimuthal-angle distance between the anti-muon and the electron (bottom right).
    All results contain YFS soft-photon resummation.
    For the \MEPSatNLO calculation we present results including
    approximate NLO \EW\ corrections in the additive and multiplicative approach.
  }
\efig
Finally, in Figs.~\ref{fig:merged_ratio_1} and \ref{fig:merged_ratio_2}
we present predictions for the ratio of differential distributions
between $\Pp\Pp \to \mu^+ \nu_\mu \Pe^- \bar \nu_\Pe$ and
$\Pp\Pp \to \mu^+ \nu_\mu \Pe^- \bar \nu_\Pe \Pj$, that have been
studied at LO \QCD, NLO \QCD, NLO \QCDpEW, and NLO \QCDtEW\ in
\refse{se:results:fo:ratios}.

In all these plots in the upper panels we display the \MEPSatNLO
predictions, now including YFS soft-photon resummation, as green dashed line including its
7-point scale variation uncertainty band. The corresponding \MEPSatLO
predictions are indicated by the blue solid line and the hatched uncertainty
band. Furthermore, \MEPSatNLO predictions including the \EW\ virtual
approximation in its additive (dotted red) and
multiplicative (solid black) manner, are provided. In the lower panels
we compile the ratios with respect to the \MEPSatNLOQCD prediction.

We begin the discussion with the results for the zero-jet event selection. In
Figs.~\ref{fig:merged-0j-1} and \ref{fig:merged-0j-2} we recognise, that
the inclusion of the NLO \QCD\ matrix elements for the zero- and one-jet
processes increases the fiducial cross section by about $15\%$
as already seen in \refta{tab:xs-j0-merged} but has a comparably 
mild impact on the shapes of differential distributions. Most notably, for
the transverse momentum of the anti-muon the shape distortion reaches about
$15\%$, and the \QCD\ corrections decrease with increasing transverse
momentum. 

Concerning the impact of the approximate NLO \EW\ corrections, two patterns emerge.
For the transverse-momentum-type observables, as well as the
invariant masses $m_{\mu^+\Pe^-}$ and $m_{2\ell2\nu}$, \EW\ corrections suppress
the high-$p_{\mathrm{T}}$ and high-mass tails, up to about $-20\%$ for the
considered observable ranges as a consequence of enhanced \EW\ 
logarithmic corrections. 
For the anti-muon rapidity distributions, as well as the two considered
angular observables, \emph{i.e.}\ $\cos\theta_{\mu^+\Pe^-}$ and $\Delta\phi_{\mu^+\Pe^-}$,
the \EW\ corrections are very small and essentially flat, consistent with the
observation for the fixed-order calculations.

However, in contrast to the fixed-order
results presented in \refse{se:results:fo:WW}, the additive and
multiplicative approach of combining \QCD\ and \EW\ corrections here
yield very similar results, with the size of the corrections being
close to the multiplicative fixed-order scheme. 
\change{The small difference between the three results} is due to several reasons. First of all, the \QCD\ corrections are
much smaller in the merged calculation since more contributions are
incorporated in the corresponding LO results and on top of that even
tend to decrease with increasing transverse momenta where \EW\ 
corrections are sizeable. Second, the additive combination as defined
in Eq.~\refeq{eq:mepsnlo_qcdpew} takes \QCDtEW\ corrections into
account via the explicit higher-multiplicity processes and the parton
showers.

\bfig
  \center
  \includegraphics[width=0.45\textwidth]{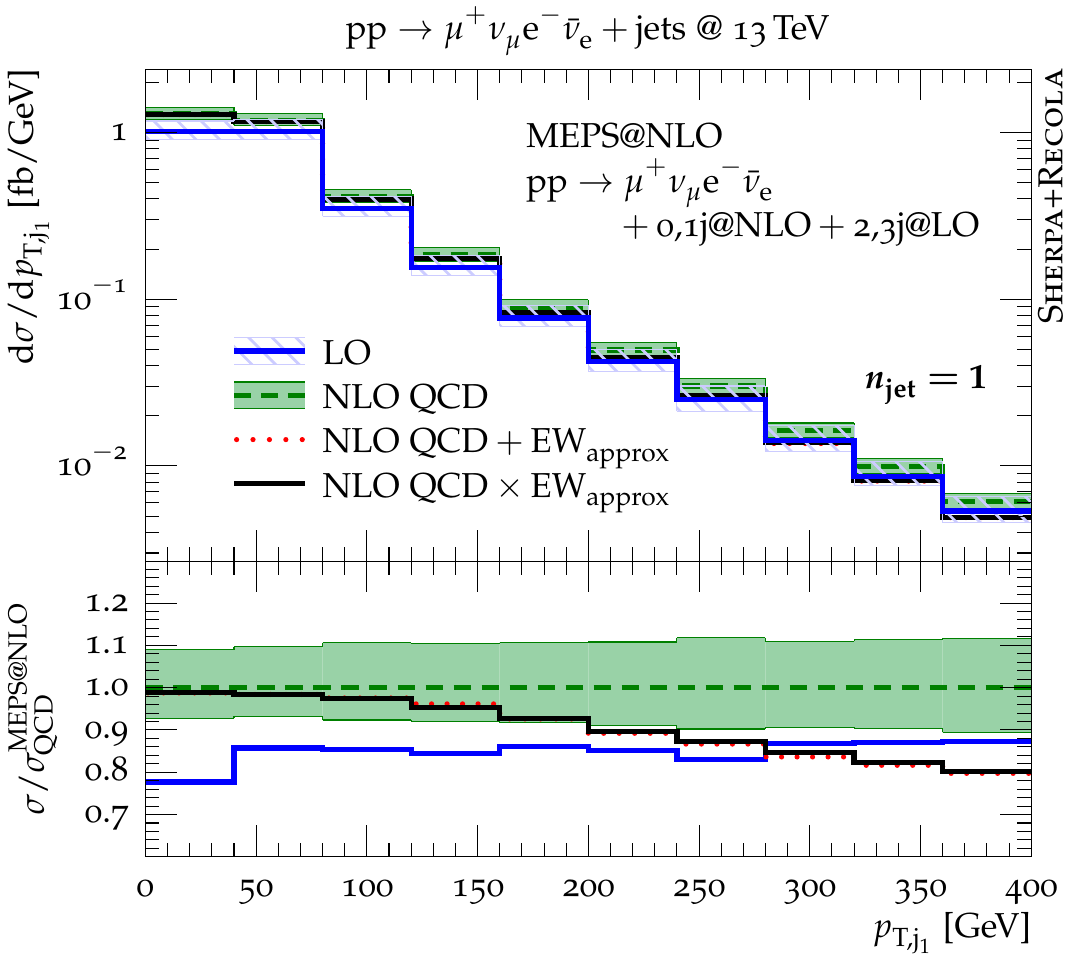}\hfill
  \includegraphics[width=0.45\textwidth]{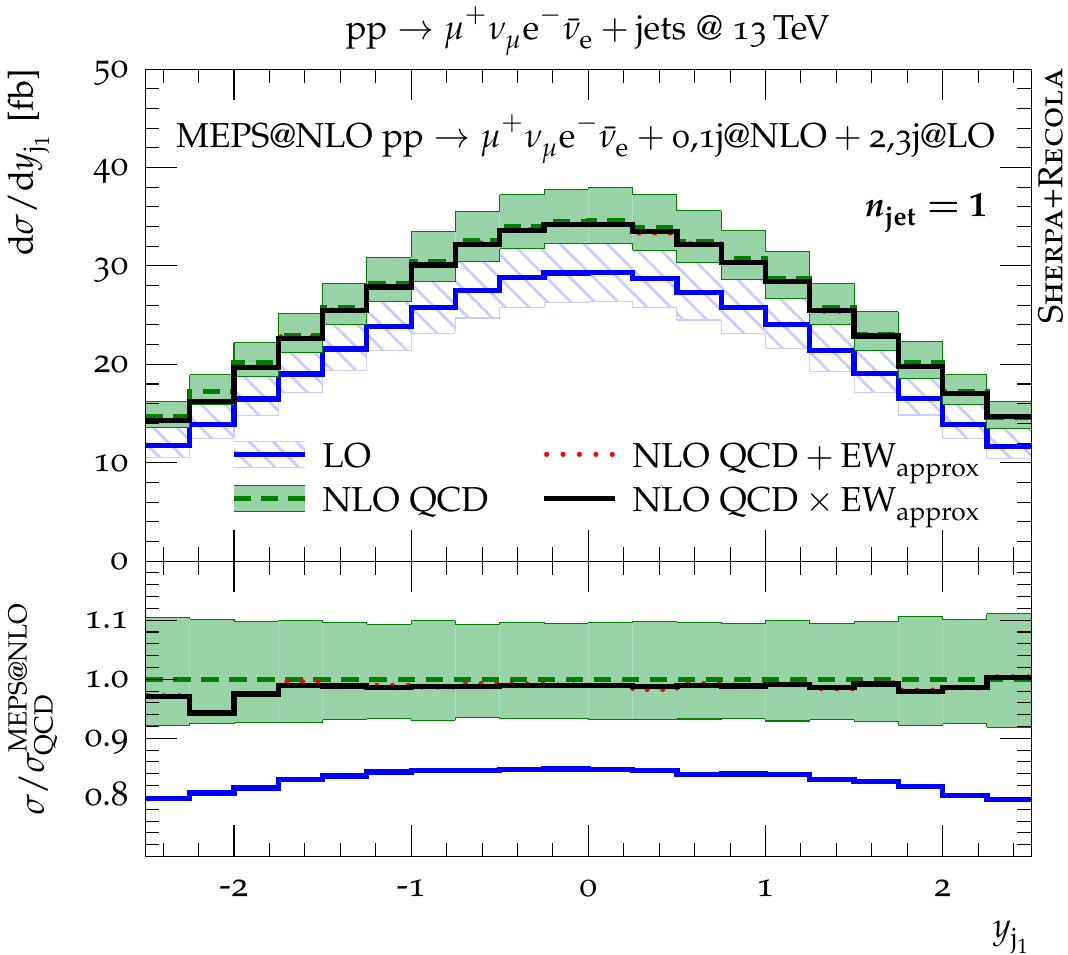}
  \includegraphics[width=0.45\textwidth]{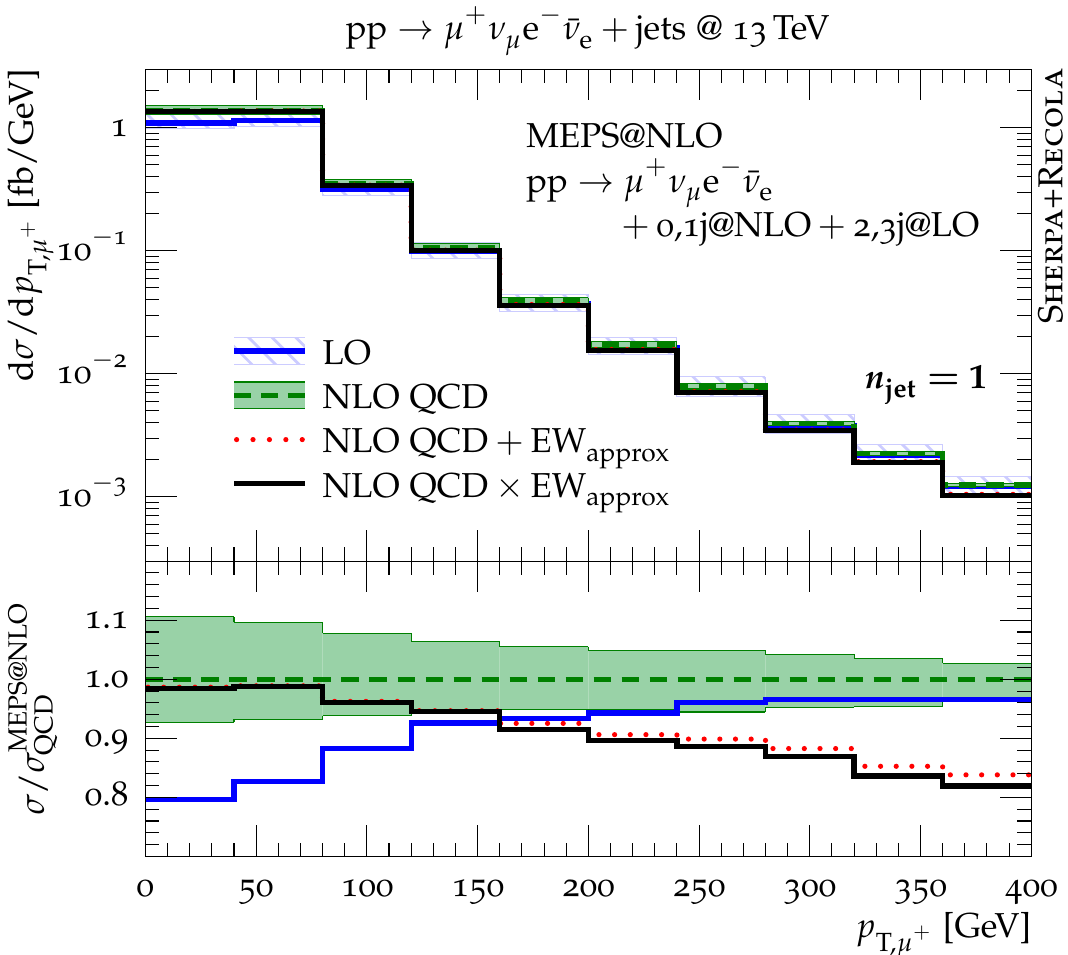}\hfill
  \includegraphics[width=0.45\textwidth]{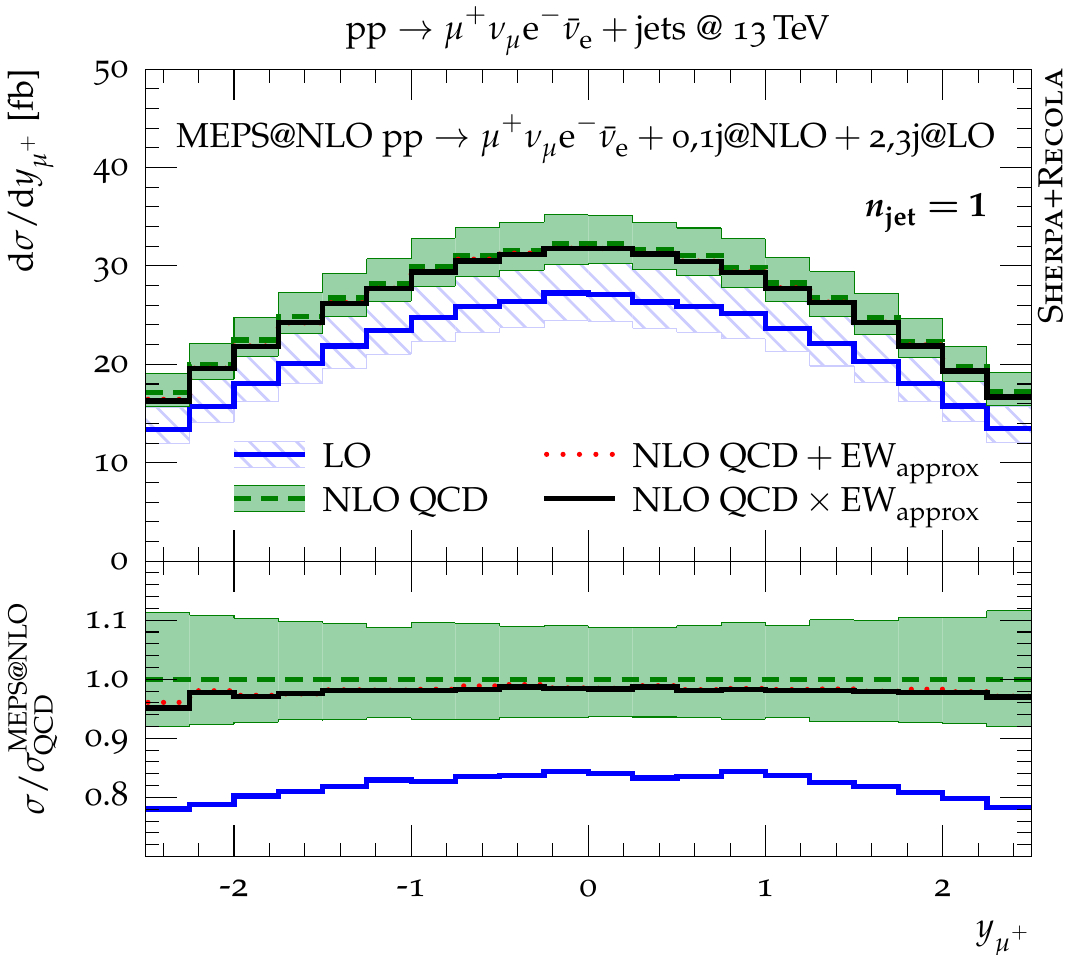}
  \includegraphics[width=0.45\textwidth]{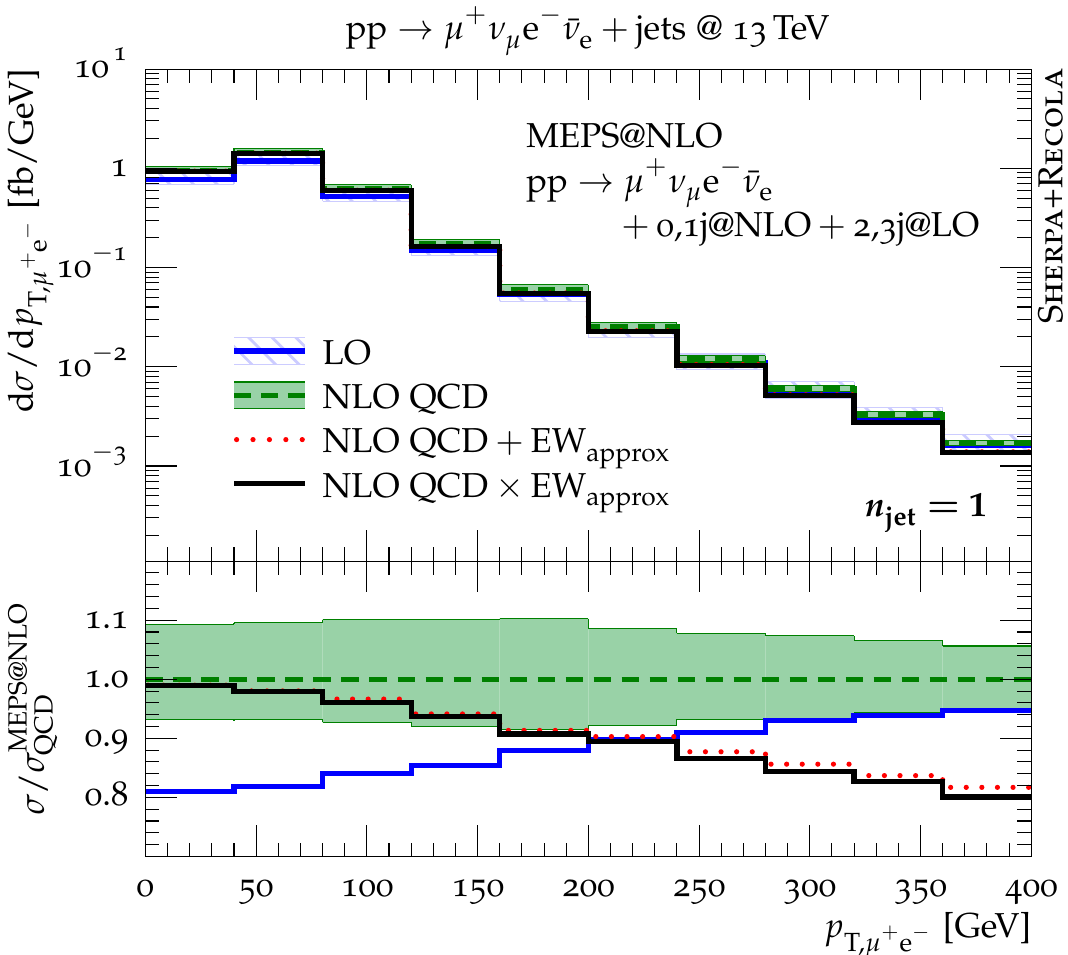}\hfill
  \includegraphics[width=0.45\textwidth]{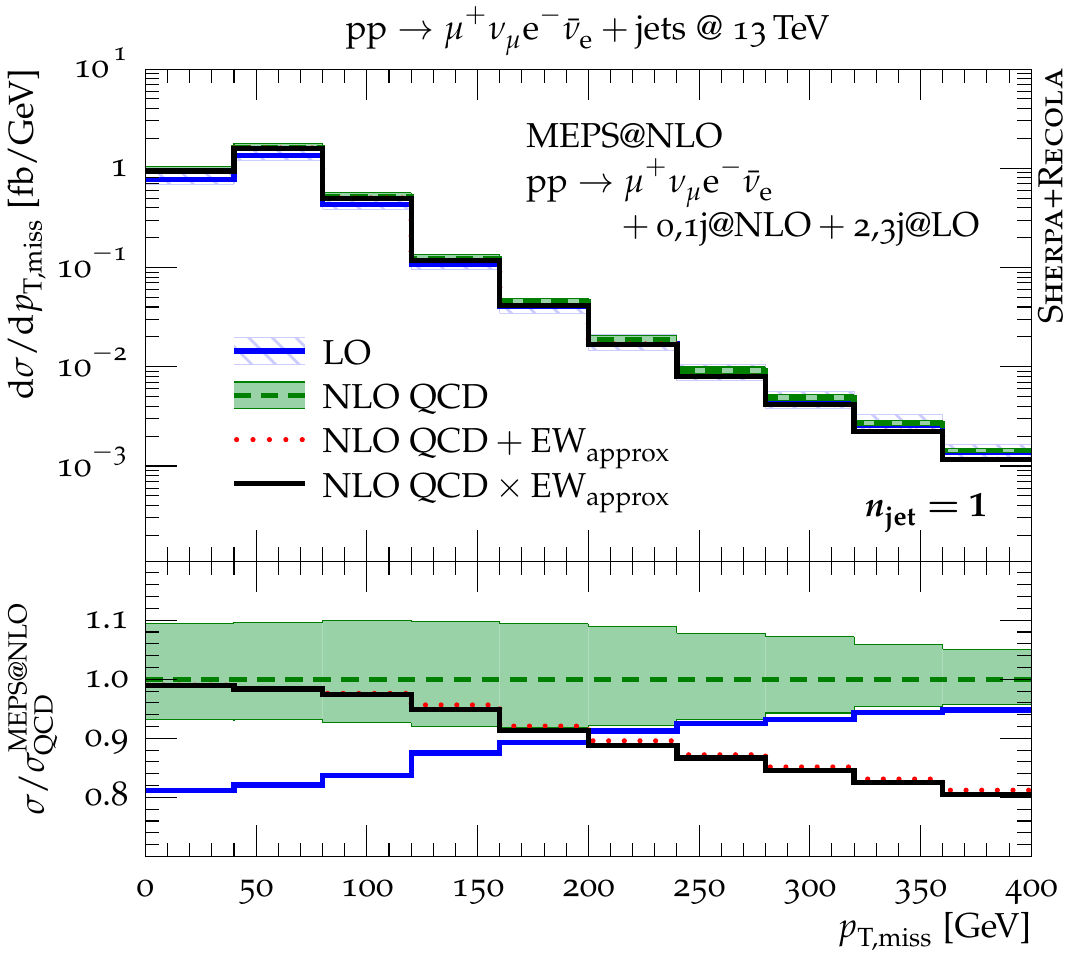}
  \caption{
    \label{fig:merged-1j-1}
    Predictions from multi-jet merged parton-shower simulations for the
    $n_\text{jet}=1$ event selection: 
    Transverse momentum of the jet (top left),
    rapidity of the jet (top right),  
    transverse momentum of the anti-muon (middle left),
    rapidity of the anti-muon (middle right),
    transverse momentum of the anti-muon--electron system (bottom left), and
    missing transverse momentum (bottom right).
    All results contain YFS soft-photon resummation.
    For the \MEPSatNLO calculation we present results including
    approximate NLO \EW\ corrections in the additive and multiplicative approach.
  }
\efig
\bfig
  \center
  \includegraphics[width=0.45\textwidth]{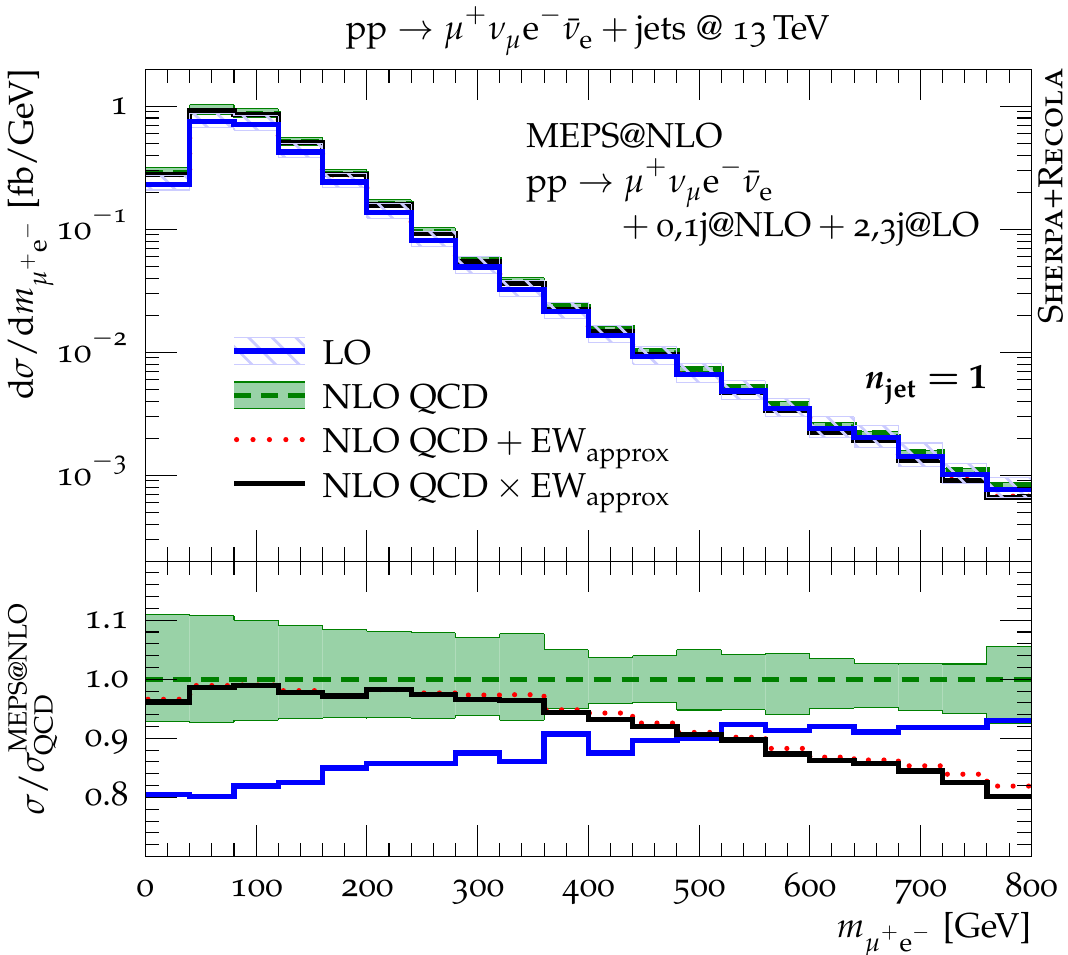}\hfill
  \includegraphics[width=0.45\textwidth]{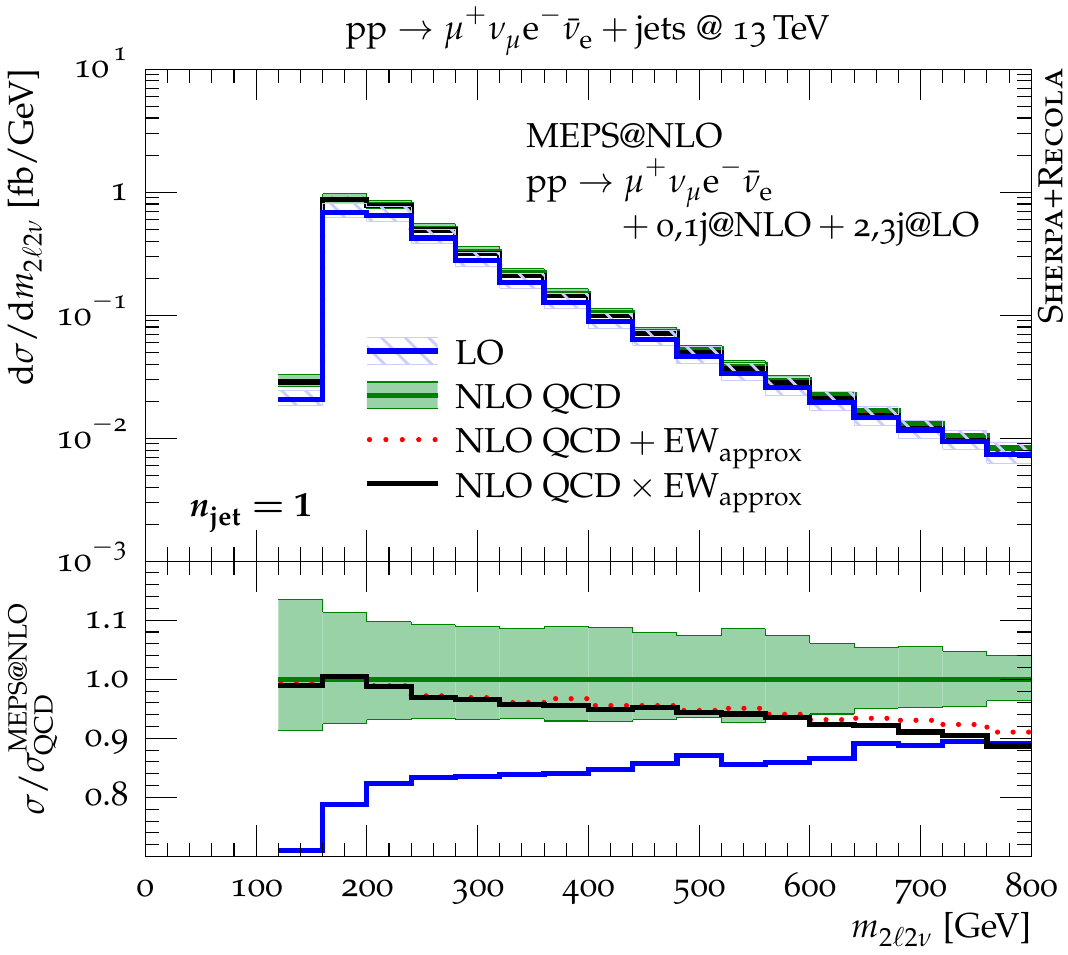}
  \includegraphics[width=0.45\textwidth]{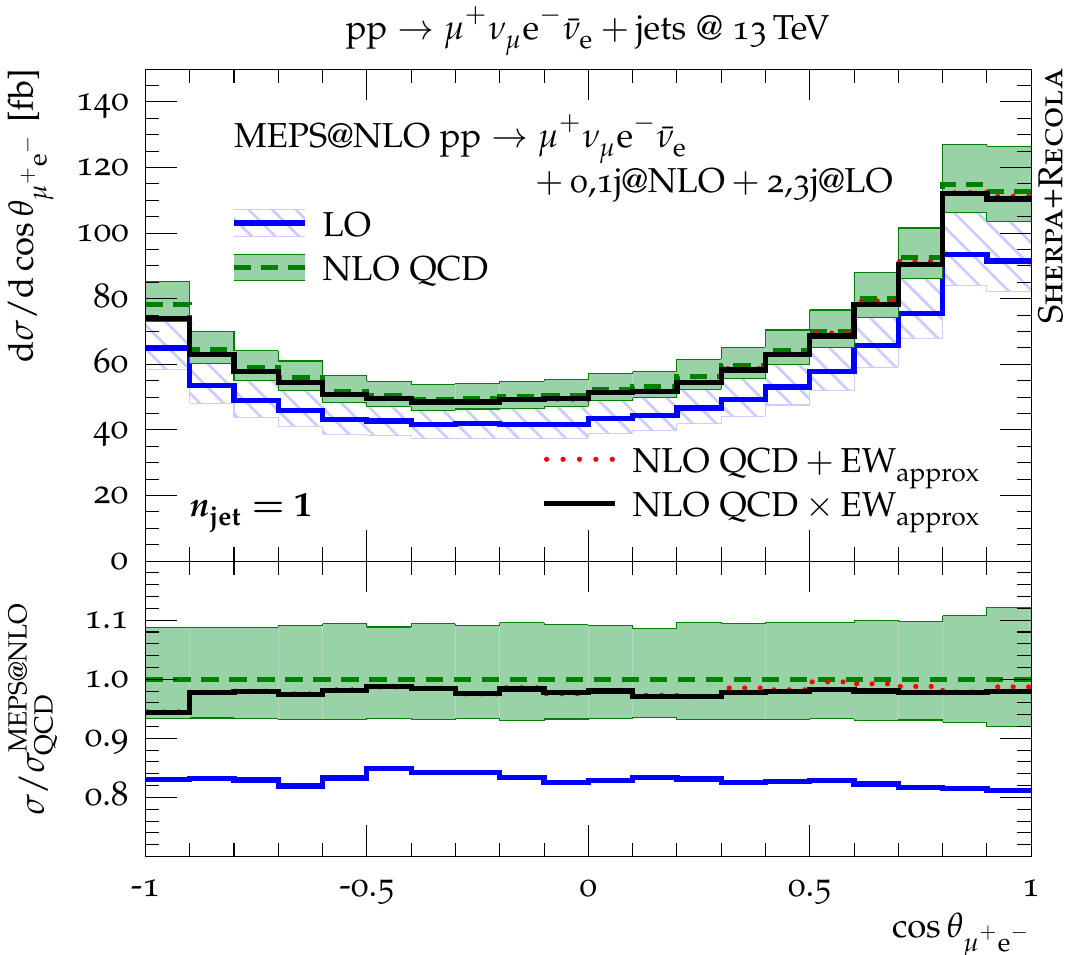}\hfill
  \includegraphics[width=0.45\textwidth]{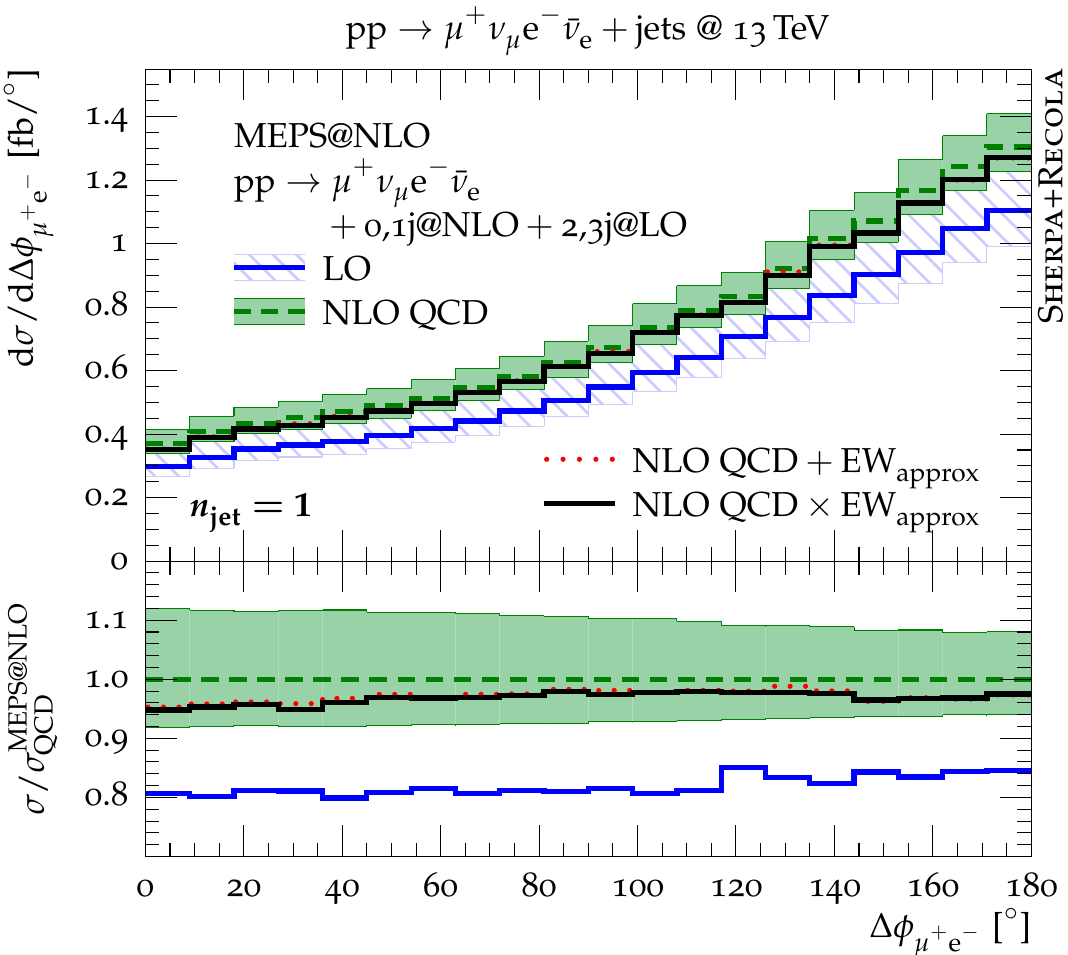}
  \caption{
    \label{fig:merged-1j-2}
    Predictions from multi-jet merged parton-shower simulations for the
    $n_\text{jet}=1$ event selection: 
    Invariant mass of the anti-muon and electron (top left),
    invariant mass of the four leptons (top right),
    cosine of the angle between the anti-muon and the electron (bottom left), and
    azimuthal-angle distance between the anti-muon and the electron (bottom right).
    All results contain YFS soft-photon resummation.
    For the \MEPSatNLO calculation we present results including
    approximate NLO \EW\ corrections in the additive and multiplicative approach.
  }
\efig
For the observables in the one-jet event selection, presented in
Figs.~\ref{fig:merged-1j-1} and \ref{fig:merged-1j-2}, similar conclusions
apply. The inclusion of the NLO \QCD\ corrections in the \MEPSatNLO calculations
increases the fiducial cross section by about $21\%$ with respect to \MEPSatLO,
cf.\ \refta{tab:xs-j1-merged}.
At the same time the systematic uncertainties get reduced by almost a
factor two.
In particular, for the transverse-momentum and invariant-mass distributions the
NLO \QCD\ corrections have significant impact on the distributions' shape, however,
much smaller than for the fixed-order evaluation of the
observables. This \change{smaller impact} is
caused by the inclusion of additional real-radiation processes through the
parton shower and the higher-multiplicity matrix elements, modelling in
particular the jet-veto process more reliably. In fact, for the jet transverse-momentum distribution,
the shape is only very mildly affected by the NLO \QCD\ corrections.

For the jet and anti-muon rapidity distribution, as well as the two angular
observables, approximative \EW\ corrections are of $1$--$2\%$ size only, well within the
\MEPSatNLO uncertainty bands, and essentially flat. For the $p_{\mathrm{T}}$-type and the
invariant-mass distributions sizeable \EW\ Sudakov-logarithmic suppression
effects are found, compatible with the observations for the fixed-order results
in \refse{se:results:fo:WWj}. 

The large deviations seen between the
NLO \QCDpEW\ and NLO \QCDtEW\ predictions at fixed order are not
present in the merged calculations. As in the case of $\PW\PW$ production,
this is due to the fact that the \MEPSatLO calculation incorporates already
a sizeable fraction of the \QCD\ corrections, and that the merged NLO
\QCDpEW\ predictions include \QCDtEW\ corrections.

\subsubsection{Ratios of \texorpdfstring{$\PW\PW$}{WW} and \texorpdfstring{$\PW\PW\Pj$}{WWj}}
\label{sec:results:merged:ratios}

\begin{table}
\begin{center}
\begin{tabular}{c|c|c|c}
  \MEPSatLO &\multicolumn{3}{c}{\MEPSatNLO}\\
 \rule[-2ex]{0ex}{3ex}%
  \QCD & \QCD & \QCDpEWapprox & \QCDtEWapprox\\
\hline  \rule{0ex}{3ex}%
0.388$^{+20.6\%}_{-13.9\%}$ & 0.408$^{+4.4\%}_{-3.2\%}$ & 0.405 & 0.405 \\
\end{tabular}
\end{center}
\caption{\label{tab:ratio_merged} Ratios of fiducial cross sections between
  $\Pp\Pp \to \mu^+ \nu_\mu \Pe^- \bar \nu_\Pe \Pj$ and 
  $\Pp\Pp\to\mu^+ \nu_\mu \Pe^- \bar \nu_\Pe$  at $\sqrt{s}=13\TeV$ 
  at \MEPSatLO, \MEPSatNLOQCD, \MEPSatNLOQCDpEWapprox and \MEPSatNLOQCDtEWapprox.}
\end{table}

Given the \MEPSatNLOQCD predictions with and without the inclusion of
approximate \EW\ NLO corrections for the exclusive zero- and one-jet event selections,
we can now proceed to study ratios of fiducial cross sections
and differential distributions. Corresponding fixed-order predictions have
been presented in \refse{se:results:fo:ratios}.

In Table~\ref{tab:ratio_merged} we compile the cross-section ratios between
$\Pp\Pp \to \mu^+ \nu_\mu \Pe^- \bar \nu_\Pe \Pj$ and
$\Pp\Pp \to \mu^+ \nu_\mu \Pe^- \bar \nu_\Pe$ for the  \MEPSatLO,
\MEPSatNLOQCD, \MEPSatNLOQCDpEWapprox, and \MEPSatNLOQCDtEWapprox calculations.
In particular for the \MEPSatLO predictions the ratio is significantly smaller
than at LO \QCD\ and closer to the NLO result, cf.~Table~\ref{tab:ratio}. 
As discussed before, pure LO calculations
do not address the applied jet vetoes, while in the \MEPSatLO approach these
as well as many higher-order contributions are addressed by the parton
shower off the respective Born process and
higher-multiplicity matrix elements. In contrast, the \MEPSatNLOQCD agrees
with its fixed-order equivalent within $1\%$. The inclusion of \EW\ corrections
in the  \MEPSatNLOQCDpEWapprox and \MEPSatNLOQCDtEWapprox approach amounts to
a reduction of the ratio by less than $1\%$, respectively, somewhat less than
at fixed order.

\bfig
  \center
  \includegraphics[width=0.45\textwidth]{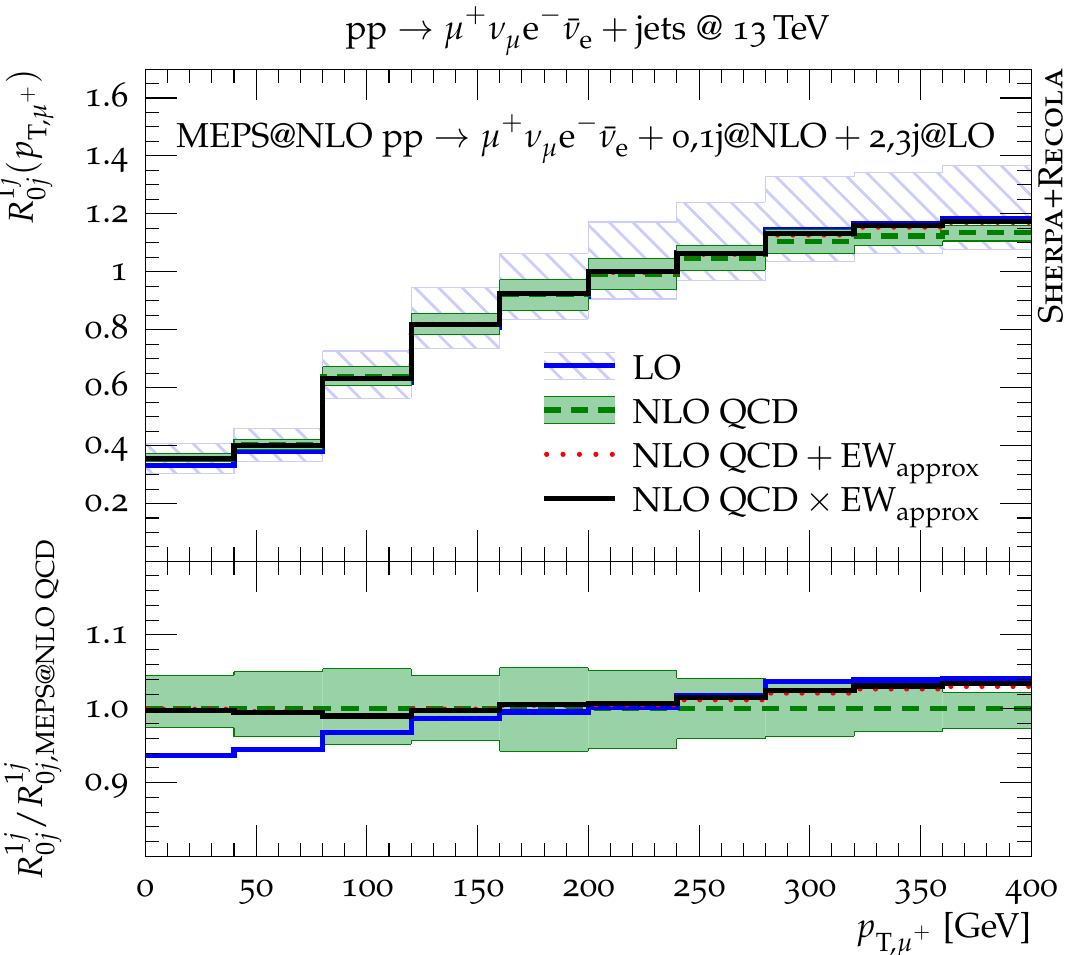}\hfill
  \includegraphics[width=0.45\textwidth]{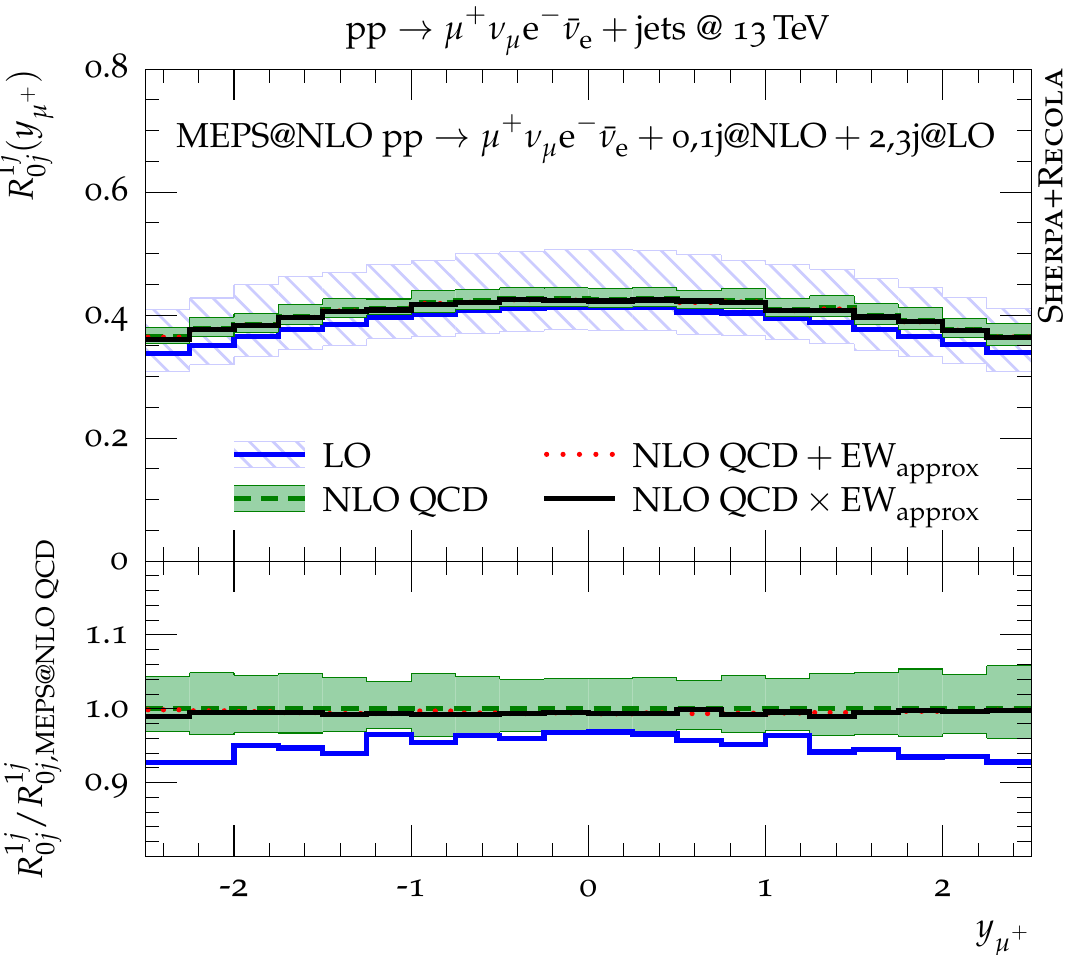}
  \includegraphics[width=0.45\textwidth]{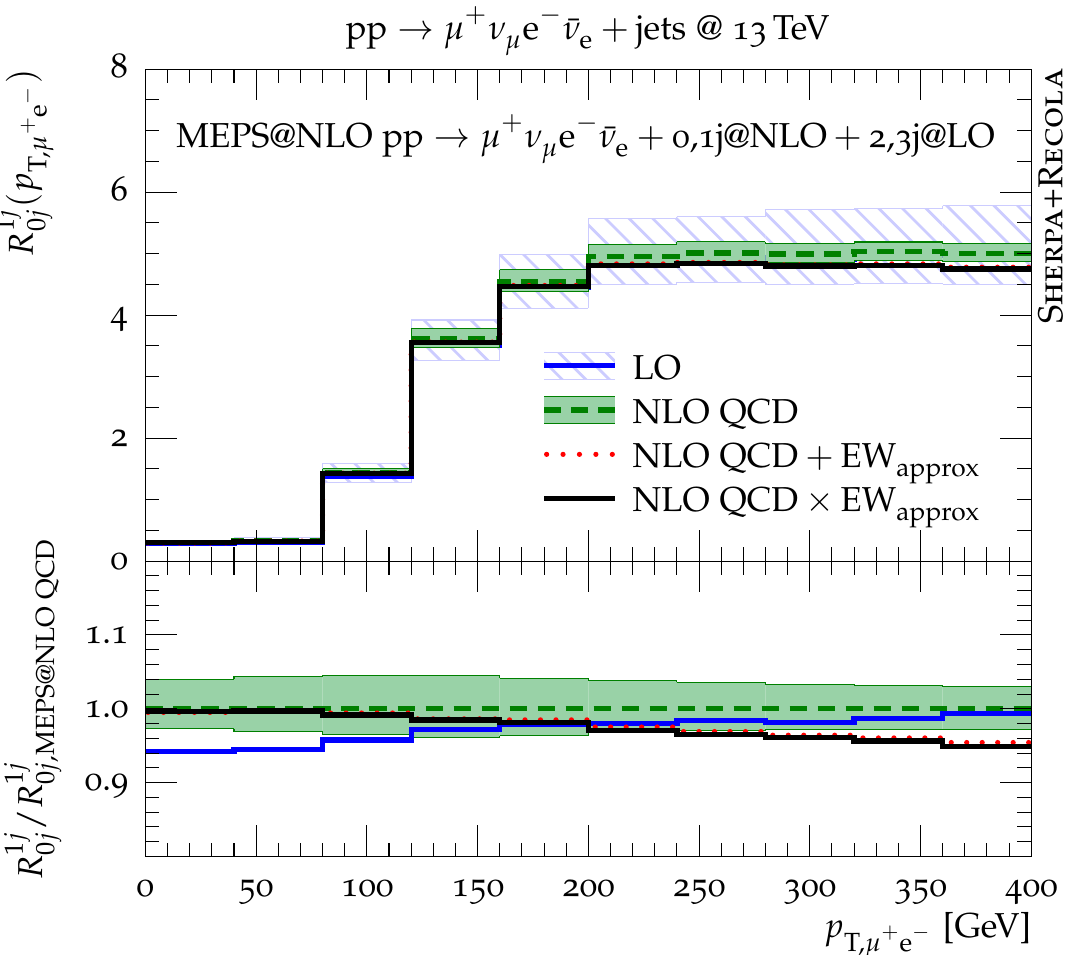}\hfill
  \includegraphics[width=0.45\textwidth]{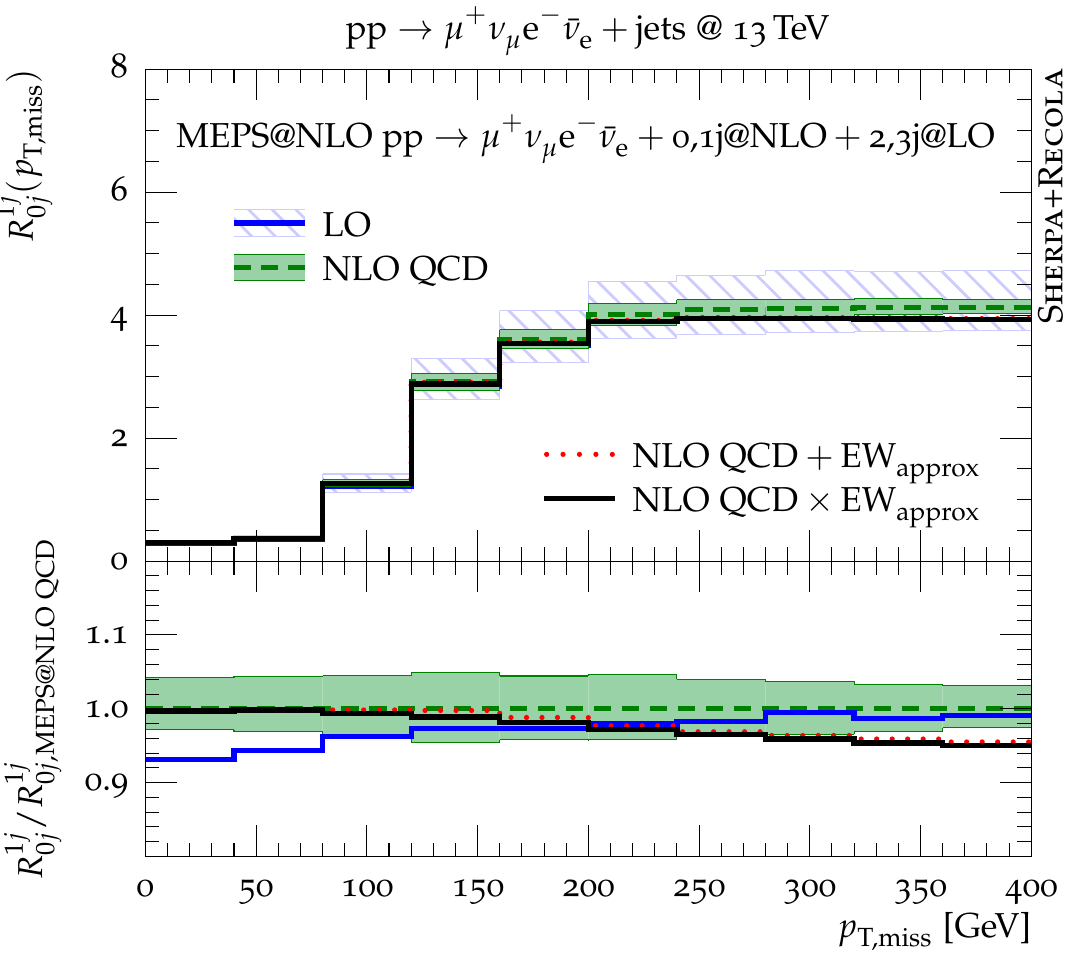}
  \caption{
    Predictions from multi-jet merged parton-shower simulations for the
    ratios between differential distributions in the one- and zero-jet event
    selection:
    Transverse momentum of the anti-muon (top left),
    rapidity of the anti-muon (top right),
    transverse momentum of the anti-muon--electron system (bottom left), and
    missing transverse momentum (bottom right).
    All results contain YFS soft-photon resummation.
    For the \MEPSatNLO calculation we present results including
    approximate NLO \EW\ corrections in the additive and multiplicative approach.
    \label{fig:merged_ratio_1}
 }
\efig
\bfig
  \center
  \includegraphics[width=0.45\textwidth]{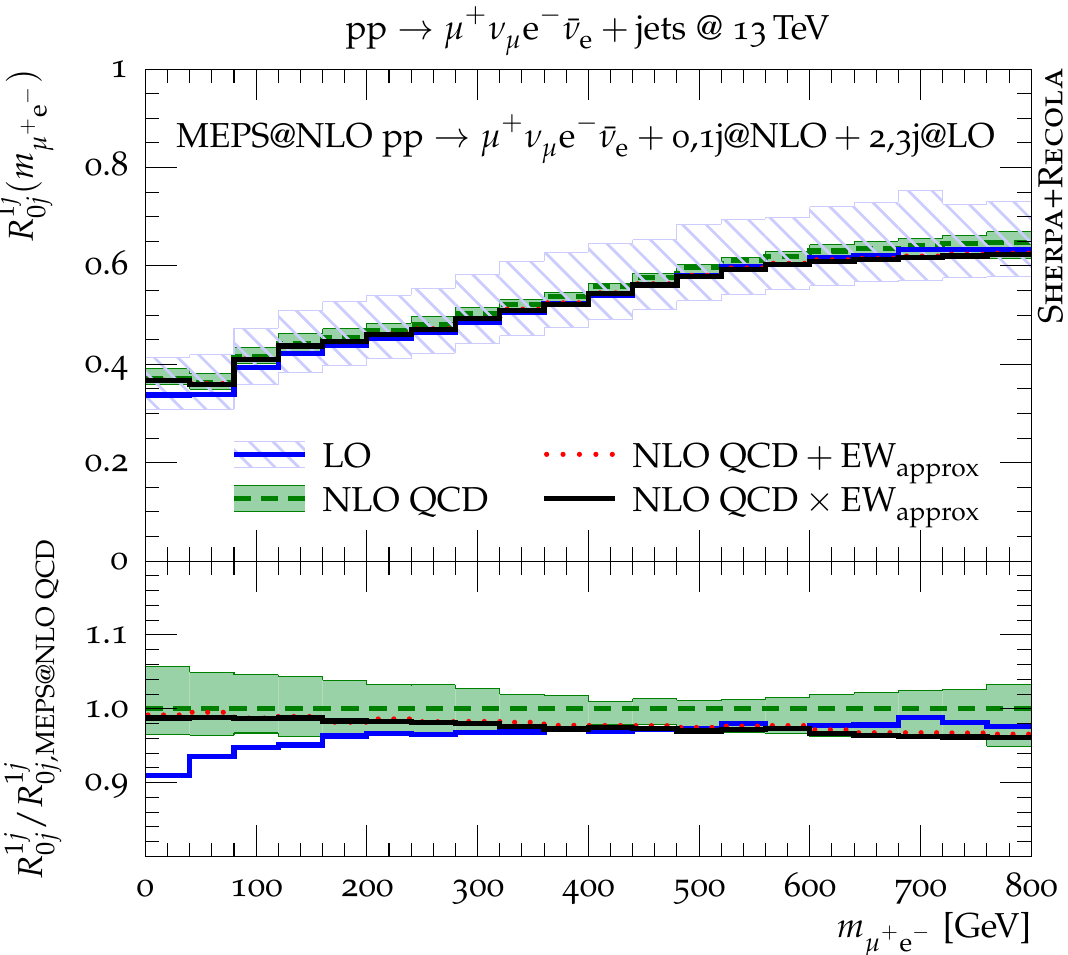}\hfill
  \includegraphics[width=0.45\textwidth]{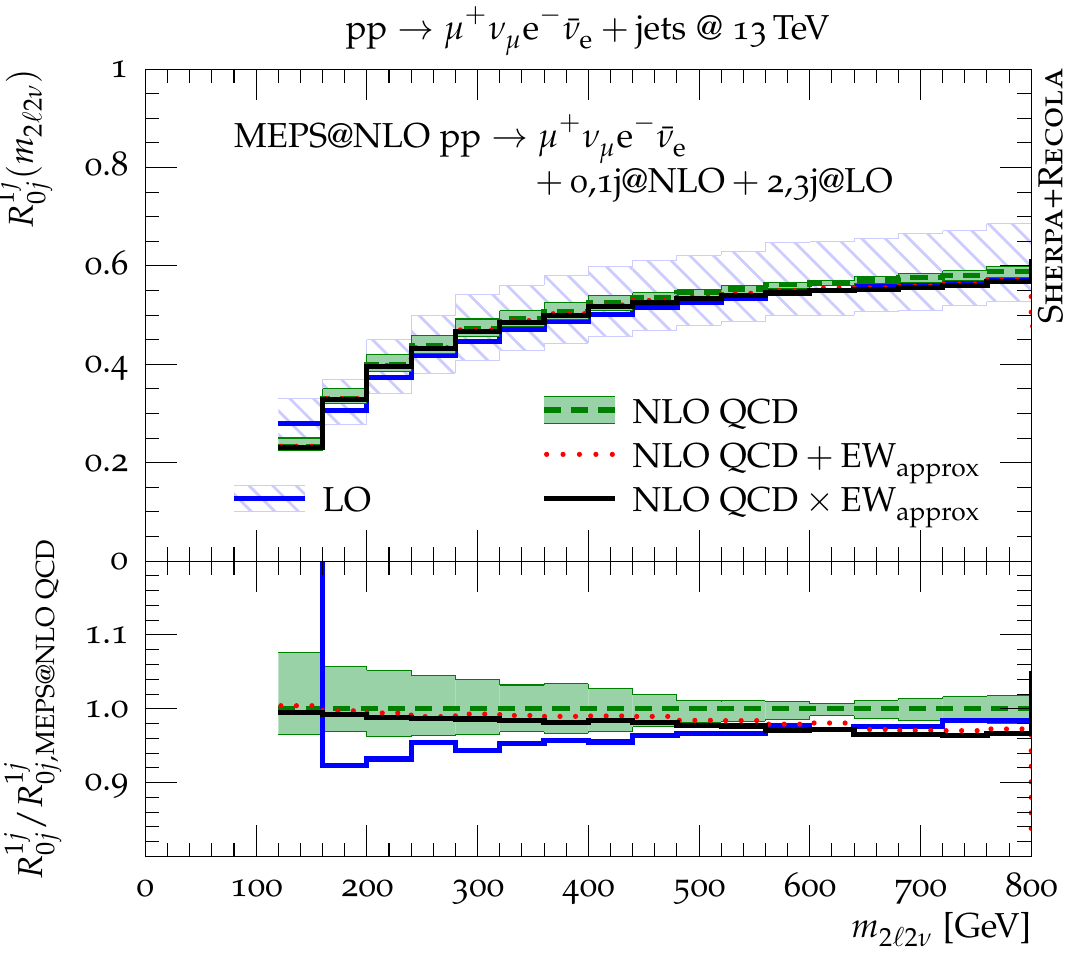}
  \includegraphics[width=0.45\textwidth]{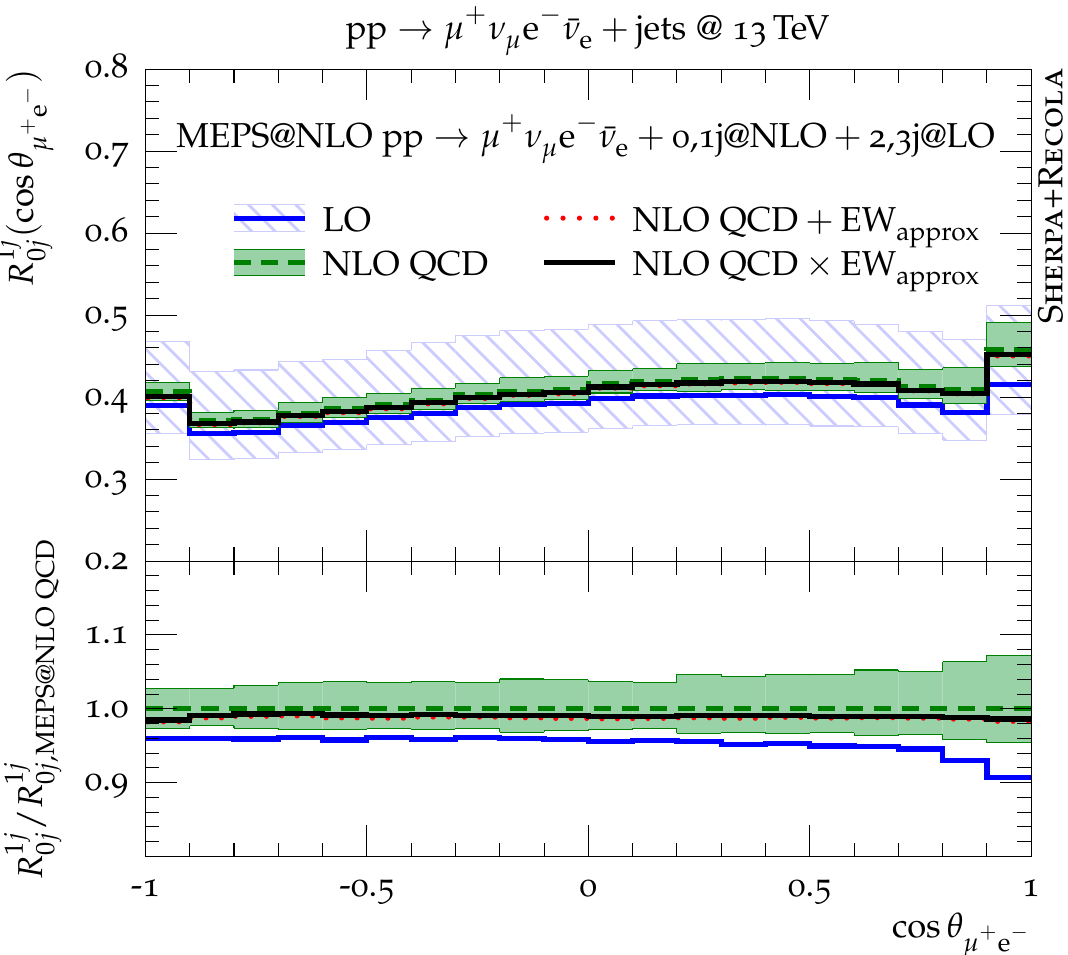}\hfill
  \includegraphics[width=0.45\textwidth]{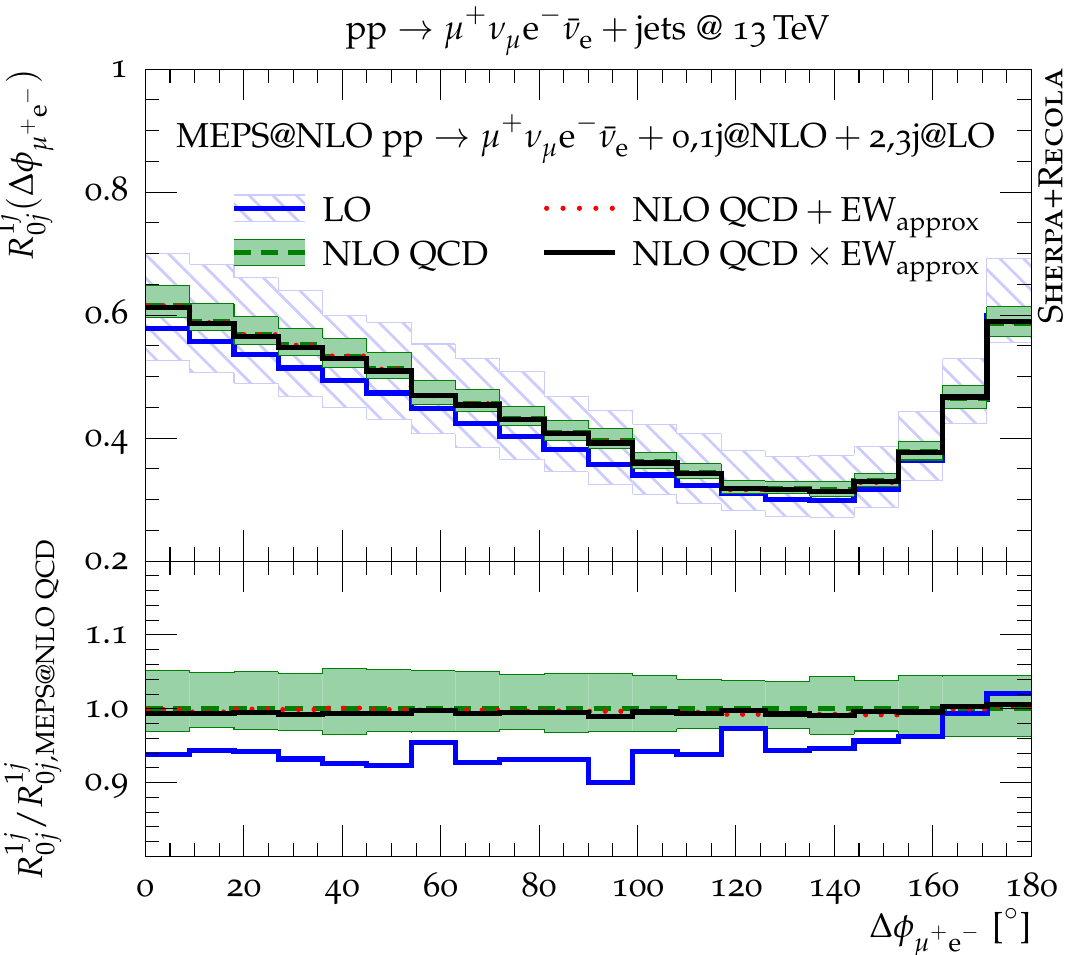}
  \caption{
    Predictions from multi-jet merged parton-shower simulations for the
    ratios between differential distributions in the one- and zero-jet event
    selection: 
    Invariant mass of the anti-muon and electron (top left),
    invariant mass of the four leptons (top right),
    cosine of the angle between the anti-muon and the electron (bottom left), and
    azimuthal-angle distance between the anti-muon and the electron (bottom right).
    All results contain YFS soft-photon resummation.
    For the \MEPSatNLO calculation we present results including
    approximate NLO \EW\ corrections in the additive and multiplicative approach.
    \label{fig:merged_ratio_2}
 }
\efig
In Figs.~\ref{fig:merged_ratio_1} and \ref{fig:merged_ratio_2} we present
ratios between differential distributions for the one- and zero-jet selection
as defined in Eq.~(\ref{eq:ratios}). What we observed for the ratios
of fiducial cross sections, is even more prominent in the kinematic distributions. When comparing
to Figs.~\ref{fig:transverse_ratio} and \ref{fig:invariant_ratio} we recognise
a dramatic difference between the LO \QCD\ and the \MEPSatLO prediction.
The huge NLO \QCD\ corrections observed before get reduced to at most $10\%$
for the merged results, with the exception of the low mass region in the $m_{2\ell2\nu}$
distribution.
For $m_{2\ell2\nu}\lsim2\MW\approx 161\GeV$ the production of
two resonant W~bosons is not possible, and the effect of
singly-resonant diagrams with different kinematics and a huge 
$K$~factor becomes relevant.
In this phase-space region the LO $\PW\PW$ cross section is stronger
suppressed than the cross sections with real-jet emission.
With the exception of this phase-space region, for all the
considered distributions the ratios are considerably stabilised by the
inclusion of the \QCD\ parton shower and higher-multiplicity matrix elements.

The associated (correlated) scale uncertainties get significantly reduced when going
from \MEPSatLO to \MEPSatNLOQCD. In particular for the phase-space regions of
large transverse momenta and large invariant masses these are much smaller
than for the fixed-order evaluations. Apart from the low-mass region in
$m_{2\ell2\nu}$ the \MEPSatNLOQCD uncertainties remain of order $\pm 5\%$.

The approximate NLO \EW\ corrections have
only mild impact on the ratios of differential distributions.  They largely cancel
between numerator and denominator and stay below $5\%$ also in the
tails of the transverse-momentum and invariant-mass distributions.
Furthermore, the \MEPSatNLOQCDpEWapprox
and \MEPSatNLOQCDtEWapprox predictions yield almost identical results,
in contrast to the NLO \QCDpEW\ fixed-order prediction. This is
basically a consequence of the reduced NLO \QCD\ corrections in the merged
calculations. Overall, apart from
the high-$p_{\mathrm{T}}$ and high-mass tails, the \EW\ corrections stay within the
\MEPSatNLOQCD uncertainty band. 

The results for the \EW\ corrections in the merged calculation should be
taken with some caution. First, only virtual \EW\ corrections are included exactly,
while real \EW\ corrections are integrated out in an approximated way. 
This approximation is expected to yield good results for observables 
in the \EW\ Sudakov regime where the kinematic invariants are large with respect to
the \PW-boson mass but not for inclusive observables such as the fiducial cross section. 
Second, the difference between \MEPSatNLOQCDpEWapprox
and \MEPSatNLOQCDtEWapprox predictions does not provide a reliable
error estimate on the missing \QCDtEW\ corrections since both
prescriptions are too close to each other.

\FloatBarrier

\section{Conclusion}
\label{se:conclusion}

This article provides a combined analysis of $\PW\PW$ and $\PW\PW\Pj$
production at higher order including effects of off-shell and
non-resonant contributions, emphasising the combination of \QCD\ and \EW\ 
corrections.  It is the first time that NLO \QCD\ and NLO \EW\ corrections
to both $\Pp\Pp \to \mu^+ \nu_\mu \Pe^- \bar\nu_\Pe$ and $\Pp\Pp \to
\mu^+ \nu_\mu \Pe^- \bar\nu_\Pe \Pj$ are presented together in
consistent set-ups.  Both processes are analysed while applying jet
vetoes in order to avoid large \QCD\ corrections.  It is worth noticing
that results including NLO \EW\ corrections for the off-shell
$\PW\PW\Pj$ production are presented here for the first time.
In addition to strictly fixed-order results, merged predictions
including different jet multiplicities and parton showers are
provided.  These are
combined in an approximate way with \EW\ corrections in the virtual \EW\
approximation \change{which describes the leading-logarithmic
  corrections in the Sudakov regime well}.  
All results have been obtained with the
\SherpaRecola framework, which is completely automated.
Though we did not
study their phenomenological impact in this article, non-perturbative
corrections to the parton-shower predictions due to hadronisation and
multiple-parton interactions can be easily invoked in the \Sherpa
framework.

After discussing numerical results at fixed order for $\PW$-pair production with and
without an extra jet separately, ratios of cross sections and
differential distributions are presented.  For fixed-order
calculations, our results \change{(in particular the study of
  differential cross-section ratios)} clearly support a
preference for the multiplicative combination of \QCD\ and \EW\ 
corrections as suggested by the structure of the enhanced \EW\ 
logarithmic corrections.  This further \change{reinforces} the use of the
approximative \EW\ corrections
in merged predictions as it relies on the assumption
that processes with different jet multiplicities have similar \EW\ 
corrections.
\change{Nonetheless it should be kept in mind that the virtual
  approximation catches the dominant corrections in the high-energy
  limit but does not improve predictions for inclusive observables.}
In addition to the predictions of the multi-jet merged sample, we
present comparisons of zero- and one-jet predictions between the 
fixed-order calculations and the merged ones using two different scale
choices.
These comparisons emphasise the benefits of the calculation based on 
multi-jet merging  which do not suffer from some of the limitations 
of the fixed-order calculations.
In the end, our study shows that the merged calculations provide more 
stable predictions, in particular regarding ratios of cross sections
and distributions for $\PW\PW\Pj$ versus $\PW\PW$ production.

Finally, the results presented here are particularly relevant for the
experimental measurements at Run~2 and the upcoming high-luminosity phase
of the LHC. 
We hope that these (as well as the corresponding tools) will be fully
exploited by both the ATLAS and CMS collaborations. 

\subsubsection*{Note added}
The same day the present article appeared on \url{arXiv.org}, an independent study on matching NLO corrections to the parton shower in diboson production was made public \cite{Chiesa:2020ttl}.

\section*{Acknowledgements}

We would like to thank Jean-Nicolas Lang for supporting and
continuously improving \Recola.
SB, SS, and MS received funding from
the European Union's Horizon 2020 research and innovation programme as
part of the Marie Sk\l{}odowska-Curie Innovative Training Network
MCnetITN3 (grant agreement no.~722104).
AD acknowledges financial support by the German
Federal Ministry for Education and Research (BMBF) under contract
no.~05H18WWCA1.
The research of MP has received
funding from the European Research Council (ERC) under the European
Union's Horizon 2020 Research and Innovation Programme (grant
agreement no.~683211).
MS is funded by the Royal Society through a
University Research Fellowship. 
SS acknowledges support
through the Fulbright-Cottrell Award and from BMBF (contract
no.~05H18MGCA1).

\appendix

\section{Fragmentation function}

This appendix is devoted to estimate the numerical impact of a proper
IR-safe photon recombination.  To that end, fragmentation
functions have been implemented in \MoCaNLO following
\citeres{Denner:2009gj,Denner:2011vu,Denner:2014ina}.  The
implementation has been validated against the code used in
\citere{Denner:2014ina} for the computation of \EW\ corrections for
$\Pp\Pp\to\ell^+\ell^-\Pj\Pj$.  The photon--jet energy fraction
\begin{equation}
 z_{\gamma} = \frac{E_\gamma}{E_\gamma+E_a}
\end{equation}
has been taken to be equal to $0.7$, where $E_\gamma$ and $E_a$
denote the energies of the photon and a \QCD\ parton, respectively.
The fit parameters entering the fragmentation function are the ones 
obtained from \citere{Buskulic:1995au} and read
\begin{equation}
\mu_0 = 0.14 \GeV \qquad {\rm and} \qquad C = -13.26 \;.
\end{equation}

In this simulation, the LO only includes \QCD\ partons in the initial
state, \ie contributions with initial-state photons  are
omitted.  Concerning the \EW\ corrections, only photon radiation and \EW\ 
virtual corrections are considered, while interference contributions
in the real radiation from matrix elements at different orders in the
couplings are not taken into account.

In Table~\ref{tab:ff}, NLO \EW\ cross sections with consistent inclusion
of photon--jet separation (cons.) and in the simplified set-up of
\refse{se:setup} (simp.) are given.  The difference between these two
prescriptions is about a per mille only.  
\begin{table}
\begin{center}
\begin{tabular}{c|c|c|c}\rule[-2ex]{0ex}{3ex}%
$\sigma^{\mathrm{LO}}$ [$\fb$] & 
$\sigma^{\mathrm{NLO}}_{\mathrm{EW,cons.}}$ [$\fb$] & 
$\sigma^{\mathrm{NLO}}_{\mathrm{EW,simp.}}$ [$\fb$] & 
$\sigma^{\mathrm{NLO}}_{\mathrm{EW,simp.}}/\sigma^{\mathrm{NLO}}_{\mathrm{EW,cons.}}-1$ [$\%$] \\
\hline \rule[-2ex]{0ex}{5ex}%
$162.545(3)$ & $155.696(5)$ & $155.883(5)$ & $0.12$ \\
\end{tabular}
\end{center}
\caption{\label{tab:ff} Fiducial cross sections for $\Pp\Pp \to \mu^+
  \nu_\mu \Pe^- \bar \nu_\Pe\Pj$ at $\sqrt{s}=13\TeV$ at LO
  and NLO \EW\ with (cons.) and without (simp.) proper photon--jet separation.
  In addition the percentage difference between the latter results is shown.}
\end{table}
In addition, in
Fig.~\ref{fig:ff} two differential distributions are shown in both
set-ups: the distribution in the transverse momentum of the hardest
jet (left) and the distribution in the invariant mass of the
anti-muon--electron system (right).  
\bfig
  \center
  \includegraphics[width=0.49\textwidth]{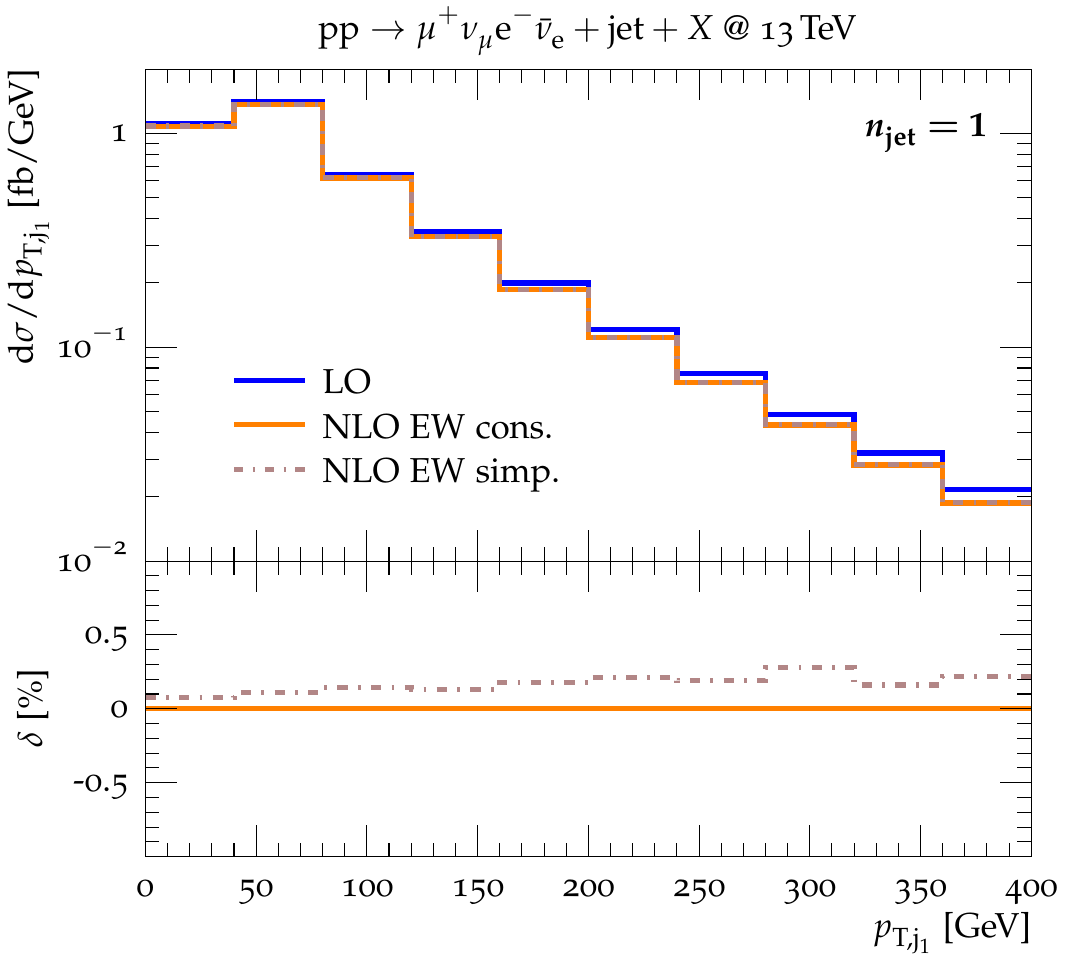}\hfill
  \includegraphics[width=0.49\textwidth]{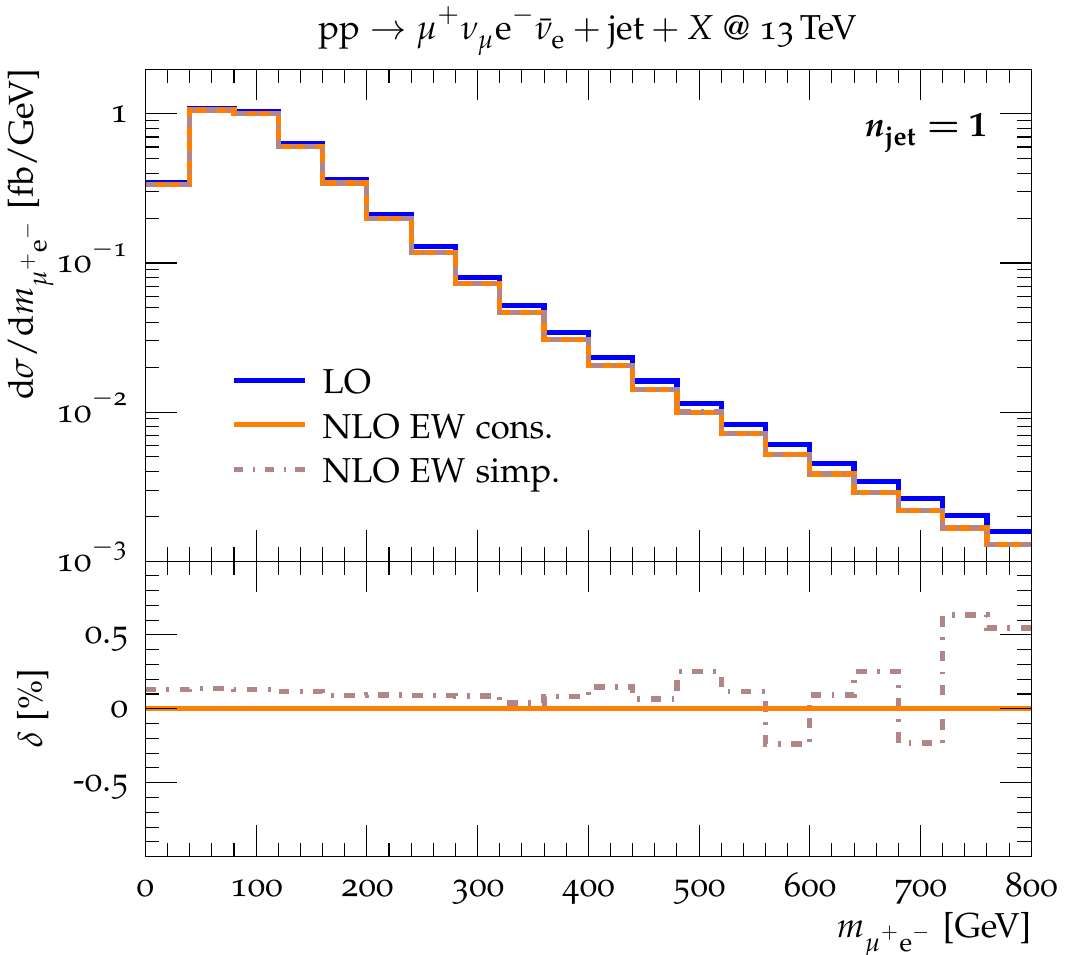}
  \caption{\label{fig:ff} Differential distributions for $\Pp\Pp \to \mu^+ \nu_\mu \Pe^- \bar \nu_\Pe\Pj$ at $\sqrt{s}=13\TeV$ at LO and NLO \EW:
    transverse momentum of the hardest jet (left) and invariant mass
    of the anti-muon--electron system (right).  The NLO \EW\ predictions
    are obtained with (cons.) and without (simp.) proper photon--jet
    separation. In the lower panels, the relative difference between
    the NLO \EW\ predictions without and with proper photon--jet
    separation in per cent is shown.}  
\efig

For large transverse momentum or
large invariant mass, the shape of the ratio of the distributions in
both methods is dominated by the Monte Carlo statistical error.
Disregarding these fluctuations, one notices differences of a few per
mille for the transverse-momentum distribution and of only about one
per mille for the invariant-mass distribution.  This reduced effect
can be explained by the fact that leptonic observables are only
indirectly sensitive to effects from photon--jet separation.  Overall,
this analysis indicates that the effect of a consistent treatment of
photon--jet separation is rather small for our calculational set-up.
This justifies the simplified approach that we have taken.

Moreover, we investigated the dependence of the simplified approach on the
technical cuts used in the \MoCaNLO generator. We did not observe a
dependence beyond the per-mille level for reasonable parameter values.

\FloatBarrier

\bibliographystyle{utphys.bst}
\bibliography{wwj}
\end{document}